\documentclass[aps,prd,superscriptaddress,showkeys,showpacs,twocolumn,nofootinbib,notitlepage,10pt]{revtex4-2}
\usepackage{hyperref}
\usepackage{amsmath,amssymb,slashed}
\usepackage{graphicx}
\usepackage{bm} % puts greek and math symbols in boldface using \bm
\usepackage{color} % {\color{red} ... }
\usepackage{tensor} %\indices{^a_b} produces the properly spaced the tensor indices 
\usepackage{bm,bbold}
\usepackage{orcidlink}
\usepackage[normalem]{ulem}
\usetikzlibrary{arrows.meta, calc, patterns}
\newcommand{\D}{{\rm d}}
\newcommand{\I}{{\rm i}}
\newcommand{\E}{{\rm e}}
\newcommand{\de}{\partial}
\DeclareMathOperator*{\sumint}{%								%sum and integral symbol
\mathchoice%
  {\ooalign{$\displaystyle\sum$\cr\hidewidth$\displaystyle\int$\hidewidth\cr}}
  {\ooalign{\raisebox{.14\height}{\scalebox{.7}{$\textstyle\sum$}}\cr\hidewidth$\textstyle\int$\hidewidth\cr}}
  {\ooalign{\raisebox{.2\height}{\scalebox{.6}{$\scriptstyle\sum$}}\cr$\scriptstyle\int$\cr}}
  {\ooalign{\raisebox{.2\height}{\scalebox{.6}{$\scriptstyle\sum$}}\cr$\scriptstyle\int$\cr}}
}

\definecolor{purple}{rgb}{0.8,0,0.6}
\definecolor{PURPLE}{rgb}{0.8,0,0.6}
\definecolor{orange}{rgb}{1,0.55,0}

\begin{document}
%%%%    TITLE
\title{Rotating synchrotron radiation: Photon emission from magnetized and rotating quark-gluon plasma}
%%%%    AUTHORS
\author{Matteo Buzzegoli\orcidlink{0000-0002-2114-5431}}\email{matteo.buzzegoli@e-uvt.ro}
\affiliation{Department of Physics, West University of Timișoara, Bulevardul Vasile Pârvan 4, Timișoara 300223, Romania}

\author{Sergiu Busuioc\orcidlink{0000-0003-1437-0988}}\email{sergiu.busuioc@e-uvt.ro}
\affiliation{Department of Physics, West University of Timișoara, Bulevardul Vasile Pârvan 4, Timișoara 300223, Romania}

\author{Jonathan D. Kroth\orcidlink{0009-0002-2040-9965}}\affiliation{Department of Physics and Astronomy, Iowa State University, Ames, Iowa 50011, USA}

\author{Nandagopal Vijayakumar\orcidlink{0000-0002-1225-1267}}\affiliation{Department of Physics and Astronomy, Iowa State University, Ames, Iowa 50011, USA}

\author{Kirill Tuchin\orcidlink{0000-0002-6850-7698}}\affiliation{Department of Physics and Astronomy, Iowa State University, Ames, Iowa 50011, USA}

%%%%    ABSTRACT
\begin{abstract}
This paper investigates the production of nonprompt photons originating from rotating synchrotron radiation (RoSyRa), specifically the emission of photons by a rigidly rotating quark-gluon plasma in thermal equilibrium, in the presence of an external magnetic field. We compute the nonprompt photon spectrum and its elliptic flow ($v_2$) at midrapidity. In particular, we investigate the finite-volume effects. We find that at low transverse momentum, the magnetic field induces a significant $v_2$, while the plasma rotation boosts the synchrotron radiation of negatively charged quarks. These findings make RoSyRa a viable candidate mechanism to resolve the ``direct photon puzzle.''
\end{abstract}

\maketitle

\section{Introduction}
The experimental program of relativistic heavy-ion collisions at facilities such as the Relativistic Heavy Ion Collider (RHIC) and the Large Hadron Collider (LHC) has provided compelling evidence for the creation of a hot and dense state of matter known as the quark-gluon plasma (QGP) \cite{Bazavov:2011nk, Borsanyi:2013bia, BRAHMS:2004adc, PHENIX:2004vcz, STAR:2005gfr}. Understanding the properties and space-time evolution of this deconfined medium is a central goal of nuclear physics. Among the various probes available, electromagnetic observables, namely photons and leptons, hold a unique position \cite{Shuryak:1978ij, David:2019wpt, Gale:2021emg, Paquet:2015lta}. Unlike strongly interacting hadrons, photons and leptons experience minimal final-state interactions as they traverse the medium, thus carrying direct, undistorted information from their point and time of emission to the detectors \cite{Shuryak:1978ij,McLerranToimela1985, KapustaLichardSeibert1991,  SrivastavaSinha1994, TurbideRappGale2004,ChatterjeeEtAl2006,Gale:2021emg, Paquet:2015lta, David:2019wpt}. This makes them invaluable tools for tomographically probing the entire space-time history of the collision, from initial hard scattering to the final decoupling stage \cite{Shuryak:1978ij, Gale:2021emg, Paquet:2015lta}.

Direct photons, defined as photons not originating from the decay of hadrons, are particularly sensitive to the thermal properties and collective motion of the QGP \cite{David:2019wpt, Shuryak:1978ij}. They provide insights into the effective temperature of the medium and the dynamic development of collectivity throughout the collision's evolution \cite{Shen:2013vja, David:2019wpt}. Despite their diagnostic potential, direct photon measurements pose a persistent challenge to standard theoretical models, often referred to as the direct photon puzzle \cite{David:2019wpt, Gale:2021emg, Paquet:2015lta}. A key aspect of this puzzle is the unexpectedly large number of photons at low transverse momentum ($k_T$) and the surprisingly large azimuthal anisotropy, quantified by the second Fourier coefficient $v_2$ (elliptic flow), observed for direct photons \cite{PHENIX_yields_2014, PHENIX_yields_2022, PHENIX_v2_2012, PHENIX_v2_2016}, which significantly deviates from theoretical expectations. Measurements by the PHENIX Collaboration at RHIC showed a pronounced positive $v_2$ signal for direct photons at low $k_T$, which was comparable in magnitude to the flow of pions \cite{PHENIX_v2_2012, PHENIX_v2_2016, PHENIX_vn_2016}. This magnitude was significantly underestimated by conventional hydrodynamic and transport models, which typically predict a much smaller anisotropy for thermal photons \cite{Shen:2013cca, Chatterjee:2013naa, vanHees:2011vb, Dion:2011pp, Linnyk:2013wma, Monnai:2014kqa, ALICE:2018dti}. Although ALICE Collaboration measurements at the LHC generally supported this conclusion, their larger systematic uncertainties prevented a definitive confirmation \cite{ALICE:2018dti}.

One promising avenue to resolve the direct photon puzzle involves considering the role of the extremely strong magnetic fields generated in noncentral heavy-ion collisions \cite{Skokov:2009qp, Deng:2012pc, Basar:2012bp, Muller:2013ila, Tuchin:2012mf, Tuchin:2014pka, Yee:2013qma, Wang:2020dsr, Sun:2024vsz}. These transient electromagnetic fields, which can reach strengths of the order of $eB\sim m_\pi^2\sim 10^{18}$ G or even $10^{20}$ G at RHIC and LHC energies respectively \cite{Skokov:2009qp, Deng:2012pc, Tuchin:2013apa, Bloczynski:2012en}, can induce a significant azimuthal anisotropy in photon emission \cite{Basar:2012bp, Muller:2013ila, Tuchin:2012mf, Tuchin:2014pka, Yee:2013qma, Wang:2020dsr, Sun:2024vsz, Zakharov:2016mmc, Zakharov:2016kte}. Theoretical proposals suggest that these magnetic fields could lead to a large direct photon $v_2$, potentially explaining the experimental observations \cite{Muller:2013ila, Wang:2020dsr, Sun:2024vsz}. However, the abundance of photons produced via synchrotron radiation is significantly lower than that observed experimentally, and the magnetic field alone cannot explain the observed yield excess or the magnitude of $v_2$. Recently, it has been shown that the distortion of the quark distribution function caused by a weak magnetic field can also significantly affect the ellipticity of photon emission~\cite{Sun:2023pil}. Other studies have sought to investigate the effect of anomalous current couplings on synchrotron radiation \cite{Wang:2024gnh,Kroth:2026kgm}.

Furthermore, observations in heavy-ion collisions indicate that the created QGP exhibits substantial vorticity \cite{STAR:2017ckg, STAR:2018gyt, ALICE:2019aid, Becattini:2024uha, Niida:2024ntm}, with magnitudes comparable to the synchrotron frequency and aligned with the magnetic field direction \cite{Buzzegoli:2023vne}. This inherent rotation of the medium further necessitates a reevaluation of photon production mechanisms.

In this context, synchrotron radiation, which is the electromagnetic radiation emitted by charged particles moving in a magnetic field, emerges as a crucial process \cite{Sokolov_Ternov_1986, Bordovitsyn_1999, Buzzegoli:2023vne}. Our previous work has explored how the presence of rotation in the plasma can profoundly influence this radiation,\footnote{Similarly, the dilepton production in a rotating medium was recently studied in~\cite{Castano-Yepes:2025zae}.} leading to either enhancement or suppression of the emission intensity depending on the relative orientation of the angular velocity and the magnetic field \cite{Buzzegoli:2022dhw, Buzzegoli:2023vne, Buzzegoli:2023yut, Buzzegoli:2024nzd,Buzzegoli:2025qfl, Kroth:2024mfd}. At leading order in the strong coupling constant, photon emission from a strongly magnetized quark-gluon plasma occurs through processes such as quark and antiquark splitting and annihilation, which are otherwise kinematically forbidden in the absence of a magnetic field \cite{Wang:2020dsr, Wang:2021ebh}. These magnetic field-mediated mechanisms can contribute significantly to the observed direct photon $v_2$ \cite{Wang:2020dsr, Sun:2024vsz}. Theoretical calculations suggest that the ellipticity coefficient $v_2$ resulting from these processes can reach a moderately high positive value, approaching approximately 0.2 at large transverse momenta, which aligns qualitatively with experimental findings \cite{Wang:2020dsr}.

This paper builds upon these advancements to compute the nonprompt photon spectrum and its elliptic flow $v_2$ originating from synchrotron radiation in a rigidly rotating noninteracting plasma. We utilize the detailed theoretical framework previously developed \cite{Buzzegoli:2022dhw,Buzzegoli:2023vne} to quantify this particular contribution to the direct photon spectrum and its azimuthal anisotropy. By focusing on this ``exotic'' production mechanism in a highly vortical and magnetized QGP, our work aims to provide further insights into the long-standing direct photon puzzle and advance the multimessenger (MM) approach to understanding relativistic heavy-ion collisions. We employ the natural units $c=\hbar=k_B=1$.

%====================================================================
\section{Nonprompt direct photons and their anisotropy}\label{sec:DirectPhotons}
%====================================================================
In heavy-ion collisions, the measurements of direct photons are reported in terms of the transverse momentum spectra  $\D N^\gamma$. Denoting with $\omega$ the photon energy, $\bm{k}$ its momentum, $k_T$ the transverse momentum, $\phi$ the azimuthal angle, $y$ the rapidity, and $\Psi_{\rm RP}$ the reaction plane, the photon spectrum is decomposed via a Fourier expansion in the azimuthal angle:
\begin{equation}
\label{eq:FourierDecomp}
\omega\frac{\D^3N^\gamma}{\D^3\bm{k}} = \left\langle\frac{\D^3N^\gamma}{\D^2 k_T\D y} \right\rangle
    \left(1 + 2\sum_{n=1}^\infty v_n \cos\left[n\left(\phi-\Psi_{\rm RP}\right) \right] \right),
\end{equation}
where $v_n$ are the corresponding anisotropic flow coefficients; the first two coefficients $v_1$ and $v_2$ characterize the directed and elliptic flow, respectively. The experimental collaborations report the integrated yields $\left\langle \D^3N^\gamma/(\D^2 k_T\D y) \right\rangle$, i.e.\ photon spectra, and the flow coefficient $v_2$ at midrapidity, $y=0$, with respect to the photon transverse momentum $k_T$.
We obtain these quantities from their definition by computing the differential yields at midrapidity and integrating over the azimuthal angle as follows:
\begin{align}
&\left\langle\frac{\D^3N^\gamma}{\D^2 k_T\D y} \right\rangle =
    \frac{1}{2\pi}\int_0^{2\pi} \frac{\D^3N^\gamma}{\D^2 k_T\D y}\D\phi,\\
&v_2 = \frac{1}{\left\langle\frac{\D^3N^\gamma}{\D^2 k_T\D y} \right\rangle}\frac{1}{2\pi}
    \int_0^{2\pi} \frac{\D^3N^\gamma}{\D^2 k_T\D y}\cos(2\phi)\D\phi.
\end{align}

\begin{figure}
    \centering
    \includegraphics[width=0.95\linewidth]{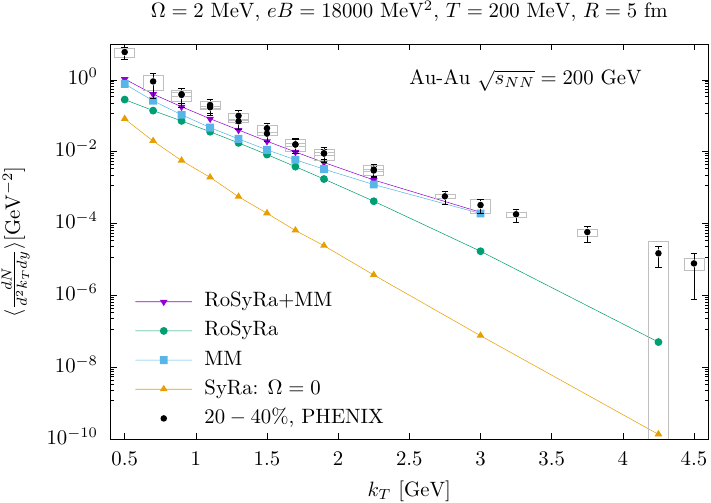}
    \includegraphics[width=0.95\linewidth]{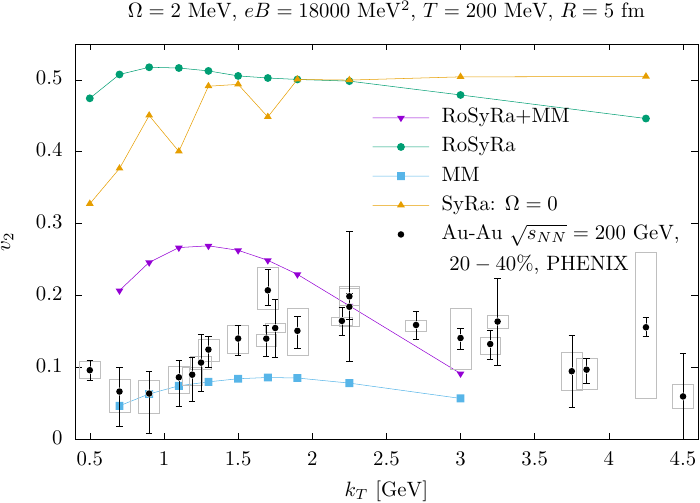}
    \caption{Top: photon spectrum, bottom: elliptic flow $v_2$,  for $eB=18000$ MeV$^2$, $\Omega=2$ MeV, $T=200$ MeV,  $R=5$ fm, two quark flavors, the cylinder height is $L=10$ fm and the QGP lifetime  $c\Delta t = 10$ fm. The data are adapted from \cite{PHENIX_yields_2014, PHENIX_yields_2022, PHENIX_v2_2012, PHENIX_v2_2016}, the multimessenger (MM) predictions are given in~\cite{Gale:2021emg}. }
    \label{fig:best_fit}
\end{figure}

State-of-the-art predictions of the nonprompt direct photon spectrum use a comprehensive multistage framework, known as multimessenger model \cite{Gale:2021emg}, that describes the entire space-time evolution of relativistic heavy-ion collisions, integrating, in addition to the thermal photons~\cite{Arnold:2001ms}, the initial state, preequilibrium dynamics, viscous fluid evolution, and a hadronic afterburner, but neglect the contribution from synchrotron radiation.
In this work, we obtain the photon spectrum resulting from synchrotron radiation by quarks in a uniformly rotating QGP, in the presence of a constant and homogeneous magnetic field $\bm B$ perpendicular to the reaction plane. The plasma is modeled as a cylindrical volume with radius $R$ and height $L$, with its symmetry axis parallel to the magnetic field. We assume that the plasma's temperature $T$ is constant and the angular velocity $\bm\Omega$ is aligned with the magnetic field $\bm B$. We call the radiation resulting from the combined effect of magnetic field and rotation ``Rotating Synchrotron Radiation'' (RoSyRa).

The anisotropy of synchrotron radiation is generated because quarks rotate in the plane perpendicular to the magnetic field, which is the reaction plane, and photons are preferentially emitted along the instantaneous velocity of the parent quark.  The enhancement of photon production due to rotation can be interpreted through the following classical analogy. The presence of rotation influences synchrotron radiation by superimposing the particle's circular motion due to the Lorentz force $q e\bm v\times \bm B$, where $qe$ is the quark's charge, with its rigid rotation~\cite{Buzzegoli:2023vne}, resulting in an effective synchrotron frequency $\omega_{\rm Eff}$. If the angular velocity $\bm\Omega$ and $q e\bm B$ point in opposite directions, the two motions combine constructively, increasing the effective synchrotron frequency, i.e. $\omega_{\rm Eff}=|q eB|/E+\Omega$, leading to a strong enhancement in radiation intensity~\cite{Buzzegoli:2022dhw,Buzzegoli:2023vne}. In heavy-ion collisions, where angular velocity and magnetic field vectors are collinear, the radiation from negative-charge quarks is enhanced, whereas the radiation from positive charges is suppressed. The net effect is an increase in photon production.

\begin{figure}
    \centering
    \includegraphics[width=0.95\linewidth]{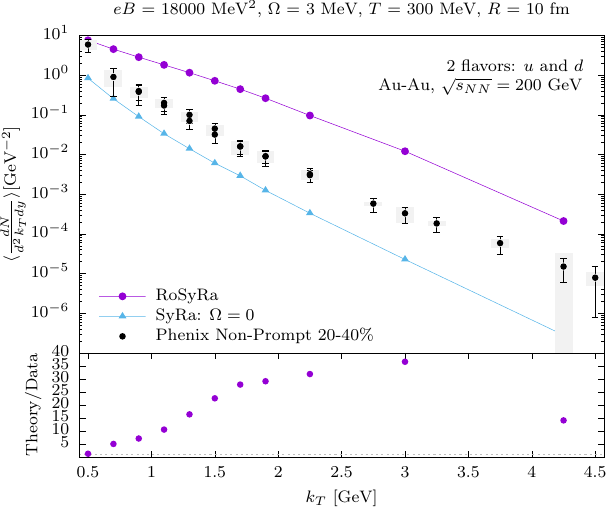}
    \caption{Photon spectrum, for  $eB=18000$ MeV$^2$, $\Omega=3$ MeV, $T=300$ MeV, $R=10$ fm, $L=10$~fm, $\Delta t =10$~fm$/c$ and two flavors. The data are adapted from~\cite{PHENIX_yields_2014, PHENIX_yields_2022}. The cyan line with triangles, $\Omega=0$. The dark violet line with circles, $\Omega\neq 0$ (RoSyRa).}
    \label{fig:large_yields}
\end{figure}

This qualitative picture is borne out by the numerical calculations which are summarized in Figs.~\ref{fig:best_fit} and~\ref{fig:large_yields}. Figure~\ref{fig:best_fit} compares the predictions of the multimessenger model \cite{Gale:2021emg} and of the RoSyRa model with the experimental data for the photon yields~\cite{PHENIX_yields_2014, PHENIX_yields_2022} and $v_2$~\cite{PHENIX_v2_2012, PHENIX_v2_2016} spectra resulting from Au-Au collisions at $\sqrt{s_{NN}}=200$ GeV and for $20\%-40$\% centrality class. The displayed results are for the following parameters: $eB=18000$ MeV$^2\simeq m_\pi^2$, $\Omega=2$ MeV, $T=200$ MeV, $R=5$ fm (the cylinder's radius), and $L=10$~fm (the cylinder's height) and the plasma's lifetime is $\Delta t= 10$~fm/c. The RoSyRa contribution to the integrated photon yield is relatively small. However, it yields a substantial increase in $v_2$ at low transverse momentum. Figure~\ref{fig:large_yields} shows the RoSyRa photon spectrum for a different set of parameters: $eB=18000$ MeV$^2\simeq m_\pi^2$, $\Omega=3$ MeV, $T=300$ MeV, $R=10$ fm, and $L=c\Delta t = 10$ fm. In this case, we have an excess of photon production compared to the experimental data. We emphasize that the present RoSyRa results are obtained within a stationary medium under idealized conditions, specifically a constant and homogeneous magnetic field and rotation lasting 10 fm$/c$. In contrast, realistic magnetic fields in heavy-ion collisions are expected to decay rapidly, and rotation is typically highly nonuniform. Consequently, the comparisons to experimental data in Figs.~\ref{fig:best_fit} and~\ref{fig:large_yields} remain qualitative and serve as an estimate of the effect. Although a comprehensive time-integrated calculation is reserved for future work, our analysis validates RoSyRa as a viable candidate mechanism for addressing the direct photon puzzle. This is supported by the fact that the rotation values employed here are lower than the late-time global averages required to explain spin polarization~\cite{STAR:2017ckg}, yet they still yield a substantial enhancement in the photon emission rate (Fig.~\ref{fig:large_yields}).

%====================================================================
\section{Electromagnetic radiation by a rotating and magnetized plasma}\label{sec:EMRPlasma}
%====================================================================
%********************************************************************
\subsection{Photon radiation by a quark}\label{sec:ByQuark}
%********************************************************************
In this section, we summarize the computation of the probability amplitude for  $q\to q\, +\, \gamma$ in the presence of a magnetic field and rotation as obtained in~\cite{Buzzegoli:2022dhw,Buzzegoli:2023vne}.
The calculation is performed using the exact solutions to the Dirac equation in the presence of both an external magnetic field $\bm{B}$ and a constant angular velocity $\bm{\Omega}$ along the same direction, taken as the $\hat{z}$ axis in this section. We consider a quark with mass $M$ and charge $q<0$ embedded in a uniformly rotating medium, assuming the regime of relatively slow rotation, $\Omega \ll \sqrt{|q eB|}$. In this context, the characteristic radial extent of the fermion wave function is significantly smaller than the light-cylinder radius $1/\Omega$, simplifying the radial boundary conditions. The electromagnetic radiation by a positive quark $q>0$ is obtained from that by a negative quark by flipping the sign of the angular velocity $\Omega \to-\Omega$~\cite{Buzzegoli:2023vne}. 

The dynamics is described by the Dirac equation,
\begin{equation}
	(\I\gamma \cdot D - M)\psi = 0,
\end{equation}
where $D_\mu = \partial_\mu + \Gamma_\mu + \I qA_\mu$ is the covariant derivative and $\Gamma$ is related to the Christoffel symbols.
To include the effect of a global rotation, the equation is solved in a reference frame rotating with angular velocity $-\bm{\Omega}$ about the $z$ axis. The rotation modifies the space-time metric, introducing a nonvanishing component $\Gamma_0 = -\Omega[\gamma_x, \gamma_y]/4$. We use the symmetric gauge for the magnetic field, $A^\mu=(0,-By/2,Bx/2,0)$.
Casting the Dirac equation into the Schrödinger form $\text{i}\partial_t\psi = \hat{H}\psi$ yields the Hamiltonian $\hat{H}$,
\begin{equation}
	\hat{H} = \gamma^0 \bm{\gamma} \cdot (\bm{p} - q\bm{A}) + \gamma^0 M + \bm{\Omega}\cdot\hat{\bm{J}}.
\end{equation}
The term $\bm{\Omega}\cdot\hat{\bm{J}}$ explicitly incorporates the effect of the rigid rotation on the quark, where $\hat{J}_z$ is the operator for the total angular momentum along the rotation axis.
The Hamiltonian $\hat{H}$ commutes with the longitudinal momentum operator $\hat{P}_z=-\I\de_z$ and the total angular momentum operator $\hat{J}_z$. Consequently, the stationary solutions $\psi$ are simultaneous eigenstates of energy $E$, longitudinal momentum $p_z$, and the magnetic quantum number $m$ (the eigenvalue of $\hat{J}_z$).

The solution to the Dirac equation in cylindrical coordinates $(t, r, \phi_0, z)$ is~\cite{Buzzegoli:2022dhw,Buzzegoli:2023vne,Sokolov_Ternov_1986}
\begin{equation}
\label{eq:QuarkWaveFunction}
\psi(t,r,\phi_0,z) = \E^{-\I E t} \frac{ \E^{\I p_z z} \E^{\I m\phi_0}}{\sqrt{L}\sqrt{2\pi}}
    \left(\begin{array}{c}
         C_1 I_{n-1,a}(\rho) \E^{-\I\phi_0/2}  \\
         \I C_2 I_{n,a}(\rho) \E^{\I\phi_0/2}  \\
         C_3 I_{n-1,a}(\rho) \E^{-\I\phi_0/2}  \\
         \I C_4 I_{n,a}(\rho) \E^{\I\phi_0/2}  
    \end{array} \right),
\end{equation}
where $\rho = |q eB|r^2/2$ is the dimensionless radial variable. Given the principal quantum number $n$ and the magnetic quantum number $m$, the radial number $a$ is an auxiliary quantum number defined by $m = n - a - \tfrac{1}{2}$.
The $I$ functions are defined in terms of the generalized Laguerre polynomials $L_n^{(\alpha)}$ as
\begin{equation}
\label{eq:DefIfunc}
\begin{split}
I_{n,a}(x) =& \sqrt{\frac{a!}{n!}}\E^{-x/2}x^{\frac{n-a}{2}} L_a^{(n-a)}(x)\\
    =& (-1)^{n+a}\sqrt{\frac{n!}{a!}}\E^{-x/2}x^{\frac{a-n}{2}} L_n^{(a-n)}(x).
\end{split}
\end{equation}
The solution~(\ref{eq:QuarkWaveFunction}) assumes that the space-time is unbounded, and it is obtained by requiring that the radial functions are exponentially suppressed at large distances ($\rho\to\infty$). Solutions in a finite volume and the resulting RoSyRa have been analyzed in~\cite{Buzzegoli:2024nzd}. To satisfy the unbounded space condition, the parameters $a$ and $n$ must be quantized as a non-negative integer $n,\,a:=0,1,2\dots$~\cite{Buzzegoli:2022dhw,Buzzegoli:2023vne}.
This quantization condition establishes the dispersion relation for the quark embedded in a rotating system in the presence of a magnetic field:
\begin{equation}
\label{eq:EnergyDisperion}
	(E - \Omega m)^2 = 2|q eB|n + p_z^2 + M^2.
\end{equation}
This relation reveals that rotation shifts the fermion energy by $-\Omega m$, thereby lifting the degeneracy of the standard Landau levels.
The coefficients $C_i$ have been chosen so that $\psi$ is also an eigenstate with eigenvalue $\zeta=\pm 1$ of the magnetic moment $\mu_z$, defined as
\begin{equation}
\bm{\mu} = \bm{\Sigma}-\frac{i \gamma^0\gamma^5}{2}\bm{\Sigma}\times(\bm{p}-q\bm{A}),
\end{equation}
and such that they normalize the wave function (\ref{eq:QuarkWaveFunction}),
\begin{equation}
    \int \psi^\dagger\, \psi\, \D^3 x = 1.
\end{equation}
This fixes the coefficients $C_i$ as
\begin{equation}
	\begin{split}
		C_{1,3} = \frac{1}{2\sqrt{2}} B_+ (A_+ \pm \zeta A_-) , \\
		C_{2,4} = \frac{1}{2\sqrt{2}} B_- (A_- \mp \zeta A_+) ,
	\end{split}
\end{equation}
where the upper signs refer to the indices 1, 2 and the lower ones to 3, 4, and
\begin{subequations}
\label{eq:DefAandB}
\begin{gather}
	A_\pm = \left(1 \pm \frac{p_z}{E}\right)^{1/2} , \\
	B_{\pm} = \left(1 \pm \frac{\zeta M}{\sqrt{E^2 - p_z^2}}\right)^{1/2} .
\end{gather}
\end{subequations}

The photon wave function $\bm{A}(\bm{x})$  is determined by solving the wave equation in the radiation gauge $A^0 = 0$, $\nabla \cdot \bm{A} = 0$. Solutions possessing a defined energy $\omega$ must satisfy the vector Helmholtz equation,
\begin{equation}
	(\nabla^2 + \omega^2)\bm{A}(\bm{x}) = 0.
\end{equation}

The general divergenceless solution $\bm{A}(\bm{x})$ is spanned by the toroidal ($\bm{T}$) field and the poloidal ($\bm{P}$) field. These basis states, often referred to as the Chandrasekhar-Kendall states, are derived from the solution $u(\bm{x})$ to the scalar Helmholtz equation $(\nabla^2 + \omega^2)u(\bm{x}) = 0$.
In cylindrical coordinates $(\phi_0, r, z)$, the scalar eigenfunction $u(\bm{x})$ is given by,
\begin{equation}
	u(\bm{x}) = J_l(k_\perp r)e^{\I (k_z z+l\phi_0)},
\end{equation}
where $k_z$ and $k_\perp$ are the longitudinal and transverse components of the photon momentum, and $k^2 = k_z^2 + k_\perp^2 = \omega^2$ and $J$ is the Bessel function. Choosing $\hat{\bm{z}}$ as the quantization axis, the toroidal and poloidal fields are explicitly defined as
\begin{align}
\bm{T}_{l, k_\perp, k_z}(\bm{x}) &= \!\!\left[ \frac{\I l}{kr} J_l(k_\perp r)\hat{\bm{r}} - \frac{k_\perp}{k} J'_l(k_\perp r)\hat{\bm{\phi}} \right]\! e^{\I (k_z z+l\phi_0)} ,\label{eq:toroidal} \quad \\
\bm{P}_{l, k_\perp, k_z}(\bm{x}) &= \left[ \I \frac{k_z k_\perp}{k^2} J'_l(k_\perp r)\hat{\bm{r}} - \frac{l k_z}{k^2 r} J_l(k_\perp r)\hat{\bm{\phi}}\right.\nonumber\\
&\hphantom{=\Big[}\left. +\frac{k_\perp^2}{k^2} J_l(k_\perp r)\hat{\bm{z}} \right] e^{\I (k_z z+l\phi_0)} ,\label{eq:poloidal}
\end{align}
where $l$ is the integer eigenvalue of the total angular momentum along the $z$ axis.
A specific linear combination of these fields defines the circularly polarized photon states,
\begin{equation}
\begin{split}
\bm{\Phi}_{h, l, k_\perp, k_z}(\phi_0, r, z) \equiv \frac{k}{k_\perp} \frac{1}{\sqrt{2}} &\left[ h\bm{T}_{l, k_\perp, k_z}(\phi_0, r, z) \right.\\
    &\left.+ \bm{P}_{l, k_\perp, k_z}(\phi_0, r, z) \right],
\end{split}
\end{equation}
where $h = \pm 1$ labels the right- or left-handed photon helicity states.
The final normalized photon wave function, representing one particle per unit volume $V$, is then
\begin{equation}
	\bm{A}_{h, l, k_\perp, k_z}(\bm{x}) = \frac{1}{\sqrt{2\omega V}} \bm{\Phi}_{h, l, k_\perp, k_z}(\phi_0, r, z)e^{-\I \omega t}.
\end{equation}

Moving now to the splitting process $q_i\to q_f\, +\, \gamma$, the initial quark $q_i$ has momentum $p_z$, energy $E$, quantum numbers $n$ and $a$, and polarization $\zeta$, while the final state $q_f$ has momentum $p'_z$, energy $E'$, quantum numbers $n'$ and $a'$, and polarization $\zeta'$; the photon has momentum $k=(\omega,\,\bm{k})$ and $h=\pm 1$ denotes its helicity and $l$ its angular momentum along $z$. The $\mathcal{S}$ matrix element of the process is
\begin{equation}
\begin{split}
\mathcal{S} =& (2\pi)\delta(E'+\omega-E)\frac{-\I q e}{\sqrt{2\omega V}}
    \langle \bm{j} \cdot \bm{\Phi} \rangle
    \delta_{m',m-l} \\ &\times\frac{2\pi}{L}\delta(p_z-p_z'-k_z),
\end{split}
\end{equation}
where $q$ is the electric charge of the particle in units of $e$ and we introduced the notation
\begin{multline}
\label{eq:OverlapInt}
\langle \bm{j} \cdot \bm{\Phi} \rangle  \delta_{m',m-l} \frac{2\pi}{L}\delta(p_z-p_z'-k_z) =\\
\int \overline{\psi}_{n',a',p_z',\zeta'}(\bm{x})\bm{\Phi}^*_{h,l,k_\perp,k_z}(\bm{x})\cdot \bm{\gamma}\,\psi_{n,a,p_z,\zeta}(\bm{x}) \D^3 x.
\end{multline}
The differential probability of the process is obtained by multiplying $|\mathcal{S}|^2$ by the phase space and the quantum states of the photon; the probability of the process is
\begin{equation}
w = \sum_{h,l} |\mathcal{S}|^2 \frac{V \D^3k}{(2\pi)^3}.
\end{equation}
Reminding that
\begin{equation*}
    \left[ \delta(E'+\omega-E) \right]^2 = \frac{T}{2\pi}\delta(E'+\omega-E)
\end{equation*}
with $T$ the process duration, we obtain the rate of the process per unit of time,
\begin{equation}
\label{eq:RateSingleProcess}
\begin{split}
\frac{\D\dot{w}_{n,a,n',a',\zeta,\zeta'}}{\D^3k} = &
    \frac{q^2 e^2}{4\pi} \frac{1}{2\pi} \frac{\delta(E-E'-\omega)}{\omega}\\
    &\times\left|\frac{2\pi}{L}\delta(p_z-p_z'-k_z) \right|^2
\sum_h |\langle \bm{j} \cdot \bm{\Phi} \rangle|^2.
\end{split}
\end{equation}
The last quantity was obtained in \cite{Buzzegoli:2022dhw,Buzzegoli:2023vne} and is given by
\begin{equation}
\label{eq:jPhiAmplitudeSq}
\begin{split}
|\langle \bm{j} \cdot \bm{\Phi} \rangle|^2 = & \frac{1}{2} I_{a,a'}^2(x)\Big|
        \sin\theta [K_4 I_{n-1,n'-1}(x) - K_3 I_{n,n'}(x)]\\
        &+ K_1 (h-\cos\theta) I_{n,n'-1}(x)\\
        &- K_2 (h+\cos\theta) I_{n-1,n'}(x) \Big|^2,
\end{split}
\end{equation}
where the argument $x$ of the $I$ function is
\begin{equation}
x = \frac{k_\perp^2}{2|q e B|},
\end{equation}
with $k_\perp$ the photon momentum transverse to the magnetic field.
The coefficients $K_i$ in Eq.~(\ref{eq:jPhiAmplitudeSq}) are defined in terms of the $C_i$ constants appearing in the wave function~(\ref{eq:QuarkWaveFunction}) of the initial and final quarks as follows:
\begin{equation}
	\begin{split}
		K_1 = C_1' C_4 + C_3' C_2 , \qquad K_2 = C_4' C_1 + C_2' C_3 , \\
		K_3 = C_4' C_2 + C_2' C_4 , \qquad K_4 = C_1' C_3 + C_3' C_1 .
	\end{split}
\end{equation}
The explicit expressions for the $K_i$'s are given in Appendix~\ref{sec:AppA}, and the products of $K_i$ pairs appearing upon taking the square in Eq.~(\ref{eq:jPhiAmplitudeSq}) are given explicitly in Eq.~(\ref{eq:KCoeffProducts}).
The sum over the spin polarizations $\zeta$ and $\zeta'$ and over the photon helicity $h=\pm 1$ are done explicitly in Appendix~\ref{sec:AppA}, yielding
\begin{equation}
\label{eq:RateSingleProcessSpinSum}
\begin{split}
\frac{\D\dot{w}_{n,a,n',a'}}{\D^3k} = & \sum_{\zeta,\zeta'}\frac{\D\dot{w}_{n,a,n',a',\zeta,\zeta'}}{\D^3k}\\
    =&\frac{q^2 e^2}{4\pi} \frac{1}{2\pi} \frac{\delta(E-E'-\omega)}{\omega}
    2  \Gamma_{n,a}(n',a',\bm{k})\\
    &\times \left|\frac{2\pi}{L}\delta(p_z-p_z'-k_z) \right|^2,
\end{split}
\end{equation}
where we defined
\begin{equation}
\begin{split}
 \Gamma_{n,a}&(n',a',\bm{k}) =  I_{a,a'}^2 \Big\{ 2 \overline{K_1^2} \left[ I_{n,n'-1}^2 + I_{n-1,n'}^2 \right] \\
        &+ \sin^2\theta  \overline{K_4^2} \left( I_{n-1,n'-1}^2 + I_{n,n'}^2 \right)\\
        &- \sin^2\theta\overline{K_1^2} \left( I_{n-1,n'}^2 + I_{n,n'-1}^2 \right)  \\
        &- 2 \overline{K_1K_2} \sin^2\theta \left[ I_{n-1,n'-1} I_{n,n'} + I_{n-1,n'} I_{n,n'-1} \right] \\
        &- 2 \sin\theta \cos\theta \Big[ \overline{K_1K_4}  I_{n-1,n'-1} I_{n,n'-1} +\\
        & +\overline{K_1K_4} I_{n,n'} I_{n-1,n'} \\
        & + \overline{K_2K_4} \left( I_{n-1,n'-1} I_{n-1,n'} + I_{n,n'} I_{n,n'-1} \right) \Big] \Big\} ,
\end{split}
\label{eq:Gamma}
\end{equation}
and
\begin{subequations}
\label{eq:Kappas}
\begin{gather}
\overline{K_1^2} = \overline{K_2^2} = \frac{(E-m\Omega) (E'-m'\Omega) - p_z p_z' - M^2}{4 (E-m\Omega) (E'-m'\Omega)},
    \label{eq:K1} \\
\overline{K_3^2} = \overline{K_4^2} = \frac{(E-m\Omega) (E'-m'\Omega) + p_z p_z' - M^2}{4 (E-m\Omega) (E'-m'\Omega)}, \\
 \overline{K_1K_2} = \overline{K_3K_4} = \frac{\sqrt{2n|q eB|}\sqrt{2n'|q eB|}}{4 (E-m\Omega) (E'-m'\Omega)}, \\
\overline{K_1K_3} = -\overline{K_2K_4} = \frac{-p_z \sqrt{2n'|q eB|}}{4 (E-m\Omega) (E'-m'\Omega)}, \\
\overline{K_1K_4} = -\overline{K_2K_3} = \frac{p_z' \sqrt{2n|q eB|}}{4 (E-m\Omega) (E'-m'\Omega)}.
    \label{eq:K1K4}
\end{gather}
\end{subequations}
The argument of all $I$ functions is $x$. 

\begin{figure}[t!h]
    \centering

\begin{tikzpicture}[
    scale=1.6,
    axis/.style={very thick, -Latex, black},
    vector/.style={very thick, -Latex, line cap=round},
    guide/.style={densely dotted, black, thin},
    angle/.style={thick, -}
]

    % Define the coordinate system basis to match the perspective
    % x points down-left (out of page), y points up, z points left
    \coordinate (O) at (0,0);
    \coordinate (X_dir) at (-0.6, -0.6); % Direction for x-axis
    \coordinate (Y_dir) at (0, 1);       % Direction for y'-axis
    \coordinate (Z_dir) at (-1, 0);      % Direction for z-axis (pointing left)
    % Define lengths for the vector components
    \def\xcomp{1.2} % Component along x (depth)
    \def\ycomp{2.2} % Component along y (height)
    \def\zcomp{-1.5} % Component along z (length to the right)

    % Note: Since Z_axis points left, a vector to the right has a negative coefficient relative to Z_dir,
    % or we can just calculate coordinates manually for the "positive" space of the vectors.

    % Calculate key points manually for precise control
    % Projection on "floor" (x-z plane) - The Red Vector Tip
    % We go Right (2.0) and Out (1.2)
    \coordinate (P_z) at (1.5, 0); % Point on z-line (right side)
    \coordinate (P_x) at ($ (0,0) + \xcomp*(X_dir) $); % Point on x-line
    \coordinate (Tip_Red) at ($ (P_z) + \xcomp*(X_dir) $); % Vector sum for Red

    % Projection on "wall" (y-z plane) - The Green Vector Tip
    % We go Right (2.0) and Up (2.2)
    \coordinate (P_y) at (0, \ycomp);
%    \coordinate (Tip_Green) at (2, \ycomp);
     \coordinate (Tip_Green) at ($ (P_x) + \ycomp*(Y_dir) $);

    % The Main Vector (Blue) - Combination of all three
%    \coordinate (Tip_Blue) at ($ (Tip_Green) + \xcomp*(X_dir) $);
    \coordinate (Tip_Blue) at ($ (Tip_Green) + \zcomp*(Z_dir) $);

    % --- Draw Dotted Projection Box ---
    % Wall rectangle (y-z plane)
%    \draw[guide] (O) -- (P_y) -- (Tip_Green) -- (P_z);
%    \draw[guide] (P_z) -- (O); % Re-draw baseline if covered

    % Floor rectangle (x-z plane)
    \draw[guide] (O) -- (P_x) -- (Tip_Red);

    % Connecting lines to the Blue Tip
    \draw[guide] (Tip_Green) -- (Tip_Blue); % Top connector
    \draw[guide] (Tip_Red) -- (Tip_Blue);   % Vertical connector

    % Top/Side phantom lines to complete the 3D box look
    \coordinate (Corner_Top_Front) at ($ (P_y) + \xcomp*(X_dir) $);
    \draw[guide] (P_y) -- (Corner_Top_Front) -- (Tip_Blue);
    \draw[guide] (P_x) -- (Corner_Top_Front);

    % --- Draw Axes ---
    % Z-axis (Left)
    \draw[axis] (O) -- (-2.5, 0) node[above left] {$z$};
    \path (O) -- (-2.5, 0) node[midway, below] {Beam};

    % Y-axis (Up)
    \draw[axis] (O) -- (0, 3) node[right] {$y$};

    % X-axis (Down-Left)
    \draw[axis] (O) -- ($ 2.8*(X_dir) $) node[below right] {$x$};

    % Green Vector (kT)
    %\draw[vector, green!80!black] (O) -- (Tip_Green) node[midway, above left, xshift=-15pt,yshift=15pt] {$\vec{k}_T$};
    \draw[vector, green!80!black] (O) -- (Tip_Green) node[midway, right] {$\vec{k}_T$};

    % Red Vector (k_perp)
    \draw[vector, red] (O) -- (Tip_Red) node[right, xshift=2pt] {$\vec{k}_{\perp}$};

    % Blue Vector (k)
    \draw[vector, blue] (O) -- (Tip_Blue) node[above right] {$\vec{k}$};

    % --- Angles ---
    % Phi (Green) - Angle from Z-line (right) to Green Vector
    % Note: Drawing 2D arcs to mimic the 3D perspective
    \draw[green!80!black] (-0.15, 0.3) arc (0:17:-1.3) node[midway, left, xshift=1pt, yshift=2pt] {$\phi$};

    % Phi_0 (Red) - Angle from Z-line (right) to Red Vector
    \draw[red] (0.15, -0.15) arc (-70:-112:0.4) node[midway, left, yshift=-6pt, xshift=8pt] {$\phi_0$};

    % Theta (Blue) - Angle from Y-axis to Blue Vector
    \draw[red] (0, 0.6) arc (89:58:0.5) node[midway, above] {$\theta$};

    % Drawn below the Z-axis
    %\draw[-Latex, black] (-0.5, -0.4) -- (-1.8, -0.4) node[midway, below] {Beam};

    % Floating vector above the green one
    \draw[vector, red] (-1, 2.0) -- (-1, 2.8) node[right, midway] {$B$};
    \draw[vector, purple] (-1.3, 2.0) -- (-1.3, 2.8) node[left, midway] {$\Omega$};

\end{tikzpicture}
    \caption{The rapidity-azimuthal $(y-\phi)$ hyperbolic coordinate system compared with the cylindrical coordinates. A generic vector $\bm{k}$, e.g.\ the photon momentum, is decomposed into a transverse (to the beam) $k_T$ and longitudinal component $k_z$; its component along the magnetic field $\bm{B}$ is $k_y$, and $k_\perp$ is its component perpendicular to $\bm{B}$.}
    \label{fig:Coords}
\end{figure}
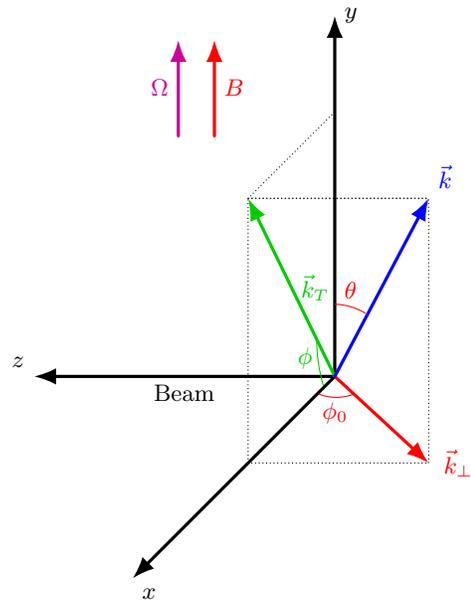

Before deriving the photon emission rate in a plasma, it is convenient to change the coordinate system from the cylindrical coordinates $(t,\,r,\,\phi_0,\,z)$ with the angular velocity and the magnetic field along $\hat{z}$ to the rapidity-azimuthal $(y-\phi)$ coordinate system used in experiments, see Fig.~\ref{fig:Coords}. In the latter reference system, the $\hat{z}$ points along the beam direction, and the magnetic field and rotation are directed along the $\hat{y}$ axis, not to be confused with the rapidity $y$. The four-momentum of the photon $k^\mu=(\omega,\,\bm{k})$ in the experimental reference system is
\begin{subequations}
\begin{align}
    k_x =& k_T \cos\phi = \omega \sin\theta \sin\phi_0 = k_\perp \sin\phi_0,\\
    k_y =& k_\parallel=k_T \sin\phi = \omega \cos\theta,\\
    k_z =& k_T \sinh y = \omega \sin\theta \cos\phi_0 = k_\perp \cos\phi_0,\\
    \omega =& k_T \cosh y,
\end{align}
\end{subequations}
where $k_T$ is the transverse momentum to the beam axis, $k_y=k_\parallel$ is the momentum parallel to magnetic field, $\theta$ is the angle between the magnetic field and $\bm{k}$.
To obtain the conversions between $(\omega,\theta)$ and $(k_T,\phi)$, we use the above identities for $k_y$ and $\omega$ to conclude that
\begin{equation}
    \cos\theta = \frac{\sin\phi}{\cosh y},    
\end{equation}
and from the momentum transverse to magnetic field
\begin{multline}
k_\perp^2=k_x^2+k_z^2 = \omega^2\sin^2\theta= k_T^2\cosh^2y\sin^2\theta\\
    =k_T^2(\cos^2\phi+\sinh^2 y) = k_T^2(1+\sinh^2 y-\sin^2\phi)\\
    =k_T^2(\cosh^2 y-\sin^2\phi),
\end{multline}
we obtain
\begin{equation}
    \sin\theta = \pm \sqrt{1 - \frac{\sin^2\phi}{\cosh^2 y}} .
\end{equation}
The indeterminate form of $\sin\theta$ comes from needing to cover the entire space with the spherical polar coordinate. This can be implemented by putting $|\sin\theta|$ wherever $\sin\theta$ appears. These equations also give the differential volume element as
\begin{equation}
\label{eq:d3krapidity}
    \D^3k = (k_T^2 \cosh y) \D k_T \D y \D\phi = (k_T \cosh y) \D^2k_T \D y .
\end{equation}
The experimental data for the direct photons are plotted at midrapidity ($y=0$), where $\bm{k}$ lies in the plane normal to the beam axis. In this case, the transformation between the coordinates has the simpler form
\begin{gather}
    \omega = k_T \\
    \cos\theta = \sin\phi,\quad
    |\sin\theta| = \sqrt{1-\sin^2\phi} = |\cos\phi| .
\end{gather}

%********************************************************************
\subsection{Photon rate per unit volume by a plasma}\label{sec:ByPlasma}
%********************************************************************
Given the photon production rate by a single quark $\D\dot{w}_{n,a,n',a'}$, as expressed by (\ref{eq:RateSingleProcessSpinSum}), we can determine the photon spectrum emitted by a rotating plasma at thermal equilibrium in a magnetic field. The differential photon emission rate of a unit volume is derived from Eq.~(\ref{eq:RateSingleProcess}) by summing and averaging over the initial and final states of the quark \cite{Kajantie:1986dh}:
\begin{equation}
\label{eq:PhotonInPlasma}
\begin{split}
\D R =& \frac{\D N_\gamma}{\D V \D t } = N_{\rm c} \sum_f\sumint\frac{\D^3 p}{(2\pi)^3}
V' \sumint\frac{\D^3 p'}{(2\pi)^3} \\
&\times \frac{\D\dot{w}_{n,a,n',a'}}{\D^3k} n_F(E) (1-n_F(E'))\D^3k,
\end{split}
\end{equation}
where $N_{\rm c}$ is the number of colors,  $f$ runs over the quark and antiquark flavors, and $V'$ is the spatial volume associated with the final quarks, as explained below in detail. The thermal distribution functions of the quarks in the plasma are given by the Fermi-Dirac distribution function
\begin{equation}
n_F(E) = \frac{1}{\E^{E/T}+1},
\end{equation}
where $T$ is the temperature of the plasma. 
Unlike the free quarks whose motion is quasiclassical and hence their momentum is continuous, in the presence of a magnetic field, the quark's transverse momentum is quantized as reflected in the Landau levels. Furthermore, the normal modes depend on the radial quantum number. As the quarks are confined inside the plasma, which is a cylinder in our case, the radial quantum number $a$ is constrained by the radial size of the cylinder, specifically $a\leq |q eB|R^2/2$, implying that modes with large negative $m$ are excluded. Accordingly, the phase space integration in (\ref{eq:PhotonInPlasma}) becomes a sum over the quantum numbers  $n$, $a$, and an integral in the longitudinal momentum, giving \cite{Chen:2015hfc}
\begin{equation}\label{eq:SumFiniteR}
\sumint\frac{\D^3 p}{(2\pi)^3} \to \sum_{n=0}^\infty \frac{1}{S} \sum_{a=0}^{|q eB|R^2/2}
    \int \frac{\D p_y}{2\pi},
\end{equation}
where $S=\pi R^2$ is the area of the cylinder's base (in the plane transverse to the magnetic field and rotation axis). Similarly, we generally allow the final quark states to have a different finite volume $V'=S' L$, with $S'=\pi R^{\prime 2}$.

Experimental measurements report the photon spectrum as $\D N_\gamma/\D^2k_T \D y$ at midrapidity ($y=0$) with $N_\gamma$ the number of photons observed in this differential region of $k$-space. From Eq.~(\ref{eq:d3krapidity}) it follows that the photon spectrum in the rapidity-azimuthal coordinate system at midrapidity is
\begin{equation}
\label{eq:raterapidity}
\begin{split}
\frac{\D\dot{w}_{n,a,n',a'}}{\D^2k_T \D y}\Big|_{y=0} =&
    \left[(k_T \cosh y) \frac{\D\dot{w}_{n,a,n',a'}}{\D^3k}\right]_{y=0}\\
=& k_T \frac{\D\dot{w}_{n,a,n',a'}}{\D^3k}\Big|_{y=0}.
\end{split}
\end{equation}
We define the differential rate of photon production as the photon spectrum produced in a unit time, per unit volume, at midrapidity,
\begin{equation}
\begin{split}
&\frac{\D R}{\D^2 k_T\D y}\Big|_{y=0} =
    \frac{\D N_\gamma}{\D V\D t \D^2 k_T\D y}\Big|_{y=0} = \sum_f \mathcal{R}_f\\
&= \sum_f \sum_{n=0}^\infty \frac{N_{\rm c}}{S} \sum_{a=0}^{|q eB|R^2/2} \int \frac{\D p_y}{2\pi}
\sum_{n'=0}^\infty \sum_{a'=0}^{|q eB|R^{\prime 2}/2} \int \frac{\D p_y'L}{2\pi}\\
 &\hphantom{=}\times k_T \frac{\D\dot{w}_{n,a,n',a'}}{\D^3k}\Big|_{y=0}
n_F(E) (1-n_F(E')).
\end{split}
\end{equation}
Consider the contributions $\mathcal{R}_f$ of a single flavor of quark with $q_f<0$.
Replacing the differential rate in Eq.~(\ref{eq:RateSingleProcessSpinSum}) and recalling the coordinate change, we then perform the integral over $p_y'$ with
\begin{equation}
    \int\left|\frac{2\pi}{L}\delta(p_y-p_y'-k_y) \right|^2\frac{\D p_y' L}{2\pi} =1,
\end{equation}
that sets $p_y' = p_y-k_y$ and gives
\begin{equation}
\label{eq:RateWithDelta}
\begin{split}
\mathcal{R}_f=& N_{\rm c} \frac{q_f^2 e^2}{4\pi} \frac{2}{2\pi}
\sum_{n=0}^\infty \frac{1}{S} \sum_{a=0}^{|q_f eB|R^2/2}
\int\frac{\D p_y}{2\pi} \sum_{n'=0}^\infty \sum_{a'=0}^{|q_f eB|R^{\prime 2}/2}\\ &\times
\delta(E-E'-\omega)\Gamma_{n,a}(n',a',\bm{k})
n_F(E) (1-n_F(E')).
\end{split}
\end{equation}

The integral over $p_y$ can be done by taking advantage of the Dirac $\delta$. First, we need to rewrite the Dirac $\delta$ in terms of $p_y$,
\begin{equation}
\delta(E-E'-\omega) = \sum_{p^0_y}\frac{\delta(p_y-p_y^0)}{|\frac{\partial}{\partial p_y} (E-E'-\omega)|},
\end{equation}
where $p_y^0$ are the roots of the energy conservation $E-E'-\omega$.
The derivative in $p_y$ is
\begin{equation}
\begin{split}
\frac{\partial (E-E'-\omega)}{\partial p_y}  =& \frac{p_y(\omega - (m-m')\Omega)}{(E-m\Omega)(E'-m'\Omega)} \\
& - \frac{\omega\cos\theta (E-m\Omega)}{(E-m\Omega)(E'-m'\Omega)},
\end{split}
\end{equation}
yielding the Dirac $\delta$
\begin{multline}
\label{eq:ExpansionDiracDelta}
\delta(E-E'-\omega) = \sum_{p^0_y}\delta(p_y-p_y^0)\\
\times\frac{(E-m\Omega)(E'-m'\Omega)}{|p_y(\omega - (m-m')\Omega) - \omega\cos\theta (E-m\Omega)|} .
\end{multline}

Recalling Eq.~(\ref{eq:EnergyDisperion}), the energy conservation inside the Dirac $\delta$ can be cast in the form
\begin{equation}
\sqrt{a^2+p_y^2} -\sqrt{b^2+(p_y-c)^2} = -\Delta,
\end{equation}
where, with $M_f$ the quark mass, we defined
\begin{subequations}
\begin{align}
a^2 =& 2|q_f eB|n + M_f^2,\\
b^2 =& 2|q_f eB|n' + M_f^2,\\
c =& \omega\cos\theta,\\
\Delta =& (n-a-n'-a')\Omega-\omega.
\end{align}
\end{subequations}
A careful examination of these equations reveals the conditions to have zero, one, or two solutions,
\begin{equation}\label{eq:EnergyConditions}
\begin{cases}
-\bar{s} \Delta> \bar{s} P,\,-\bar{s}\Delta<-|c| & \text{No roots}\\
|c|<-\bar{s}\Delta<\bar{s} P & \text{2 roots }p^{0\mp}_y\\
-|c|<-\Delta<|c| & \text{1 root }p^{0-}_y
\end{cases}
\end{equation}
where we defined $\bar{s}=$sign$(a-b)$, and $P$ is the extremum of the $p_y$ function $\sqrt{a^2+p_y^2} -\sqrt{b^2+(p_y-c)^2}$ given by
\begin{equation}
P = (a-b)\sqrt{1+\frac{c^2}{(a-b)^2}}.
\end{equation}
The two roots are given by
\begin{equation}
\label{eq:pyRoots}
p^{0\mp}_y = \frac{c(b^2-a^2+c^2-\Delta^2) \mp \textrm{sgn}(c \Delta) \sqrt{\Delta^2\alpha}}{2(c^2-\Delta^2)},
\end{equation}
with
\begin{equation}
\alpha = a^4+(b^2+c^2-\Delta^2)^2-2a^2(b^2-c^2+\Delta^2).
\end{equation}
When $a=b$, which corresponds to $n'=n$, there cannot be two roots, and the only root $p^{0-}_y$ is obtained when $\Delta^2-c^2\leq 0$ and it simplifies into
\begin{equation}
p^0_y  = \frac{c(c^2-\Delta^2)-\text{sgn}(c)\Delta\sqrt{(c^2-\Delta^2)(4a^2+c^2-\Delta^2)}}{2(c^2-\Delta^2)}.
\end{equation}

Then, using the representation (\ref{eq:ExpansionDiracDelta}) in Eq.~(\ref{eq:RateWithDelta}), we can easily integrate over $p_y$ to obtain
\begin{equation}
\begin{split}
\mathcal{R}_f =& \frac{q_f^2 e^2}{4\pi} \frac{2N_{\rm c}}{(2\pi)^2}
\sum_{n=0}^\infty \frac{1}{S} \sum_{a=0}^{|q_f eB|R^2/2}
\sum_{n'=0}^\infty \sum_{a'=0}^{|q_f eB|R^{\prime 2}/2} \sum_{p^0_y}\\
&\times \frac{(E-m\Omega)(E'-m'\Omega) \Gamma_{n,a}(n',a',\bm{k})}{|p^0_y(\omega - (m-m')\Omega) - \omega\cos\theta (E-m\Omega)|}\\
&\times n_F(E) (1-n_F(E')),
\end{split}
\end{equation}
where the $\Gamma$ function is evaluated at $p_y=p^0_y$.
Note that the conditions (\ref{eq:EnergyConditions}) limit the phase space of the parameters $n,\, n',\, a$, and $a'$ to the region where one or two roots of the energy conservation are possible. The corresponding values of the parameters can be obtained analytically and are given in Appendix~\ref{sec:PhaseSpace}. Avoiding regions of the parameter space that are prohibited by the kinematics is essential for minimizing computation time. The following formulas are assumed to be subject to these constraints, although we do not explicitly indicate them to avoid bulky expressions.  

Since in Eqs. (\ref{eq:K1})-(\ref{eq:K1K4}), each coefficient has a common denominator, we define
\begin{equation}
\tilde{\Gamma}_{n,a}(n',a',\bm{k}) = 4 (E-m\Omega) (E'-m'\Omega) \Gamma_{n,a}(n',a',\bm{k})
\end{equation}
and obtain
\begin{equation}\label{eq:FinalResult}
\begin{split}
\mathcal{R}_f =& \frac{q_f^2 e^2}{4\pi} \frac{N_{\rm c}}{(2\pi)^2S}
\sum_{n=0}^\infty \sum_{a=0}^{|q_f eB|R^2/2}
\sum_{n'=0}^\infty \sum_{a'=0}^{|q_f eB|R^{\prime 2}/2} \sum_{p^0_y}\\
&\times \frac{1}{2}\frac{\tilde{\Gamma}_{n,a}(n',a',\bm{k})n_F(E) (1-n_F(E'))}{|p^0_y(\omega - (m-m')\Omega) - \omega\cos\theta (E-m\Omega)|}.
\end{split}
\end{equation}
This is the photon rate produced by negatively charged quarks in QGP. As mentioned earlier, the contribution of positive quarks $q>0$ can be obtained by flipping the sign of the angular velocity, $\Omega \to-\Omega$, as shown in \cite{Buzzegoli:2023vne}. [As the solution of the Dirac equation for a positive charge differs from Eq.~(\ref{eq:QuarkWaveFunction}), it is more convenient to flip $\Omega$ than derive the expression for the rate for the positive charge anew.] We note that photon radiation by positive charges is strongly suppressed, and its contribution to the total photon rate is negligible, as confirmed numerically in Sec.~\ref{sec:Rotating}. 

The total rate is obtained by summing over the flavors
\begin{equation}
\label{eq:PosAndNeg}
\begin{split}
\frac{\D R}{\D^2 k_T\D y}\Big|_{y=0} =& \sum_f \mathcal{R}_f
    = \sum_{f,q<0} \mathcal{R}_f^{q<0}(\Omega) +  \sum_{f,q>0}\mathcal{R}_f^{q>0}(\Omega)\\
    =& \sum_{f,q<0} \mathcal{R}_f^{q<0}(\Omega) +  \sum_{f,q>0}\mathcal{R}_f^{q<0}(-\Omega),
\end{split}
\end{equation}
where the expression for $\mathcal{R}_f^{q<0}(\Omega)$ is given by Eq.~(\ref{eq:FinalResult}).
In Eq.~(\ref{eq:FinalResult}) the quantum numbers $n$, $n'$, $a$, and $a'$ run over all values allowed by energy and momentum conservation. The quantum numbers that do not preserve energy and momentum do not have solutions for $p_y^0$ and their contribution is set to zero, see  Appendix~\ref{sec:PhaseSpace} for details. Equation~(\ref{eq:Rf_constrainedKinematics}) represents the explicit formula for the photon rate.

To obtain the total photon spectrum produced in a heavy-ion collision, the rate spectrum in~(\ref{eq:PosAndNeg}) is integrated over time and space,
\begin{equation}
\label{eq:Photon_yield_1}
\frac{\D N_\gamma}{\D^2k_T \D y}\Big|_{y=0}  = \int \D t\int_V \D^3x\, \frac{\D R_\gamma}{\D^2k_T \D y}\Big|_{y=0}.
\end{equation}
In our simplified case, with a constant magnetic field in a homogeneous system, we obtain
\begin{equation}
\label{eq:FinalTotPhoton}
\frac{\D N_\gamma}{\D^2k_T \D y}\Big|_{y=0} = \Delta t L\pi R^2\,  \frac{\D R_\gamma}{\D^2k_T \D y}\Big|_{y=0}.
\end{equation}

The rate averaged over the azimuthal angle is obtained by integration over $\phi$ as follows:
\begin{equation}
\left\langle \frac{\D R_\gamma}{\D^2k_T \D y}\Big|_{y=0} \right\rangle =
  \frac{1}{2\pi} \int_0^{2\pi} \D\phi\, \frac{\D R_\gamma}{\D^2k_T \D y}\Big|_{y=0},
\end{equation}
or, equivalently, we can integrate over $\theta$,
\begin{equation}
\label{eq:IntYields1}
\left\langle \frac{\D R_\gamma}{\D^2k_T \D y}\Big|_{y=0} \right\rangle
=\frac{1}{2\pi} \int_0^{2\pi} \D\theta\, \frac{\D R_\gamma}{\D^2k_T \D y}\Big|_{y=0}.
\end{equation}
Using the reflection symmetries of the rate,  discussed in Appendix~\ref{sec:Reflection}, (\ref{eq:IntYields1}) reduces to
\begin{equation}
\label{eq:RIntegral}
\left\langle \frac{\D R_\gamma}{\D^2k_T \D y}\Big|_{y=0} \right\rangle
=\frac{4}{2\pi} \int_0^{\pi/2} d\theta\, \frac{\D R_\gamma}{\D^2k_T \D y}\Big|_{y=0}.
\end{equation}
This is the equation used, together with Eqs.~(\ref{eq:FinalTotPhoton}), (\ref{eq:PosAndNeg}), and  (\ref{eq:FinalResult}), to obtain Figs.~\ref{fig:best_fit} and~\ref{fig:large_yields}.
Similarly, the elliptic flow coefficient $v_2^\gamma$ at midrapidity is obtained as
\begin{equation}
\label{eq:v2Integral}
\begin{split}
v_2^\gamma\Big|_{y=0} =& \frac{1}{\left\langle \frac{\D R_\gamma}{\D^2k_T \D y}\Big|_{y=0} \right\rangle}
    \frac{1}{2\pi}\int_0^{2\pi}\D\phi \,\frac{\D R_\gamma}{\D^2k_T \D y}\Big|_{y=0}\cos(2\phi)\\
=& \frac{(-1)}{\left\langle \frac{\D R_\gamma}{\D^2k_T \D y}\Big|_{y=0} \right\rangle}
    \frac{4}{2\pi}\int_0^{\pi/2}\D\theta\,\frac{\D R_\gamma}{\D^2k_T \D y}\Big|_{y=0}\cos(2\theta).
\end{split}
\end{equation}
%

%********************************************************************
\subsection{Constrained and unconstrained emission and the thermodynamic limit}\label{sec:Constrained}
%********************************************************************
The RoSyRa assumes that plasma performs rigid rotation with constant angular velocity $\Omega$. In such a system, causality restricts the radial extent of the space-time to the light-cylinder radius, $R_\Omega = 1/\Omega$. While the wave functions (\ref{eq:QuarkWaveFunction}) formally extend to spatial infinity, they must be confined to the region $r \leq R_\Omega$ to remain physically valid.

Strictly speaking, a well-posed quantum theory in a rotating frame requires the imposition of boundary conditions (BCs) at $R \leq R_\Omega$ to ensure the self-adjointness of the Hamiltonian and the vanishing of conserved fluxes at the surface~\cite{Buzzegoli:2024nzd}. However, the relevance of these BCs depends on the ratio between the magnetic length and the light-cylinder radius:
\begin{equation}
\rho_\Omega = \frac{|q_f eB|}{2\Omega^2}.
\end{equation}
In the slow rotation regime ($\rho_\Omega \gg 1$), which is the focus of this analysis, BCs are considered negligible~\cite{Buzzegoli:2023vne,Buzzegoli:2024nzd}. In this limit, the causality requirement is instead satisfied by constraining the quantum numbers of the fermion states. Specifically, the average square radius of an unpolarized state, given by~\cite{Buzzegoli:2023vne}
\begin{equation}
\langle r^2 \rangle = \frac{2}{|q_f eB|} \left(n + a + \frac{1}{2}\right),
\end{equation}
must satisfy $\langle r^2 \rangle \Omega^2 \ll 1$. This implies a cutoff for the principal ($n$) and radial ($a$) quantum numbers: $n, a \ll \rho_\Omega$. If these values are exceeded, the states would occupy the causality-violating region $r > 1/\Omega$.

When evaluating the radiation rate from Eq.~(\ref{eq:FinalResult}), we consider a plasma of radius $R < 1/\Omega$. This means that the quantum numbers of the initial quark satisfy the condition
\begin{equation}\label{eq:rsqr}
 \frac{2}{|q_f eB|} \left(n + a + \frac{1}{2}\right)\le R^2.
\end{equation}
While the initial state is confined by (\ref{eq:rsqr}), the final quark may escape the initial volume unless it is confined by a condition similar to (\ref{eq:rsqr}),
\begin{equation}\label{eq:rsqr2}
 \frac{2}{|q_f eB|} \left(n' + a' + \frac{1}{2}\right)\le R'^2.
\end{equation}
A \textit{priori}, the two parameters $R$ and $R'$ do not have to coincide. To account for this, we consider two scenarios involving different implementations of the cutoff in the sum over $a'$:
\begin{itemize}
    \item \textbf{Unconstrained emission} $(R<R'=R_\Omega)$: The final quark is not confined to the initial volume and can occupy the entire causally connected phase space. The rate is obtained by setting the effective emission radius to $R' = 1/\Omega$.
    
    \item \textbf{Constrained emission} $(R=R'<R_\Omega)$: The final quark is forced to remain within the initial plasma volume. The rate is obtained by setting $R' = R$. 
\end{itemize}

The constraints (\ref{eq:rsqr}) and (\ref{eq:rsqr2}) significantly simplify the process of setting the boundary conditions. However, caution is necessary when taking the thermodynamic limit. For a rigidly rotating system, it is impossible to take the large-volume limit ($R \to \infty$) at a fixed $\Omega$ without violating causality. At most, the initial volume can encompass the entire light cylinder ($R = 1/\Omega$), in which case the two scenarios coincide. 

As we will show in Sec.~\ref{sec:nonrotating}, the unconstrained emission recovers the standard synchrotron radiation results and a well-defined thermodynamic limit as $\Omega \to 0$. Conversely, the constrained emission scenario is used to investigate finite-volume effects and the validity of the thermodynamic hypothesis. In this context, the system reaches the thermodynamic limit if the total radiation spectrum $\D N_\gamma /\D^2 k_T \D y$ scales linearly with the plasma volume $V = L \pi R^2$. Equivalently, the system can be considered in the thermodynamic limit if the spectrum rate~(\ref{eq:FinalResult}) converges to finite values at large enough radii $R$.

%====================================================================
\section{Synchrotron Radiation: the nonrotating plasma}\label{sec:nonrotating}
%====================================================================
As discussed in the previous sections, the RoSyRa framework incorporates the effects of both the external magnetic field and the bulk rotation of the plasma. Before addressing the full complexity of the rotating medium, we first investigate the photon emission in the limit of a nonrotating plasma ($\Omega \to 0$).

%********************************************************************
\subsection{The nonrotating limit}
%********************************************************************
We start by performing the nonrotating limit of the photon production rate~(\ref{eq:FinalResult}). With the only exception of the $I_{a,a'}^2(x)$ function factorized in the quantity $\Gamma_{n,a}(n',a',\bm{k})$ defined in Eq.~(\ref{eq:Gamma}), the rate $\mathcal{R}_f$ depends on $a$ and $a'$ only through terms coupled to the angular velocity in the form $m\,\Omega$ and $m'\,\Omega$. With this observation, the limit of vanishing rotation is trivial: the sums over $a$ and $a'$ are factorized. Furthermore, the sums can be performed analytically in the large-volume limit. To this end, we introduce the degeneracy factor $n_B$ coming from the summation of the radial quantum number. In the constrained scenario, where $R'=R$, it is obtained as follows:
\begin{equation}
\label{eq:DegeneracyFactor}
\begin{split}
n_B^{\rm Con}(x;\,R)=\!\! \sum_{a,a'=0}^{\frac{|q_f eB|R^2}{2}} \frac{I^2_{a,a'}(x)}{S}
    =\!\! \sum_{a=0}^{\frac{|q_f eB|R^2}{2}} \sum_{a'=0}^{\frac{|q_f eB|R^2}{2}} \frac{I^2_{a,a'}(x)}{\pi R^2},
\end{split}
\end{equation}
where  $x=\omega^2\sin^2\theta/(2|qeB|)$. Using the known property of the $I$ function~\cite{Sokolov_Ternov_1986,Buzzegoli:2022dhw}
\begin{equation}
\label{eq:ISquareSum}
\sum_{a'=0}^{\infty} I^2_{a,a'}(x) = 1,
\end{equation}
the degeneracy factor in the unconstrained emission becomes
\begin{equation}
\label{eq:DegeneracyFactorUnc}
\begin{split}
n^{\rm Unc}_B(R)=&\lim_{R'\to\infty} \sum_{a=0}^{|q_f eB|R^2/2} \sum_{a'=0}^{|q_f eB|R'^2/2} \frac{I^2_{a,a'}(x)}{\pi R^2}\\
    =&\sum_{a=0}^{|q_f eB|R^2/2} \frac{1}{\pi R^2}
    =\frac{2+|q_f eB|R^2}{2\pi R^2}.
\end{split}
\end{equation}
In the nonrotating case, the thermodynamic limit ($R\to \infty$) is well defined and the degeneracy factor reduces to the usual Landau one,
\begin{equation}
\label{eq:DegeneracyFactorLandau}
\begin{split}
n^{R=\infty}_B=&\lim_{R\to\infty} n^{\rm Unc}_B(R)
    = \frac{|q_f eB|}{2\pi}.
\end{split}
\end{equation}
It thus emerges that the system can be considered in the thermodynamic limit when $|q_f eB|R^2\gg 1$.
From Eqs.~(\ref{eq:FinalResult}) and~(\ref{eq:DegeneracyFactor}), the photon production rate in magnetic field without rotation is
\begin{equation}\label{eq:FinalResultNoOmega}
\begin{split}
\mathcal{R}_f(\Omega=0) =& \frac{q_f^2 e^2}{4\pi} \frac{N_c}{4\pi^2}
n_B(x;\,R,\,R')
\sum_{n=0}^\infty \sum_{n'=0}^\infty \sum_{p^0_y}\\
&\times \frac{1}{2}
\frac{\tilde{\Gamma}_{n,n'}(\bm{k})n_F(E) (1-n_F(E'))}{|p^0_y\,\omega - \omega\cos\theta E|},
\end{split}
\end{equation}
where the energy $E$ and $E'$ of the initial and final states of the quarks at vanishing rotation are given by:
\begin{align}
E =& \sqrt{2 n|q_f eB| + p_y^2 + M_f^2},\\
E' =& \sqrt{2 n'|q_f eB| + (p_y-\omega\cos\theta)^2 + M_f^2},
\end{align}
with $M_f$ the mass of the quark, and
\begin{gather}
\tilde{\Gamma}_{n,n'}(\bm{k}) = 4\, E\, E'\, \Gamma_{n,n'}(\bm{k}),\\
\Gamma_{n,n'}(\bm{k}) = \frac{\Gamma_{n,a}(n',\,a',\,\bm{k})}{I^2_{a,a'}(x)} ,    
\end{gather}
where $\Gamma_{n,a}(n',\,a',\,\bm{k})$ is given in Eq.~(\ref{eq:Gamma}) and where, denoting $p_y'=p_y-\omega\cos\theta$, the coefficients $\overline{K_iK_j}$ in Eq.~(\ref{eq:Kappas}) at $\Omega=0$ reduce to
\begin{subequations}
\begin{gather}
\overline{K_1^2} = \frac{E E' - p_y p_y' - M_f^2}{4\,E\,E' }, \\
\overline{K_4^2} = \frac{E E' + p_y p_y' - M_f^2}{4\,E\,E' }, \\
\overline{K_1K_2} = \frac{\sqrt{2n|q_f eB|}\sqrt{2n'|q_f eB|}}{4\,E\,E' }, \\
\overline{K_2K_4} = \frac{p_y \sqrt{2n'|q_f eB|}}{4\,E\,E' }, \\
\overline{K_1K_4} = \frac{p_y' \sqrt{2n|q_f eB|}}{4\,E\,E' }.
\end{gather}    
\end{subequations}
We also note that the energy conservation equation $E-E'-\omega=0$ at vanishing rotation $\Omega=0$ admits solutions $p_y^0$ only for $n>n'$. The sums of $n$ and $n'$ can be reduced to the region where energy conservation admits solutions, as shown in Eq.~(\ref{eq:Rf_constrainedKinematicsNoRotation}).
For the different scenarios of constrained, unconstrained, and thermodynamic limit, the rate is obtained from Eq.~(\ref{eq:FinalResultNoOmega}), substituting the degeneracy factor $n_B$ from  Eq.~(\ref{eq:DegeneracyFactor}), (\ref{eq:DegeneracyFactorUnc}), and (\ref{eq:DegeneracyFactorLandau}), respectively. The observable quantities are obtained from the nonrotating rate~(\ref{eq:FinalResultNoOmega}) in the same fashion as the rotating one, following Eqs.~(\ref{eq:Photon_yield_1})-(\ref{eq:v2Integral}).
The numerical code used for the computation is similar to the one used for the RoSyRa and is described in Sec.~\ref{sec:Rotating}. 

%********************************************************************
\subsection{Comparison with previous results}\label{nonrotatingComparison}
%********************************************************************
%
\begin{figure}[t!h]
    \centering
    \includegraphics[width=0.95\linewidth]{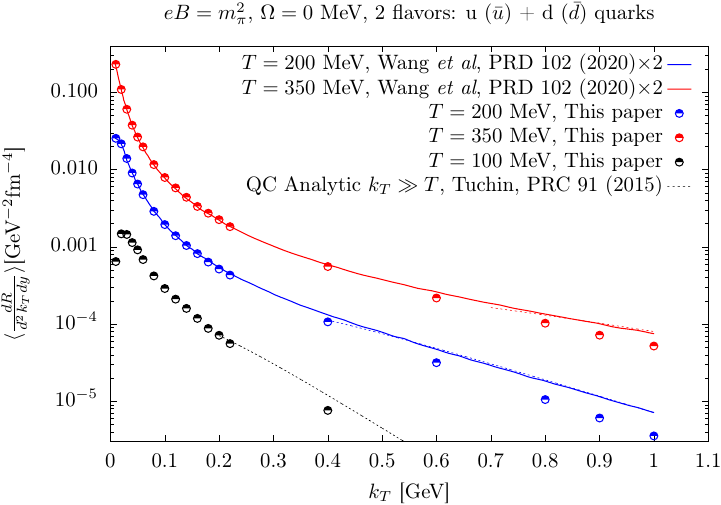}
    \includegraphics[width=0.95\linewidth]{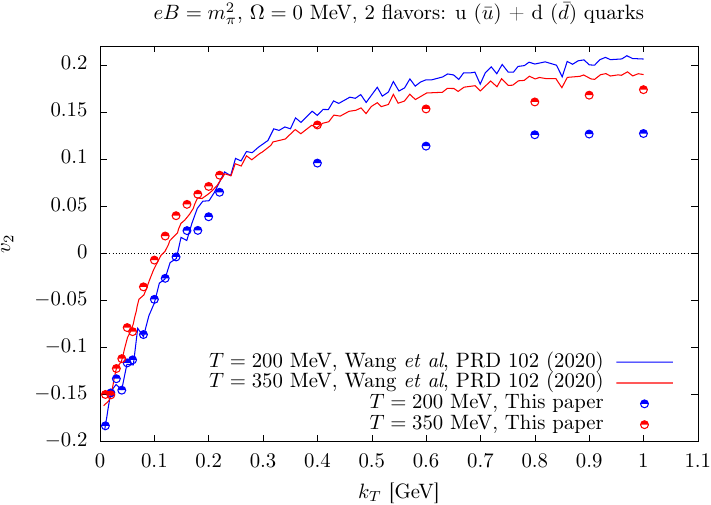}
    \caption{The numerical results for the integrated rates~(\ref{eq:RIntegral}) and the elliptic flow~(\ref{eq:v2Integral}) in comparison with~\cite{Wang:2020dsr} and with the high $k_T$ quasiclassical (QC) analytical value~\cite{Tuchin:2014pka}.}
    \label{fig:Comparison}
\end{figure}
The photon emission from a magnetized plasma in the absence of rotation has been studied in~\cite{Tuchin:2012mf,Tuchin:2014pka,Tuchin:2013apa} within the quasiclassical approach and in~\cite{Wang:2020dsr,Wang:2021ebh,Wang:2021eud,Wang:2023fst} within a full quantum regime using a different approach compared to this work. In~\cite{Wang:2020dsr} the photon production rate is obtained with the inclusion, in addition to the splitting processes studied here, of the annihilation processes where $q+\bar{q}\to\gamma$.
To validate and check the results for the rate in Eqs~(\ref{eq:FinalResult}) and~(\ref{eq:FinalResultNoOmega}) and our numerical code, we computed the integrated rates~(\ref{eq:RIntegral}) and the elliptic flow~(\ref{eq:v2Integral}) spectra for the same parameters as the spectra reported in~\cite{Wang:2020dsr}. We used Eq.~(\ref{eq:FinalResultNoOmega}) in the thermodynamic limit with $n_B$ given by~(\ref{eq:DegeneracyFactorLandau}) and we set $eB=m_\pi^2 =18225$ MeV$^2$ and temperatures $T=100$, $200$, and $350$ MeV and we considered two quark flavors: $u$, $\bar{u}$, $d$, and $\bar{d}$ with $M_u=2.16$ and $M_d=4.7$ MeV.
\begin{figure}[t!b]
    \centering
    \includegraphics[width=0.95\linewidth]{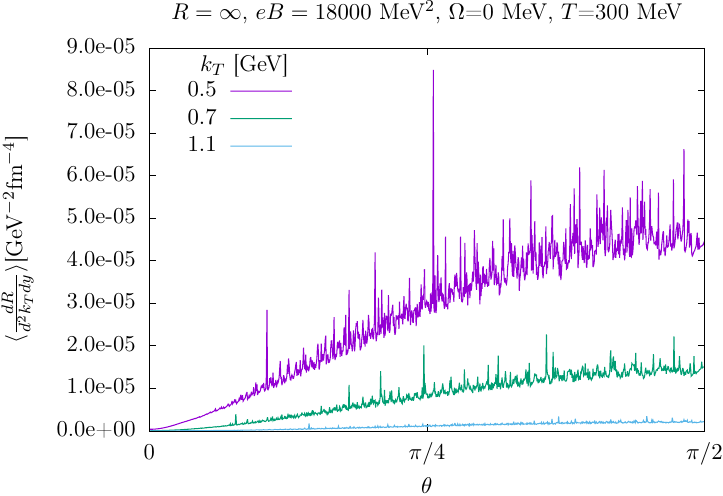}
    \caption{Nonrotating plasma: thermodynamic limit $R=\infty$ of rates as a function of $\theta$ for various values of photon transverse momenta $k_T$. }
    \label{fig:nonrotating_rates_unc_vs_theta_var_kT}
\end{figure}

Figure~\ref{fig:Comparison} shows agreement of our numerical results with those of~\cite{Wang:2020dsr}.\footnote{After a reanalysis done with the authors, we concluded that the integrated rates reported in~\cite{Wang:2020dsr} missed a factor 2; Fig.~\ref{fig:Comparison} shows the corrected values.} Note that discrepancies at higher $k_T$ are expected because annihilation processes start to provide a noticeable contribution~\cite{Wang:2020dsr}. As we are interested in the low $k_T$ region, in this work we neglected the contributions of annihilation processes, which can be computed following the same procedure described in the previous sections. Figure~\ref{fig:Comparison} also shows the high $k_T$ behavior expected for a quasiclassical (QC) emission as computed analytically in~\cite{Tuchin:2014pka} and given as dotted lines. The quantum calculations yield, as expected, a lower intensity as it includes the recoil of the photon.

As mentioned, in~\cite{Wang:2020dsr} and in Fig.~\ref{fig:Comparison} the masses of the quarks are set to their current masses $M_{0f}$, which are much smaller than the magnetic field, that is $|q_f eB|\gg M_f^2$. However, inside the QGP quarks acquire a thermal mass that can be estimated as $M_T=\sqrt{g_s^2/3}T$, where $T$ is the temperature of the plasma and $g_s$ is the strong coupling constant, which is about $2$. Lattice calculations estimate the thermal mass to be between 0.7 $T$ and 0.9 $T$, depending on the temperature. Henceforth in this work, we simply set the mass of the quarks to
\begin{equation}
\label{eq:QuarkMass}
M_f = M_{0f} + T.
\end{equation}
This describes a more realistic scenario and drastically changes the scaling between the magnetic field and mass, being now $|q_f eB|< M_f^2$. Even though most of the processes are dominated by temperature and thermal fluctuations, there might be qualitative differences between the two regimes, for example the appearance of strong anisotropies characterized by a difference between transverse and longitudinal pressure when $|q_f eB|\gg M_f^2$. For this reason, before moving to the rotating case, we analyze the nonrotating case when quarks acquire a thermal mass as in Eq.~(\ref{eq:QuarkMass}).

%********************************************************************
\subsection{Results including the thermal mass}
%********************************************************************
This section focuses on the nonrotating limit with thermal mass of quarks, analyzing the photon emission rates~(\ref{eq:FinalResultNoOmega}) to establish a baseline for the RoSyRa effects. The photon spectrum was calculated for various values of $T$ and $eB$ in both the constrained and the thermodynamic ($R\to\infty$) limits. These scenarios are abbreviated as ``Con'' and ``$R=\infty$'' throughout the text and figures.

\begin{figure}[t!b]
    \centering
    \includegraphics[width=0.95\linewidth]{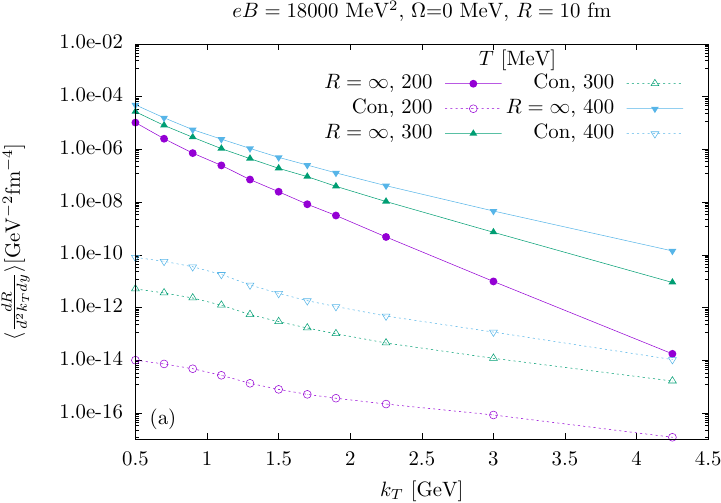}
    \includegraphics[width=0.95\linewidth]{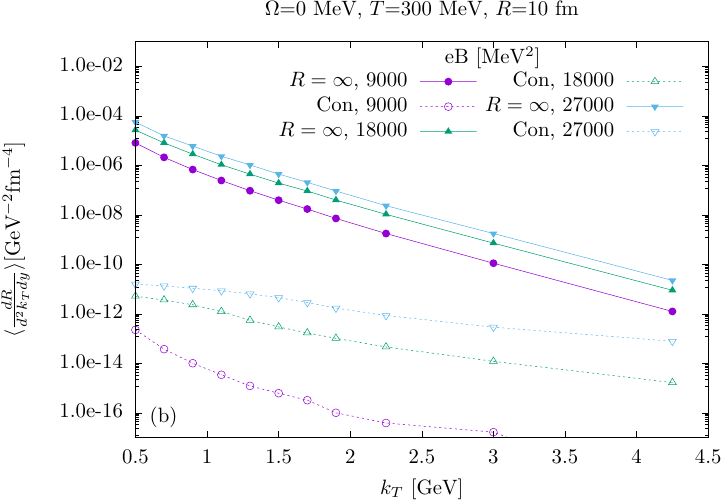}
    \caption{Nonrotating plasma: integrated rates with respect to $k_T$ for various values of (a) temperature and (b) magnetic field for Con emission at $R=10$ fm (empty symbols) and for $R=\infty$ (full symbols). }
    \label{fig:nonrotating_rates_unc_vs_T_and_eB}
\end{figure}

Figure \ref{fig:nonrotating_rates_unc_vs_theta_var_kT} illustrates the differential photon production rate as a function of angle $\theta$ for three values of the photon transverse momenta $k_T$. For each value of $k_T$, the maximum rate is obtained close to $\theta\approx\pi/2$.
Figure~\ref{fig:nonrotating_rates_unc_vs_T_and_eB}(a) explores the temperature and Fig.~\ref{fig:nonrotating_rates_unc_vs_T_and_eB}(b) the magnetic field dependence of the differential photon production rate as a function of the photon transverse momentum $k_T$ for a nonrotating plasma ($\Omega = 0$). The photon production rate increases with temperature as well as with magnetic field intensity for the whole range of photon momenta. The results show the comparison between the unconstrained emission scenario in the thermodynamic limit ($R=\infty$) (full symbols) and the constrained scenario (Con) (empty symbols), with a finite radius $R=10$ fm. As expected, we observe a significant divergence between the two scenarios. The constrained emission yields are noticeably suppressed compared to the unconstrained case by many orders of magnitude. In the nonrotating case, the only difference between the constrained and unconstrained scenarios is the degeneracy factor $n_B$ given respectively by Eqs.~(\ref{eq:DegeneracyFactor}) and~(\ref{eq:DegeneracyFactorLandau}).
The observation of a large discrepancy between the two scenarios raises the question whether the QGP can be considered in the thermodynamic limit; for $eB=m_\pi^2$, $q=2/3$, and $R=15$ fm, we obtain $|q eB|R^2\simeq 35$, which is not very large.

\begin{figure}[t!b]
    \centering
    \includegraphics[width=0.95\linewidth]{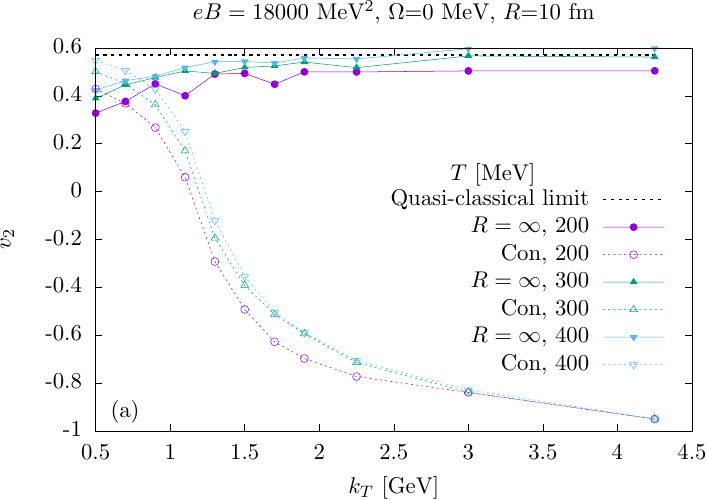}
    \includegraphics[width=0.95\linewidth]{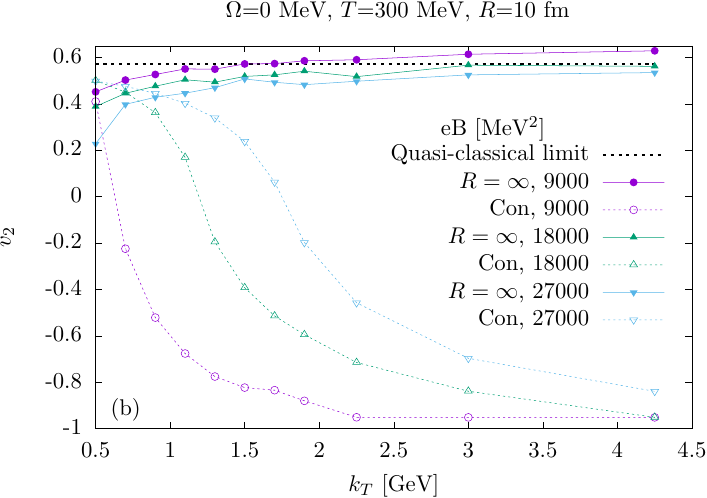}
    \caption{Nonrotating plasma: $v_2$ with respect to $k_T$ for various values of (a) temperature and (b) magnetic field for Con emission at $R=10$ fm (empty symbols) and for $R=\infty$ (full symbols).}
    \label{fig:nonrotating_v2_unc_vs_T_and_eB}
\end{figure}

In Fig. \ref{fig:nonrotating_v2_unc_vs_T_and_eB}(a), we present the elliptic flow coefficient $v_2$ for three different plasma temperatures ($T = 200, 300, 400$ MeV) while maintaining a constant magnetic field $eB =18000\text{ MeV}^2\approx m_\pi^2$, in both constrained and infinite-volume scenarios. A substantial positive $v_2$ is observed in the $R=\infty$ case for the whole range of $k_T$; this is slightly larger than the one obtained in~\cite{Wang:2020dsr} and in the previous section without thermal mass (notice that Fig.~\ref{fig:nonrotating_v2_unc_vs_T_and_eB} starts from $k_T=0.5$ GeV when $v_2$ in Fig.~\ref{fig:Comparison} is positive).
This large $v_2$ is the reason why the magnetic field is considered as a candidate explanation for the ``direct photon puzzle''. 
The obtained results for $v_2$ are in agreement with the quasiclassical limit $v_2=4/7\simeq 0.57$ obtained for large $k_T$ in~\cite{Tuchin:2014pka} and shown in the figure as a black dashed horizontal line. Quantum effects reduce the quasiclassical limit and introduce a mild temperature and magnetic field dependence.

The unconstrained rate at finite $R$ differs from the infinite-volume one by a constant factor, namely the Landau degeneracy (\ref{eq:DegeneracyFactorLandau}) is replaced by (\ref{eq:DegeneracyFactorUnc}). As this is a constant factor and $v_2$ is a normalized quantity, the resulting $v_2$ is independent of $R$ and the unconstrained $v_2$ at finite radius equals the infinite radius one. Instead, the finite radius rate is suppressed compared to the infinite radius and, as mentioned, approaches the latter at large $|q eB| R^2$.

On the other hand, the constrained case results in a positive $v_2$ for low values of $k_T$ and a rapid transition to negative values as $k_T$ increases.
This behavior is understood by analyzing the degeneracy factor $n_B(x;\,R)$ in Eq.~(\ref{eq:DegeneracyFactor}). Note that $n_B$ depends on $\theta$ and $k_T$ through $x$, which at midrapidity reads  $x=k_T^2\sin^2\theta/(2|qeB|)$. Clearly, the value of $x$ increases with $k_T$. As shown above, the degeneracy $n_B$ becomes the Landau level and yields the unconstrained results when the sum~(\ref{eq:ISquareSum}) is close to 1 and $|q eB| \gg 1$. In the constrained case, the sum on $a'$ on~(\ref{eq:ISquareSum}) stops at $a'=|q eB| R^2/2$. It can be shown that the sum~(\ref{eq:ISquareSum}) is close to one when $|q eB| R^2/2 \gg x$. For this reason, at fixed $R$ the constrained results are close to the unconstrained ones at small $k_T$. Instead, as $k_T$ is increased, the value $x$ is also increasing and the degeneracy factor $n_B$ becomes smaller, yielding a lower rate. Similarly, at sufficiently large $k_T$, due to the $\sin^2\theta$ dependence on $x$, the photon rate at small angles becomes larger than that at large angles, yielding the opposite trend of those shown in Fig.~\ref{fig:nonrotating_rates_unc_vs_theta_var_kT} and hence a negative $v_2$.
In physical terms, the constrained emission forbids quarks with large angular momentum and prevents photon emission; in particular, it suppresses emission for large $k_T$ and favors $\theta\simeq 0$ over $\theta\simeq \pi/2$.

In Fig.~\ref{fig:nonrotating_v2_unc_vs_T_and_eB}(b), where we present the elliptic flow coefficient $v_2$ for three different magnetic field intensities ($eB = 9000, 18000, 27000$ MeV$^2$) while maintaining a constant temperature $T =300$ MeV, we observe the same behavior in terms of constrained and unconstrained emission. In the constrained case, $v_2$ decays rapidly with $k_T$, while in the unconstrained case, it increases toward the high $k_T$ asymptotic limit, represented by a dashed line. We conclude the discussion of the nonrotating case by noticing that the total photon yields resulting from the synchrotron radiation, obtained from the averaged rate $\langle \D R/\D^2 k_T\D y\rangle$ using Eq.~(\ref{eq:FinalTotPhoton}) with $c\Delta t=L=10$ fm, is too small to explain the photon excess in the measurements and to provide a large enough $v_2$, see Figs.~\ref{fig:best_fit} and ~\ref{fig:large_yields}.

%====================================================================
\section{\texorpdfstring{R\lowercase{o}S\lowercase{y}R\lowercase{a}: Rotating Synchrotron Radiation}{RoSyRa: Rotating Synchrotron Radiation}}\label{sec:Rotating}
%====================================================================
The calculation of the averaged RoSyRa photon emission rates (\ref{eq:RIntegral}) and the elliptic flow coefficient $v_2$ (\ref{eq:v2Integral}) is performed using a custom-developed \textsf{c} numerical code. The framework is designed to handle the summation over the discrete Landau levels and radial quantum numbers characterizing the quark states in a rotating, magnetized plasma given in Eq. (\ref{eq:FinalResult}). To ensure computational efficiency and numerical stability, the code incorporates high-precision arithmetic and parallel processing techniques.

A critical challenge in evaluating the synchrotron radiation rates involves the computation of the overlap integrals between the initial and final quark states, see Eq.~(\ref{eq:OverlapInt}). These integrals depend on generalized Laguerre polynomials, $L_n^{(k)}(x)$, see Eqs.~(\ref{eq:jPhiAmplitudeSq}) and  \eqref{eq:DefIfunc}, which are known to exhibit numerical instability due to large oscillatory behavior and cancellation errors at high quantum numbers. To mitigate this, we employ the \textsf{MPFR} (Multiple Precision Floating-Point Reliable) library~\cite{Fousse2007}. Specifically, the functions utilize 128-bit precision variables to compute the generalized Laguerre polynomials and the associated functions $I(x)$ defined in Eq.~\eqref{eq:DefIfunc}. This ensures that the orthogonality and recurrence relations of the wave functions are preserved to a high degree of accuracy, preventing the accumulation of rounding errors that would otherwise corrupt the summation over thousands of Landau levels.

The total photon yield is obtained by summing over the initial and final principal quantum numbers ($n, n'$) and radial quantum numbers ($a, a'$). The summation over the initial Landau level $n$ constitutes the primary loop for convergence. Rather than a simple threshold cutoff, the code implements a dynamic weighted convergence criterion. The summation continues until the relative contribution of the last added term, weighted by the current iteration index $n$, falls below a specified goal precision (typically $10^{-3}$),
\begin{equation}
n \cdot \frac{\mathcal{R}_f(n)-\mathcal{R}_f(n-1)}{\mathcal{R}_f(n)} < \epsilon_{\text{goal}}    ,
\end{equation}
where $\mathcal{R}_f(n)$ is Eq.~(\ref{eq:FinalResult}) with the sum up to $n$.
This conservative condition ensures that the tail of the distribution, which decays slowly for high-energy photons, is adequately sampled. 
Furthermore, to respect the causal structure of the rotating system, the summation is bounded by the causal limit $n \le \rho_\Omega \equiv |q eB|/(2\Omega^2)$, preventing the inclusion of states that would violate causality at the cylinder boundary.
For a typical run of $\Omega=3$ MeV, $T=300$ MeV, and $eB=18000$ MeV$^2$, the convergence of the sum at $k_T=0.5$ GeV ranges from $n\sim 0.025\rho_\Omega$ for $R=5$ fm and low $\theta$, to $n\sim 0.84\rho_\Omega$ for $R=65$ fm and $\theta=\pi/2$ , the value being proportional to both $R$ and $\theta$.
The causal limit is usually reached for $\theta=\pi/2$ and large $k_T$ and $R$.

Energy conservation imposes strict kinematic constraints on the longitudinal momentum $p_y$. For every set of quantum numbers ($n, n'$) and photon angle $\theta$, the code uses the analytical solutions of the energy conservation equation $E - E' - \omega = 0$ given in Eq.~(\ref{eq:pyRoots}) to determine the allowed values of $p_y$; if no real solution exists, the transition is kinematically forbidden and is not included in the sum, see Appendix~\ref{sec:PhaseSpace} for details. Finite-volume effects are incorporated by restricting the radial quantum numbers $a$ and $a'$ as discussed in Sec.~\ref{sec:Constrained}. The code distinguishes between two emission scenarios: ``Constrained'' (Con, shown with empty symbols in the plots), with both the initial and final quarks confined within the plasma cylinder radius $R$, requiring $\{a,a'\} \le R^2 |q_f eB| / 2$, and ``Unconstrained'' (Unc, shown with full symbols in the plots), with only the initial quark confined to a cylinder with radius $R$, while the final state is allowed to extend beyond the boundary (up to the causal limit), mimicking radiation into an infinite medium. 

The code leverages OpenMP for parallel execution, distributing independent calculations for different quantum numbers across multiple processor cores to accelerate the computation.

We explored the RoSyRa photon emission for a range of parameters relevant to heavy-ion collisions while keeping $\Omega < \sqrt{|q_f eB|}$, which is the relatively slow rotation regime and the validity condition for Eq.~(\ref{eq:FinalResult}), specifically,
\begin{subequations}
\begin{align}
eB =&\{9000,\, 18000,\, 27000\}\text{ MeV}^2\\
    =&\{0.45,\, 0.9,\, 1.35\}\,m_\pi^2,\nonumber\\
\Omega =&\{0,\, 1,\, 2,\, 3,\, 4,\, 5,\,  6\}\text{ MeV},\\
T = & \{ 200,\, 300,\, 440\}\text{ MeV},\\
R =  &  \{5,\, 10,\, 20,\, 30,\, 40,\, 50,\, 65\}\text{ fm}.
\end{align}
\end{subequations}
For a fixed quark flavor the quark mass in Eq.~(\ref{eq:FinalResult}) is given by the sum of the current mass of the quark $M_{f0}$ and the thermal mass set to the temperature of the plasma $M_T=T$. We considered different quark flavors, namely $u$, $d$, $s$, and $c$, to study the effect of the current mass $M_{f0}$, which is set respectively to: 2.16, 4.7, 93.5, and 1273 MeV.
When it is not specifically stated, the results are obtained by summing the contributions of all these flavors, even though $\bar{u}$ and $d$ quarks are the only relevant ones.
To represent heavy-ion collisions, the angular velocity vector and the magnetic field vector point in the same direction, implying that the radiation from negative charges is enhanced and the radiation from positive charges is suppressed. We verified numerically that the suppression is large enough that the radiation from positive charges can be neglected. As explained in Sec.~\ref{sec:ByPlasma}, see Eq.~(\ref{eq:PosAndNeg}), the radiation from positive charges is actually calculated flipping the sign of $\Omega$, yielding the same result.

\begin{figure}[t!b]
    \centering
    \includegraphics[width=0.95\linewidth]{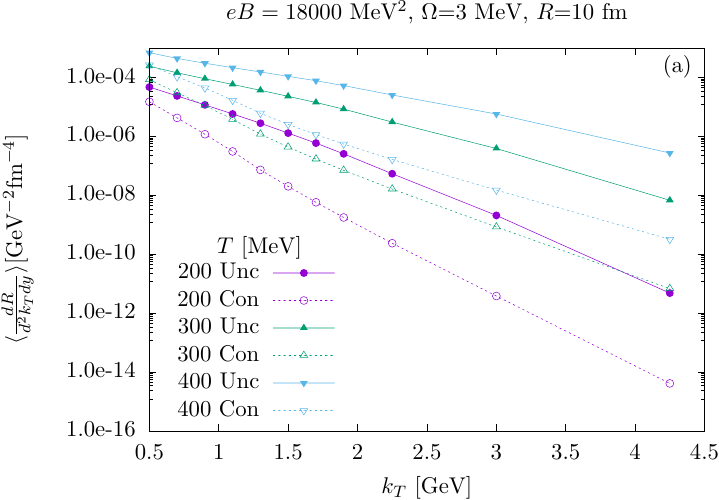}
    \includegraphics[width=0.95\linewidth]{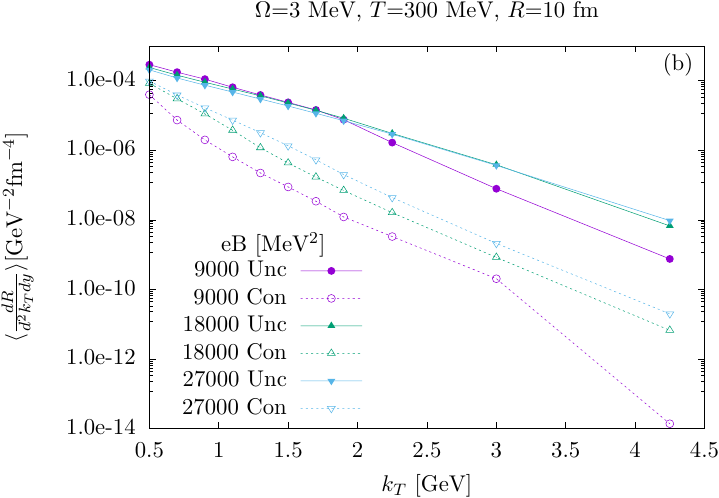}
    \caption{Rotating plasma: rates with respect to $k_T$ for various values of (a) temperature and (b) magnetic field. }
    \label{fig:rotating_rates_con_unc_vs_T_and_eB}
\end{figure}

The angular average rate $\langle \D R/\D^2k_T \D y \rangle$ and the $v_2$ coefficient at midrapidity are obtained using Eqs.~(\ref{eq:RIntegral}) and~(\ref{eq:v2Integral}) with numerical integration, specifically the trapezoidal rule, with ten points for $\theta\in[0,\,\pi/2]$. We verified for a few cases that the results are not significantly affected if the number of points is increased up to $1000$.

Figure \ref{fig:rotating_rates_con_unc_vs_T_and_eB} illustrates the differential photon emission rates, $\langle \D R/\D^2k_T \D y \rangle$, for a rotating quark-gluon plasma ($\Omega = 3$ MeV, $R = 10$ fm), specifically analyzing the sensitivity of the spectra to thermodynamic and magnetic parameters.
The upper panel illustrates the evolution of the emission spectrum as the temperature increases from $T=200$ to $400$ MeV, while the magnetic field is held constant at $eB = 18000$ MeV$^2$. A pronounced enhancement in the production rate is observed with an increase in temperature, consistent with the expected thermal scaling of the synchrotron radiation. The spectral slope hardens at higher temperatures, reflecting the increased thermal occupation numbers and phase space availability for high $ k_T$ photon emission. 
In the bottom panel, we examine the impact of varying the magnetic field strength ($eB \in \{9000, 18000, 27000\}$ MeV$^2$) at a fixed temperature of $T=300$ MeV. The results indicate a positive correlation between the magnetic field intensity and the overall yield. The enhancement is particularly notable in the low-to-intermediate $k_T$ region, suggesting that the stronger magnetic confinement of the quarks amplifies the synchrotron emission intensity. 
The Con scenario consistently exhibits a suppressed yield compared to the Unc approximation, with the deviation becoming increasingly significant at larger transverse momenta ($k_T > 2$ GeV). This suppression implies that imposing kinematic constraints or finite-size effects on the Landau level transitions effectively truncates the high-energy tail of the spectrum.

\begin{figure}[t!b]
    \centering
    \includegraphics[width=0.95\linewidth]{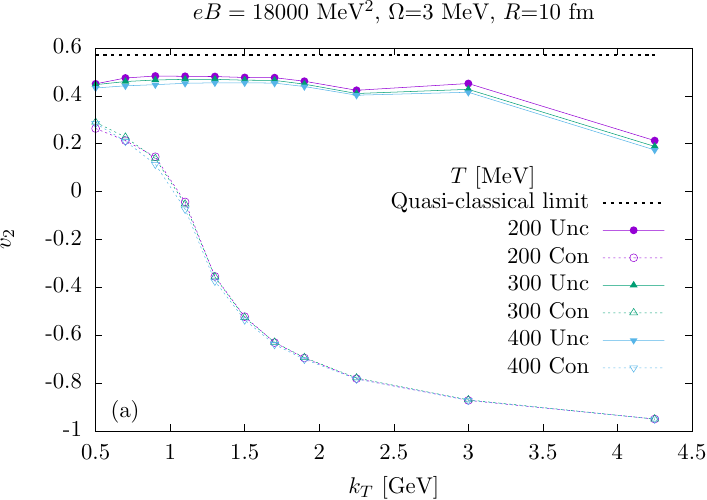}
    \includegraphics[width=0.95\linewidth]{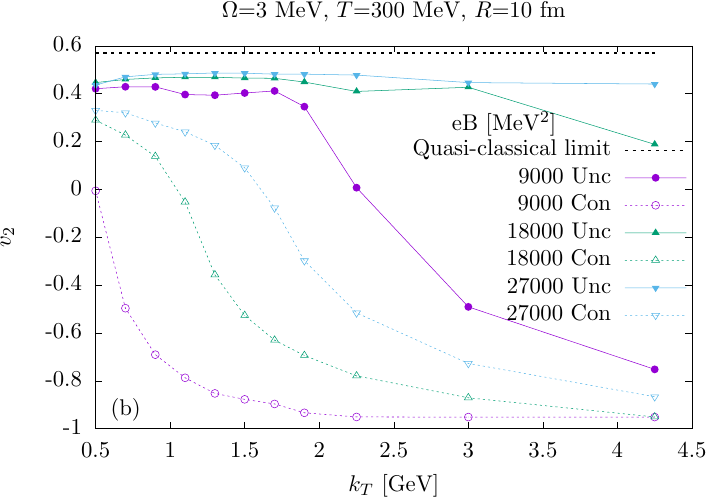}
    \caption{Rotating plasma: $v_2$ with respect to $k_T$ for various values of (a) temperature and (b) magnetic field.}
    \label{fig:rotating_v2_con_unc_vs_T_and_eB}
\end{figure}

Figure~\ref{fig:rotating_v2_con_unc_vs_T_and_eB} illustrates the elliptic flow coefficient, $v_2(k_T)$, for photons emitted from a rotating ($\Omega=3$ MeV, $R=10$ fm) and magnetized quark-gluon plasma, presenting the dependence of the anisotropy on thermodynamic and magnetic parameters. The first panel (varying temperature $T \in \{200, 300, 400\}$ MeV with fixed $eB=18000$ MeV$^2$) shows negligible variation of $v_2$ with temperature; the constrained scenario exhibits qualitative agreement with the nonrotating case, with positive values at low $k_T$ and decreasing into negative values as $k_T$ increases. In the bottom panel (varying magnetic field $eB \in \{9000, 18000, 27000\}$ MeV$^2$ with fixed $T=300$ MeV), a strong sensitivity to the magnetic field intensity for both Con and Unc scenarios is revealed, increasing the magnetic field leading to an increase in $v_2$.
Overall, the behavior with respect to $k_T$ is preserved, in the Con case we have decreases at different rates, while in the Unc case only the $eB=9000$  MeV$^2$ case reaches negative values of $v_2$ for large $k_T$. Similar to what was discussed for the nonrotating case, the fall to negative values of $v_2$ in the Con case is to be attributed to the excluded phase space, which is more relevant for high $k_T$ and for $\sin\theta\sim 1$, changing the slope of the angle dependence of the photon emission. For the same reason, in Fig.~\ref{fig:rotating_v2_con_unc_vs_T_and_eB}(b) we see that as $eB$ increases, also $|q eB| R^2$ increases, widening the phase space and moving $v_2$ toward the value it has for synchrotron radiation in the large-volume limit and in the absence of rotation.

\begin{figure}[t!b]
    \centering
    \includegraphics[width=0.95\linewidth]{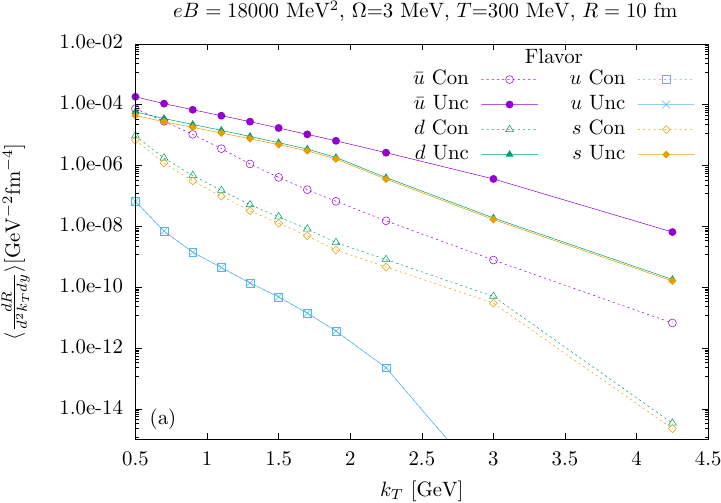}
    \includegraphics[width=0.95\linewidth]{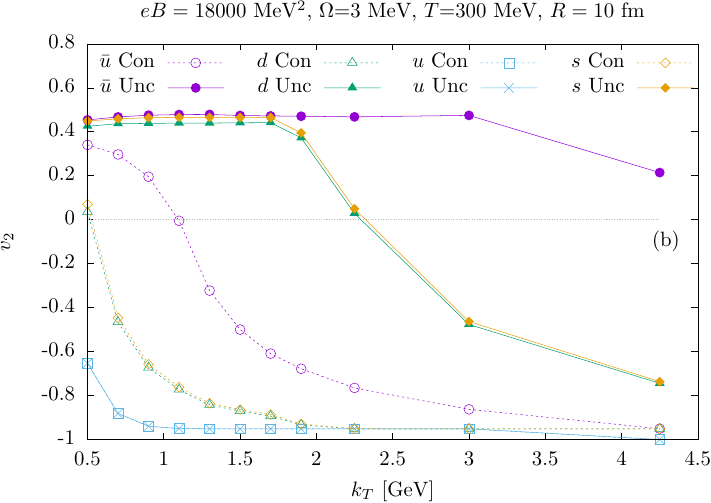}
    \caption{Flavor dependence. (a) Rates and (b) $v_2$ with respect to $k_T$ for various quark flavors, for both constrained (empty symbols) and unconstrained (full symbols) emission. }
    \label{fig:con_unc_vs_kT_var_flavor}
\end{figure}

The rates and elliptic flows from single quark flavors are shown in Fig.~\ref{fig:con_unc_vs_kT_var_flavor}. The only differences between the flavors are the charge $q_f$ and the current mass $M_{0f}$. The only positive-charge quark shown is $u$, as all the positive quark rates are significantly suppressed compared to their negative counterpart.
The main effect of rotation is to enhance the radiation intensity from negative charges, while also suppressing that of positive charges. The suppression results largely from the kinematic constraints on photon emission, resulting in a very narrow phase space, see also Appendix~\ref{sec:PhaseSpace}. The allowed transitions are so narrow that the Con emission already contains the relevant transitions and equals the results of the Unc case, as can be seen from the figure. As expected, the radiation of heavier quarks is much weaker than that of the light quarks. We observe that $v_2$ remains higher for a wider range of $k_T$ for lighter quarks.

\begin{figure}[tb]
    \centering
    \includegraphics[width=0.95\linewidth]{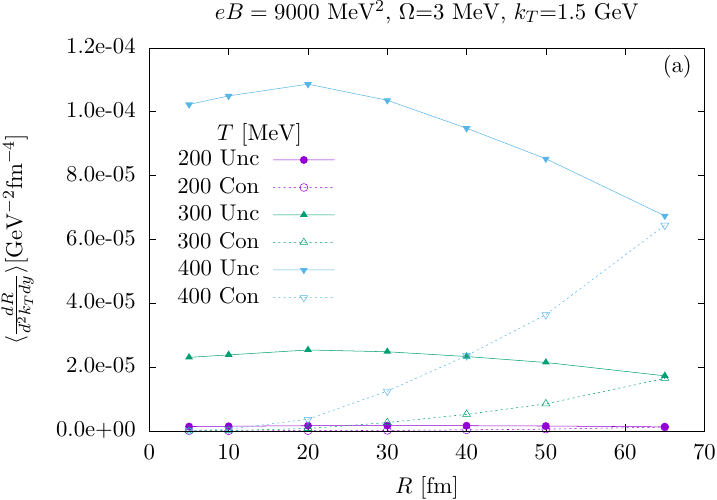}
    \includegraphics[width=0.95\linewidth]{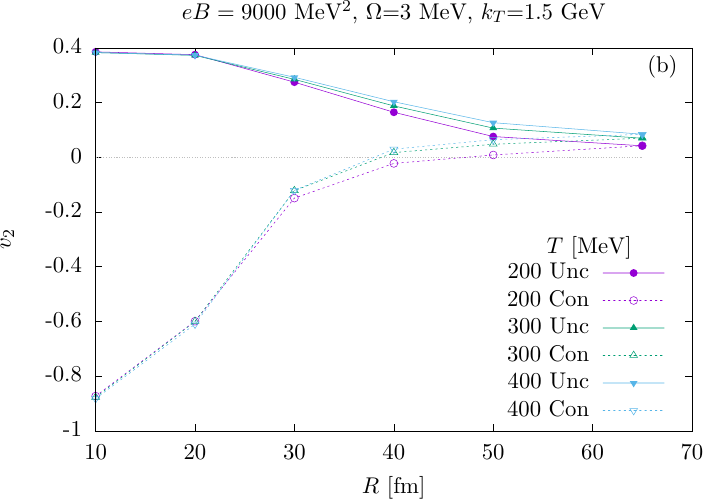}
    \caption{Radial dependence. (a) Rates and (b) $v_2$ with respect to radius $R$ for various values of temperature, for both constrained and unconstrained emission.}
    \label{fig:con_unc_vs_R}
\end{figure}

To illustrate the difference between the Con and Unc emission and how they converge once the system radius reaches the causal radius $R_\Omega=1/\Omega$, Fig.~\ref{fig:con_unc_vs_R} shows the rate and $v_2$ at a fixed transverse momentum $k_T=1.5$ GeV for both scenarios as a function of the radius and for different temperatures. The total number of photons emitted is increasing with the radius. However, the rates being the total photons divided by $R^2$, we see in Fig.~\ref{fig:con_unc_vs_R}(a) that for the Unc case the intensity of emission increases more than $R^2$ for small radii and then grows less than $R^2$ until it reaches the Con value, which grows faster with $R$.
A well-defined thermodynamic limit requires that the rates become constant past some value of $R$, like in the nonrotating case. In this case we reach the causal limit before an asymptotic value of the rates is reached.
Changing $R$ also changes the angular distribution of the photon emission.
\begin{figure}[t!b]
    \centering
    \includegraphics[width=0.95\linewidth]{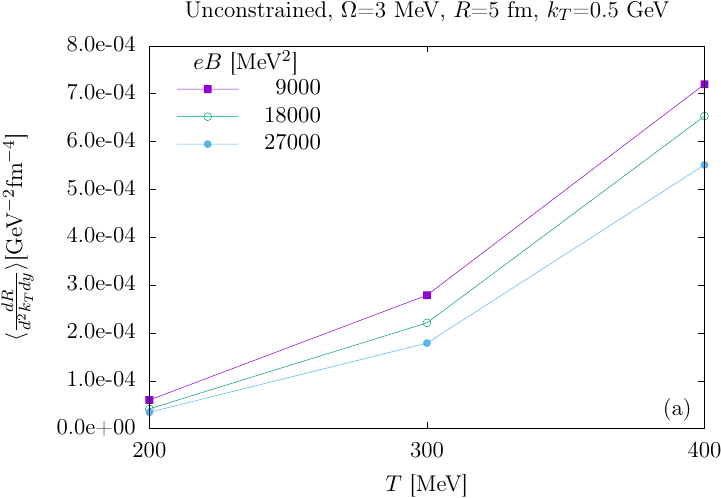}
    \includegraphics[width=0.95\linewidth]{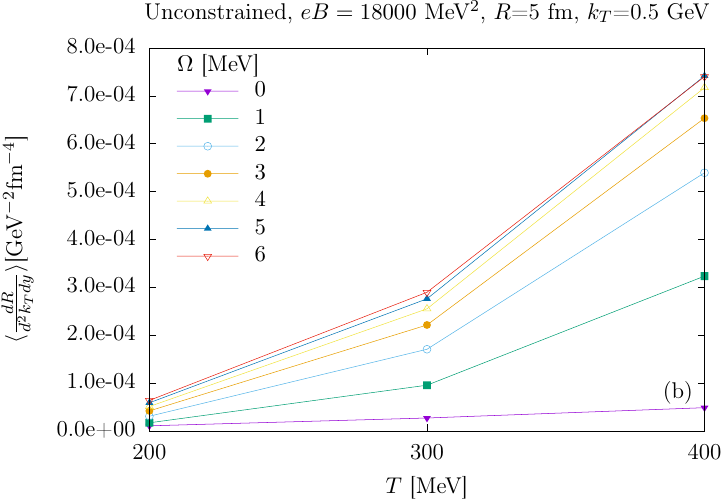}
    \includegraphics[width=0.95\linewidth]{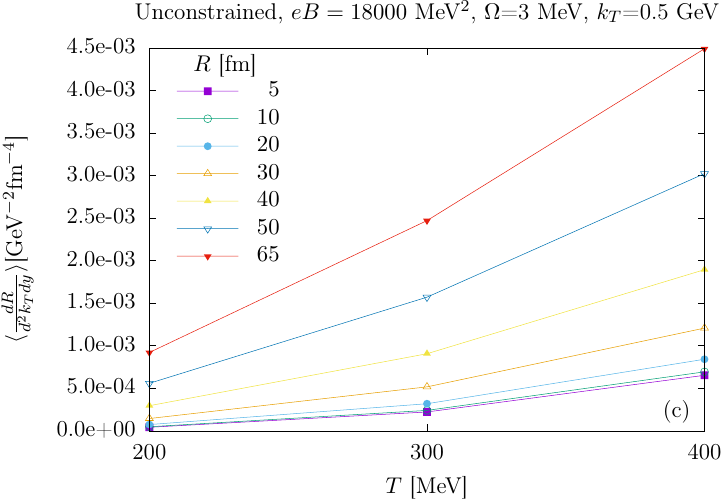}
    \caption{Temperature dependence. Rates with respect to the temperature $T$ for various values of the (a) magnetic field $eB$, (b) rotation $\Omega$, and (c) radius $R$. }
    \label{fig:rates_unc_vs_T}
\end{figure}
\begin{figure}[t!b]
    \centering
    \includegraphics[width=0.95\linewidth]{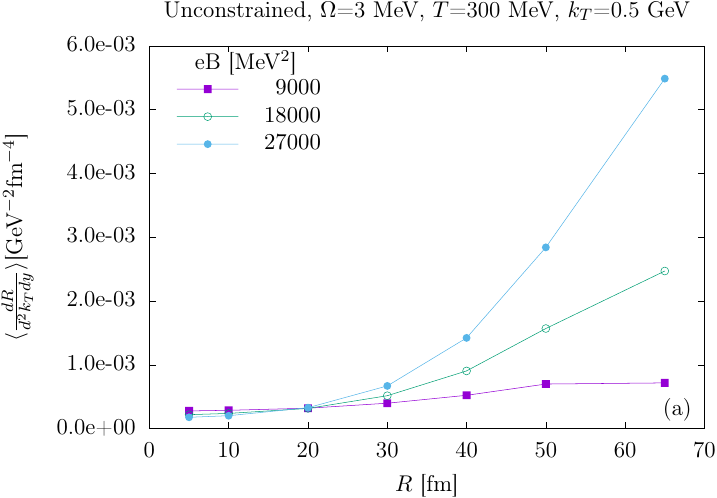}
    \includegraphics[width=0.95\linewidth]{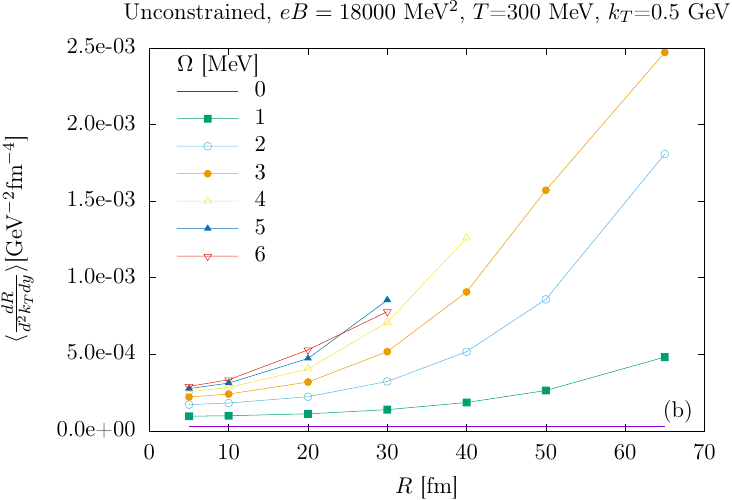}
    \includegraphics[width=0.95\linewidth]{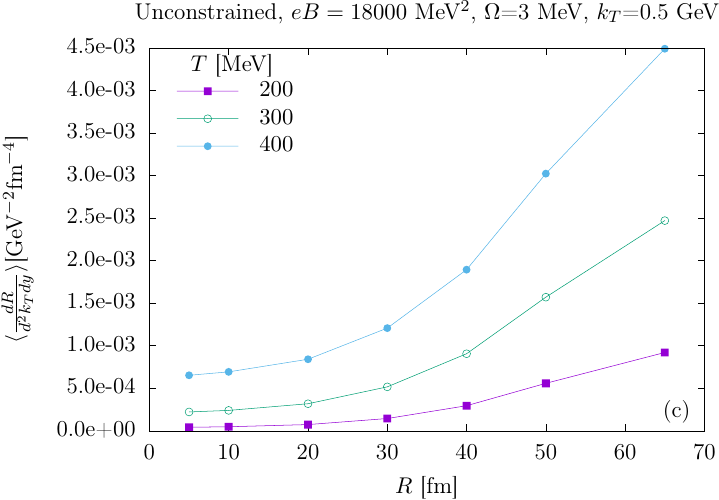}
    \caption{Radius dependence. Rates with respect to the radius $R$ for various values of the (a) magnetic field $eB$, (b) rotation $\Omega$, and (c) temperature $T$. }
    \label{fig:rates_unc_vs_R}
\end{figure}
In Fig.~\ref{fig:con_unc_vs_R}(b) we see that as $R$ approaches $R_\Omega$ the angular distribution becomes more homogeneous with small $v_2$, suggesting that rotation can reduce the angular anisotropy of synchrotron radiation and that a more rigorous treatment is needed to assess the impact of rotation on $v_2$ for large volumes. As per their definition, the Unc and Con emissions are the same for $R=R_\Omega$. From now on, the analysis of the results will only focus on the unconstrained emission.
The unconstrained photon emission rates at a fixed low transverse momentum $k_T=0.5$ GeV are depicted in Fig.~\ref{fig:rates_unc_vs_T} as a function of temperature for different values of [Fig.~\ref{fig:rates_unc_vs_T}(a)] magnetic field, [Fig.~\ref{fig:rates_unc_vs_T}(b)] rotation, and [Fig.~\ref{fig:rates_unc_vs_T}(c)] radius. The rates increase nonlinearly with temperature, magnetic field intensity, rotation, and radius. Not shown in the figure, $v_2$ is mostly insensitive to temperature.

In Fig.~\ref{fig:rates_unc_vs_R} the radial dependence of unconstrained photon emission rates at a fixed low transverse momentum $k_T=0.5$ GeV is compared for different values of [Fig.~\ref{fig:rates_unc_vs_R}(a)] magnetic field, [Fig.~\ref{fig:rates_unc_vs_R}(b)] rotation, and [Fig.~\ref{fig:rates_unc_vs_R}(c)] temperature. The rates increase nonlinearly with temperature, rotation, and radius. At varying magnetic field and radius, a peculiar effect is observed. At large radius, as expected, stronger magnetic fields result in a greater radiation intensity. Instead, in small systems, we notice an ``inverse field effect,'' where weaker magnetic fields result in a greater rate. We investigate this effect in more detail below. We also notice that the weak magnetic field seems to have reached convergence at $R=50$ fm.

\begin{figure}[t!b]
    \centering
    \includegraphics[width=0.95\linewidth]{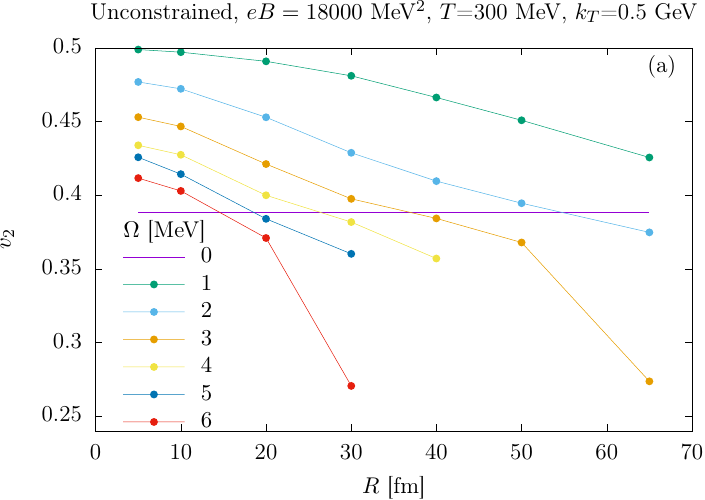}
    \includegraphics[width=0.95\linewidth]{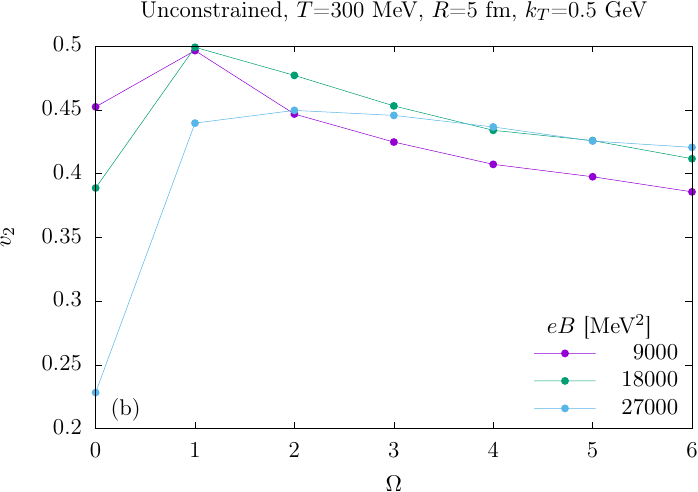}
    \caption{Radius (a) and rotation (b) dependence of $v_2$ for various values of, respectively, rotation $\Omega$ and magnetic field $eB$. }
    \label{fig:v2_unc_vs_R}
\end{figure}

From the elliptic flow at fixed $k_T=0.5$ GeV depicted in Fig.~\ref{fig:v2_unc_vs_R}, we conclude that the effect of an increasing rotation is to reduce the angular anisotropies, resulting in a lower $v_2$. Moreover, we see from Fig.~\ref{fig:v2_unc_vs_R}(b) that the relevance of rotation on $v_2$ is proportional to the ratio between rotation and magnetic field.
It is possible that including the stages of heavy-ion collisions where rotation is dominant compared to the rapidly decaying magnetic field reduces the $v_2$ obtained in Fig.~\ref{fig:best_fit}. As noted earlier, the effect of increasing the system radius is to reduce $v_2$, once again suggesting that the inclusion of a proper boundary condition to have a well-defined large-volume limit can help to explain the experimental measurements of the photon elliptic flow.

\begin{figure}[t!b]
    \centering
    \includegraphics[width=0.95\linewidth]{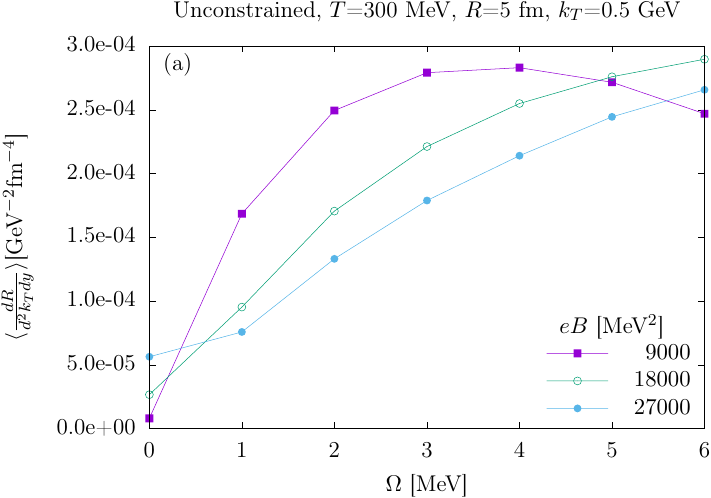}
    \includegraphics[width=0.95\linewidth]{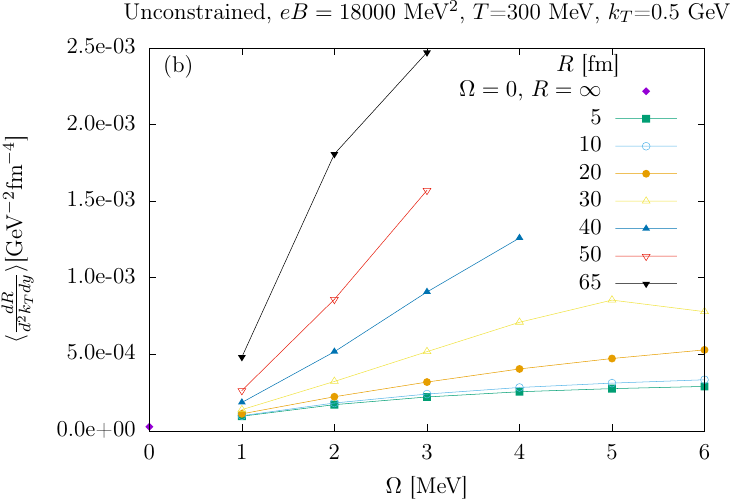}
    \includegraphics[width=0.95\linewidth]{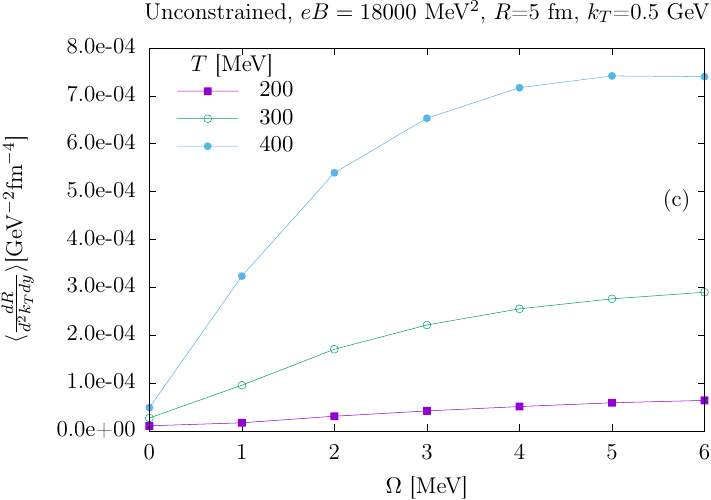}
    \caption{Rotation dependence. Rates with respect to the rotation $\Omega$ for various values of the (a) magnetic field $eB$, (b) radius $R$, (c) temperature $T$. }
    \label{fig:rates_unc_vs_Omega}
\end{figure}

The impact of rotation on the unconstrained rates is studied in Fig.~\ref{fig:rates_unc_vs_Omega} at fixed $k_T=0.5$ GeV with different values of [Fig.~\ref{fig:rates_unc_vs_Omega}(a)] magnetic field, [Fig.~\ref{fig:rates_unc_vs_Omega}(b)] radius, and [Fig.~\ref{fig:rates_unc_vs_Omega}(c)] temperature. We generally observe a nonlinear growth with rotation, with a steeper slope for larger radii. While some curves appear to reach an asymptotic value, the weak field line in Fig.~\ref{fig:rates_unc_vs_Omega}(a) demonstrates that the enhancement effect diminishes as rotation increases; this aligns with findings for synchrotron emission in fast rotation~\cite{Buzzegoli:2024nzd}.
\begin{figure}[h!tb]
    \centering
    \includegraphics[width=0.95\linewidth]{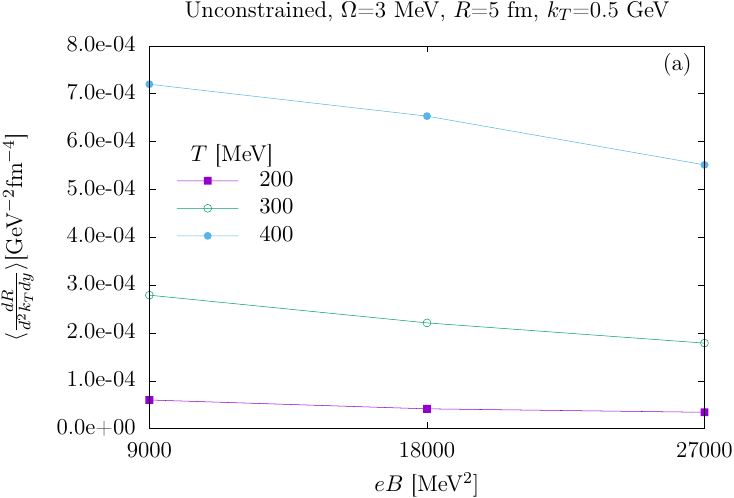}
    \includegraphics[width=0.95\linewidth]{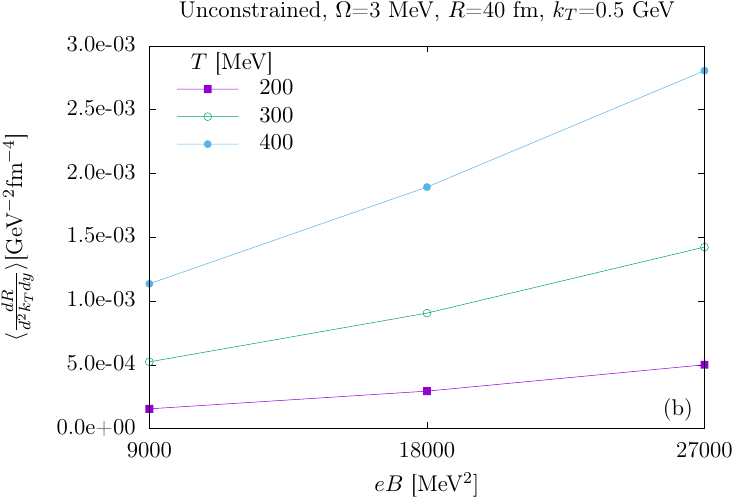}
    \includegraphics[width=0.95\linewidth]{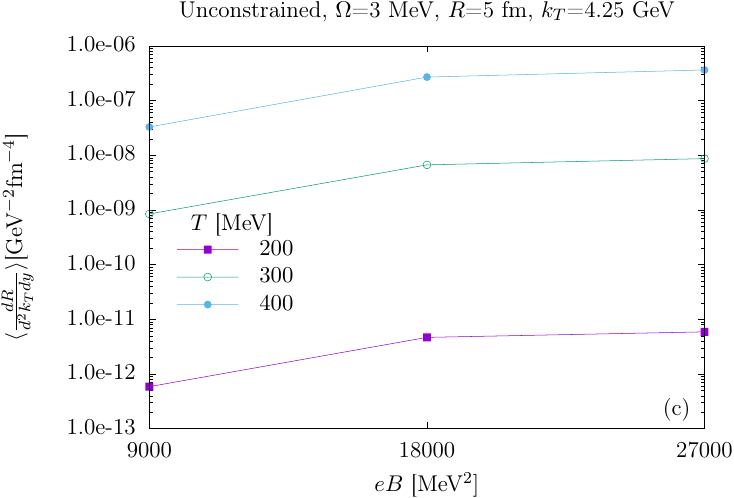}
    \caption{Rates at fixed $\Omega=3$ MeV with respect to the magnetic field $eB$. (a) Small radius $R=5$ fm and small momentum $k_T=0.5$ GeV shows the IFE. (b) A large radius $R=40$ fm and small momentum does not show the IFE. Similarly, (c) shows that, for large momentum $k_T=4.25$ GeV, there is no IFE even at small radius $R=5$ fm. }
    \label{fig:rates_unc_vs_eB_T}
\end{figure}
%

%********************************************************************
\subsection{Magnetic field dependence: Inverse field effect}\label{sec:InverseField}
%********************************************************************
It is worth analyzing the impact of the magnetic field strength and in particular the inverse field effect (IFE) observed in Fig.~\ref{fig:rates_unc_vs_R} for a small system. Fixing the angular velocity to $\Omega=3$ MeV, and taking a small radius $R=5$ fm and small transverse momentum $k_T=0.5$ GeV, Fig.~\ref{fig:rates_unc_vs_eB_T}(a) shows an unexpected IFE, where rates at a given temperature are larger for weaker magnetic field intensities.
As the radius is increased, for instance $R=40$ fm in Fig.~\ref{fig:rates_unc_vs_eB_T}(b), the IFE disappears and the rates grow with the magnetic field. The IFE disappears for large transverse momentum as well, see Fig.~\ref{fig:rates_unc_vs_eB_T}(c) showing $R=5$ fm and $k_T=4.25$ GeV. Similarly, at large transverse momentum and radius, the rates grow more than linearly with the magnetic field.

\begin{figure}[tb]
    \centering
    \includegraphics[width=0.95\linewidth]{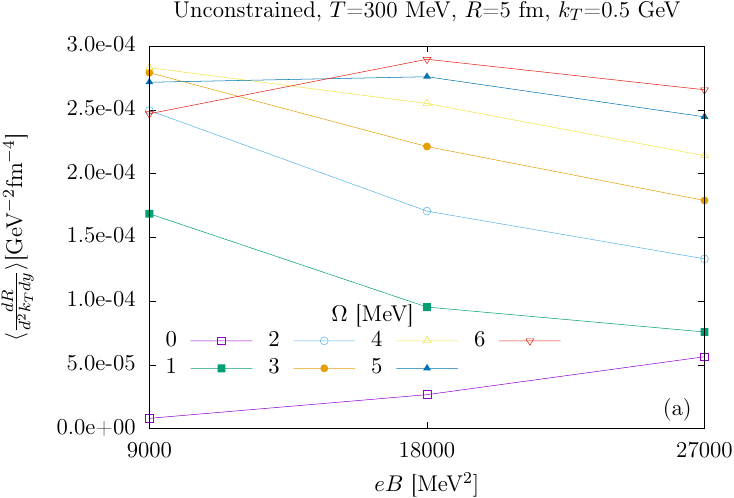}
    \includegraphics[width=0.95\linewidth]{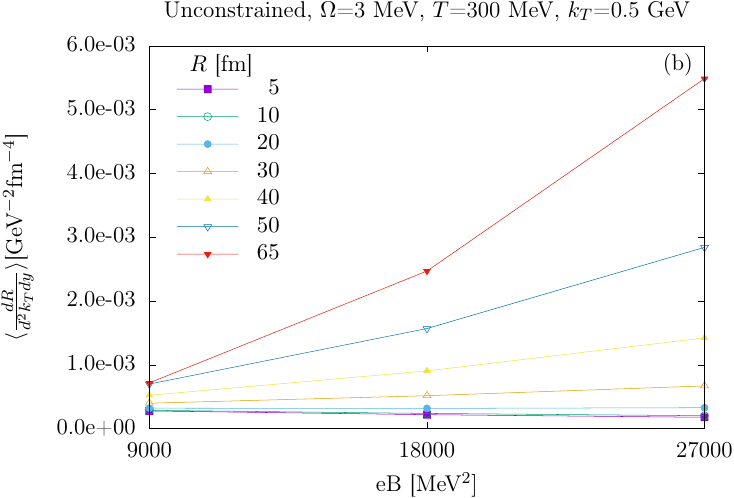}
    \caption{Magnetic field dependence. Rates with respect to the rotation $eB$ for various values of the (a) rotation $\Omega$ and (b) radius $R$. }
    \label{fig:rates_unc_vs_eB_2}
\end{figure}

Figure~\ref{fig:rates_unc_vs_eB_2} studies how the IFE changes for different values of rotation [Fig.~\ref{fig:rates_unc_vs_eB_2}(a)] and radius [Fig.~\ref{fig:rates_unc_vs_eB_2}(b)]. It is known that there is no IFE for $\Omega=0$. Instead, we observe that the IFE is present for $\Omega=1$ MeV and the IFE tends to disappear for larger angular velocities. Figure~\ref{fig:rates_unc_vs_eB_2}(b) clearly shows how the IFE disappears for sufficiently large systems. For $R=5$ and $10$ fm, the slope of the lines is negative, for $R=20$ fm is almost flat and for larger radius the slope is positive.
The Con emission, not shown in the figures, does not have the IFE; this is understood because the Con case reduces the phase space, allowing only for the less relevant transitions, therefore placing the different magnetic fields on the same footing. Instead, in the Unc case, weaker magnetic fields have access to the most relevant transitions while stronger magnetic fields cannot access their most relevant transitions that require larger angular momenta.

The reason for the appearance of the IFE is to be found in the complex interplay between the extent of the allowed phase space, constrained both by kinematics and by causality, and what are the most probable and most energetic transitions. These quantities all depend on the magnetic field intensity, the photon transverse momentum, and the radius, constraining the phase space of the initial state of the quark. As we are using a simpler solution without a boundary condition but with a causal cutoff, it is not clear from this analysis whether this IFE is a physical effect or an artifact of the approximation used.

%====================================================================
\section{Conclusions and Outlook}\label{sec:Conclusion}
%====================================================================
In summary, we studied the photon emission from a rigidly rotating hot plasma in a constant homogeneous magnetic field, which we called the RoSyRa. We solved the Dirac equation for free fermions in the presence of magnetic field and we accounted for the rotation of the medium using a rotating coordinate system. We derived solutions of the Dirac equation without boundary conditions and limited the physical volume of the system to a cylinder of radius $R=1/\Omega$ to preserve causality. From these solutions we computed the transition amplitude of the photon radiation process $q\to q+\gamma$ and the resulting differential photon yield as given by (\ref{eq:FinalResult}). In particular, using numerical methods in the slow rotation regime where $\Omega < \sqrt{|q eB|}$, we studied the angular average of the photon emission rate $\langle \D R/\D^2k_T \D y \rangle$ and the angular anisotropy characterized by elliptic flow $v_2$. We considered two scenarios, one in which the initial quarks are constrained in the plasma volume and can escape it after the emission of a photon, and one in which the final quark is also constrained to remain in the initial volume.

We showed that the plasma rotation greatly enhances the yields of synchrotron radiation while maintaining a large $v_2$. These two features are needed to explain the direct photon puzzle in heavy-ion collisions~\cite{David:2019wpt, Gale:2021emg, Paquet:2015lta}. Figures~1 and~2 show the main results comparing the predictions from RoSyRa with the experimental data~[16-19]. While this comparison remains qualitative due to the simplified system evolution, the results suggest that RoSyRa, combined with a realistic plasma evolution, could resolve the tension between photon emission theories and experimental observations.

The main limitations of the model studied here are the use of unbounded solutions of the Dirac equation and the use of static and homogeneous thermodynamic variables (temperature, rotation, and magnetic field). Proper boundary conditions can be included by extending the work of~\cite{Buzzegoli:2024nzd} for photons emitted by a single fermion to the emission by a plasma of fermions in the same way as done in this work. In this way, modes with all angular momentum can be included, extending the validity of the model to all possible values of rotation, in particular the case where it is larger than the magnetic field. This is important for applications to heavy-ion collisions, as the magnetic field is believed to decay quickly, while rotation is sustained throughout the evolution. As a consequence, only in the earlier stages is the magnetic field the dominant quantity, while at later stages rotation exceeds it. Instead, to extend this model to nonhomogeneous plasmas one can adopt the adiabatic approximation and promote the thermodynamic constant variables to fields. This requires a numerical integration over the volume of the plasma and over the time evolution. As usual, the evolution and the space dependence of the thermodynamic variables can be obtained with a hydrodynamic model or code.

The numerical evaluation of the photon emission given by (\ref{eq:FinalResult}) requires more computational time than the conventional models for direct photon production~\cite{Gale:2021emg,Arnold:2001ms}. However, the results seem to be close to what can be obtained using a quasiclassical description of the phenomena, see for instance Figs.~\ref{fig:Comparison} and~\ref{fig:rotating_v2_con_unc_vs_T_and_eB}. The quasiclassical methods developed in~\cite{Buzzegoli:2025qfl} could be an efficient way to estimate the effect, especially if a full time and space dependence is considered.

We conclude by mentioning other possible experimental signatures of RoSyRa. As the effect is driven by the magnetic field, we expect a different impact if the background magnetic field is different. This is precisely the objective of the 2018 RHIC isobar run~\cite{Voloshin:2010ut,STAR:2021mii}, with Ru-Ru and Zr-Zr at $\sqrt{s_{NN}}=200$ GeV. According to RoSyRa, we expect larger photon excess and larger $v_2$ for stronger magnetic fields, which is a simple qualitative behavior. Moreover, photons emitted by synchrotron radiation at midrapidity are linearly polarized. We expect that rotation will not wash this polarization. A linear polarization, including the one in the (virtual) photons emitted by RoSyRa, can be revealed in specific azimuthal anisotropies in the dielectron spectrum~\cite{Li:2019yzy}. For this end, a dedicated calculation of photon polarization from RoSyRa and the resulting anisotropy in the dilepton spectra must be developed and performed.

%%%%%%%%%%%%%%%%%%%%%%%%%%%%%%%%
\acknowledgments
M.B. and S.B. are grateful to V. Ambrus, A. Bandyopadhyay, and M. Chernodub for many fruitful discussions. We are indebted to Xinyang Wang and Igor Shovkovy for facilitating a detailed comparison of our results with theirs.
The work of M.B. and S.B. was supported by the European Union - NextGenerationEU through Grant No.\ 760079/23.05.2023, funded by the Romanian Ministry of Research, Innovation and Digitization through Romania's National Recovery and Resilience Plan, call no.\ PNRR-III-C9-2022-I8. The work of K.T., J.D.K., and N.V. was supported in part by the U.S. Department of Energy under Grant No.\ DE-SC0023692.

\section*{Data availability}
The data that support the findings of this article are openly available~\cite{Buzzegoli:2026sji}.
%\clearpage

\appendix
%====================================================================
\section{Differential rate of the quark splitting process}\label{sec:AppA}
%====================================================================
In this appendix, we write the explicit expressions for the differential rate of the quark splitting process $q_i\to q_f\, +\, \gamma$ and its sum over quark spin polarizations.
In Sec.~\ref{sec:ByQuark} we obtained the differential rate
\begin{equation}
\begin{split}
\frac{\D\dot{w}_{n,a,n',a',\zeta,\zeta'}}{\D^3k} = &
    \frac{q^2 e^2}{4\pi} \frac{1}{2\pi} \frac{\delta(E-E'-\omega)}{\omega} \\
&\times \left|\frac{2\pi}{L}\delta(p_z-p_z'-k_z) \right|^2
    \sum_h |\langle \bm{j} \cdot \bm{\Phi} \rangle|^2,
\end{split}
\end{equation}
with
\begin{equation}
\begin{split}
\label{eq:AppjPhiAmplitudeSq}
|\langle \bm{j} \cdot \bm{\Phi} \rangle|^2 =& \frac{1}{2} I_{a,a'}^2(x)\left|
        \sin\theta [K_4 I_{n-1,n'-1}(x) - K_3 I_{n,n'}(x)]\right.\\
        &+ K_1 (h-\cos\theta) I_{n,n'-1}(x) \\
        &\left.- K_2 (h+\cos\theta) I_{n-1,n'}(x)\right|^2,
\end{split}
\end{equation}
where the $K_i$ coefficients are given in terms of the $C_i$ coefficients of the quark wave function in Eq.~(\ref{eq:QuarkWaveFunction}) as follows:
\begin{equation}
	\begin{split}
		K_1 = C_1' C_4 + C_3' C_2 , \qquad K_2 = C_4' C_1 + C_2' C_3 , \\
		K_3 = C_4' C_2 + C_2' C_4 , \qquad K_4 = C_1' C_3 + C_3' C_1 .
	\end{split}
\end{equation}
Writing these out in $A$'s and $B$'s defined in Eq.~(\ref{eq:DefAandB}), we obtain
\begin{subequations}
\begin{align}
K_1 =& \frac{1}{4} B'_+ B_- (A_+'A_- + \zeta\zeta' A_-'A_+),\\
K_2 =& \frac{1}{4} B'_- B_+ \zeta \zeta' (A_+'A_- + \zeta\zeta' A_-'A_+) ,
\end{align}
\begin{align}
K_3 =& \frac{1}{4} B'_- B_- (A_-'A_- - \zeta\zeta' A_+'A_+) , \\
K_4 =& \frac{1}{4} B'_+ B_+ (-\zeta\zeta') (A_-'A_- - \zeta\zeta' A_+'A_+) ,
\end{align}
\end{subequations}
where $\zeta$ and $\zeta'$ are the polarization states of the initial and final state of the quark, respectively. We are interested in the products of two $K_i$ coefficients generated by the square modulus in Eq.~(\ref{eq:AppjPhiAmplitudeSq}).
There are only three combinations of $A_\pm$ needed in multiplying these together,
\begin{equation}
\begin{split}
(A_+' A_- +& \zeta\zeta' A_-'A_+)^2 = 2 \Bigg(1 - \frac{p_z p_z'}{EE'}\\
    &+ \zeta\zeta' \frac{\sqrt{E^2 - p_z^2}\sqrt{E'^2-p_z'^2}}{EE'}\Bigg) ,\nonumber \\
(A_-' A_- &- \zeta\zeta' A_+'A_+)^2 = 2 \Bigg(1 + \frac{p_z p_z'}{EE'}\\
    &- \zeta\zeta' \frac{\sqrt{E^2 - p_z^2}\sqrt{E'^2-p_z'^2}}{EE'}\Bigg) , \nonumber\\
(A_+'A_- &+ \zeta\zeta' A_-'A_+)(A_-'A_- - \zeta\zeta' A_+'A_+) =\\
-2& \left( \frac{p_z \sqrt{E'^2-p_z'^2} + \zeta\zeta' p_z' \sqrt{E^2-p_z^2}}{EE'} \right) .\nonumber
\end{split}
\end{equation}
The $B_\pm$'s multiply trivially or as
\begin{equation}
	B_+ B_- = \left(1 - \frac{M^2}{E^2 - p_z^2}\right)^{1/2} = \frac{\sqrt{2n|qeB|}}{\sqrt{E^2 - p_z^2}} .
\end{equation}

Assembling the $K_i$ products, one obtains
\begin{widetext}
\begin{subequations}
\label{eq:KCoeffProducts}
\begin{gather}
	K_1^2 = \frac{1}{8EE'} \left(1 + \frac{\zeta' M}{\sqrt{E'^2 - p_z'^2}}\right) \left(1 - \frac{\zeta M}{\sqrt{E^2 - p_z^2}}\right) \left(EE' - p_z p_z' + \zeta\zeta' \sqrt{E^2 - p_z^2}\sqrt{E'^2-p_z'^2}\right) , \\
	K_2^2 = \frac{1}{8EE'} \left(1 - \frac{\zeta' M}{\sqrt{E'^2 - p_z'^2}}\right) \left(1 + \frac{\zeta M}{\sqrt{E^2 - p_z^2}}\right) \left(EE' - p_z p_z' + \zeta\zeta' \sqrt{E^2 - p_z^2}\sqrt{E'^2-p_z'^2}\right)	, \\
	K_3^2 = \frac{1}{8EE'} \left(1 - \frac{\zeta' M}{\sqrt{E'^2 - p_z'^2}}\right) \left(1 - \frac{\zeta M}{\sqrt{E^2 - p_z^2}}\right) \left(EE' + p_z p_z' - \zeta\zeta' \sqrt{E^2 - p_z^2}\sqrt{E'^2-p_z'^2}\right) ,\\
	K_4^2 = \frac{1}{8EE'} \left(1 + \frac{\zeta' M}{\sqrt{E'^2 - p_z'^2}}\right) \left(1 + \frac{\zeta M}{\sqrt{E^2 - p_z^2}}\right) \left(EE' + p_z p_z' - \zeta\zeta' \sqrt{E^2 - p_z^2}\sqrt{E'^2-p_z'^2}\right) ,
\end{gather}
\begin{gather}
	K_1 K_2 = \frac{1}{8EE'} \zeta\zeta' \frac{\sqrt{2n|q eB|}\sqrt{2n'|q eB|}}{\sqrt{E^2 - p_z^2}\sqrt{E'^2 - p_z'^2}} \left(EE' - p_z p_z' + \zeta\zeta' \sqrt{E^2 - p_z^2}\sqrt{E'^2-p_z'^2}\right) ,\\
	K_3 K_4 = \frac{-1}{8EE'} \zeta\zeta' \frac{\sqrt{2n|q eB|}\sqrt{2n'|q eB|}}{\sqrt{E^2 - p_z^2}\sqrt{E'^2 - p_z'^2}} \left(EE' + p_z p_z' - \zeta\zeta' \sqrt{E^2 - p_z^2}\sqrt{E'^2-p_z'^2}\right) ,\\
	K_1 K_3 = \frac{-1}{8EE'} \frac{\sqrt{2n'|q eB|}}{\sqrt{E'^2 - p_z'^2}} \left(1 - \frac{\zeta M}{\sqrt{E^2 - p_z^2}}\right) \left( p_z \sqrt{E'^2-p_z'^2} + \zeta\zeta' p_z' \sqrt{E^2-p_z^2} \right) , \\
	K_1 K_4 = \frac{1}{8EE'} \frac{\sqrt{2n|q eB|}}{\sqrt{E^2 - p_z^2}} \left(1 + \frac{\zeta' M}{\sqrt{E'^2 - p_z'^2}}\right) \left( \zeta\zeta' p_z \sqrt{E'^2-p_z'^2} + p_z' \sqrt{E^2-p_z^2} \right) , \\
	K_2 K_3 = \frac{-1}{8EE'} \frac{\sqrt{2n|q eB|}}{\sqrt{E^2 - p_z^2}} \left(1 - \frac{\zeta' M}{\sqrt{E'^2 - p_z'^2}}\right) \left( \zeta\zeta' p_z \sqrt{E'^2-p_z'^2} + p_z' \sqrt{E^2-p_z^2} \right) , \\
	K_2 K_4 = \frac{1}{8EE'} \frac{\sqrt{2n'|q eB|}}{\sqrt{E'^2 - p_z'^2}} \left(1 + \frac{\zeta M}{\sqrt{E^2 - p_z^2}}\right) \left( p_z \sqrt{E'^2-p_z'^2} + \zeta\zeta' p_z' \sqrt{E^2-p_z^2} \right) .	
\end{gather}
\end{subequations}
\end{widetext}

In computing the photon emission in a plasma, one needs to sum these products over the spin polarizations $\zeta$ and $\zeta'$. Given two coefficients $K_i$, we define
\begin{equation}
\overline{K_i K_j}=\frac{1}{2}\sum_{\zeta,\zeta'}K_i K_j.
\end{equation}
and we obtain
\begin{subequations}
\begin{gather}
    \overline{K_1^2} = \overline{K_2^2} = \frac{(E-m\Omega) (E'-m'\Omega) - p_z p_z' - M^2}{4 (E-m\Omega) (E'-m'\Omega)}, \\
    \overline{K_3^2} = \overline{K_4^2} = \frac{(E-m\Omega) (E'-m'\Omega) + p_z p_z' - M^2}{4 (E-m\Omega) (E'-m'\Omega)},
\end{gather}
\begin{gather}
    \overline{K_1K_2} = \overline{K_3K_4} = \frac{\sqrt{2n|q eB|}\sqrt{2n'|q eB|}}{4 (E-m\Omega) (E'-m'\Omega)},\\
    \overline{K_1K_3} = -\overline{K_2K_4} = \frac{-p_z \sqrt{2n'|q eB|}}{4 (E-m\Omega) (E'-m'\Omega)}, \\
    \overline{K_1K_4} = -\overline{K_2K_3} = \frac{p_z' \sqrt{2n|q eB|}}{4 (E-m\Omega) (E'-m'\Omega)}.
\end{gather}
\end{subequations}
Putting all together, the polarization sums of the differential rate are
\begin{equation}
\begin{split}
\frac{\D\dot{w}_{n,a,n',a'}}{\D^3k} = &
    \frac{q^2 e^2}{4\pi} \frac{1}{2\pi} \frac{\delta(E-E'-\omega)}{\omega}\\
    \times&\left|\frac{2\pi}{L}\delta(p_z-p_z'-k_z) \right|^2
    \sum_h \sum_{\zeta,\zeta'}|\langle \bm{j} \cdot \bm{\Phi} \rangle|^2,
\end{split}
\end{equation}
with
\begin{widetext}
\begin{equation}
\begin{split}
 \sum_{\zeta,\zeta'} |\langle \bm{j} \cdot \bm{\Phi} \rangle|^2 =& I_{a,a'}^2 \Bigg\{ 2 \overline{K_1^2} \left[ I_{n,n'-1}^2 + I_{n-1,n'}^2 \right]
        + \sin^2\theta \left[ \overline{K_4^2} \left( I_{n-1,n'-1}^2 + I_{n,n'}^2 \right) - \overline{K_1^2} \left( I_{n-1,n'}^2 + I_{n,n'-1}^2 \right) \right]\\
        &- 2 \overline{K_1K_2} \sin^2\theta \left[ I_{n-1,n'-1} I_{n,n'} + I_{n-1,n'} I_{n,n'-1} \right]
        - 2 \sin\theta \cos\theta \Big[ \overline{K_1K_4} \left( I_{n-1,n'-1} I_{n,n'-1} + I_{n,n'} I_{n-1,n'} \right) \\
        & + \overline{K_2K_4} \left( I_{n-1,n'-1} I_{n-1,n'} + I_{n,n'} I_{n,n'-1} \right) \Big] \\
        &+ h \Big[ 2 \overline{K_1^2} \cos\theta \left( I_{n-1,n'}^2 - I_{n,n'-1}^2 \right)
         + 2 \sin\theta \Big( \overline{K_1K_4} \left[ I_{n-1,n'-1} I_{n,n'-1} - I_{n,n'} I_{n-1,n'} \right] \\
        & + \overline{K_2K_4} \left[ I_{n,n'} I_{n,n'-1} - I_{n-1,n'-1} I_{n-1,n'} \right] \Big) \Bigg] \Big\},
\end{split}
\end{equation}
\newpage
\end{widetext}
where the argument of the $I$ functions is always $x= k_\perp^2/(2|q e B|)$.
Notice that the only dependence on the photon helicity $h$ is a linear term that does not contribute when summed over $h$. Therefore, after defining two auxiliary functions $\Gamma$ and $\Gamma^{(1)}$ as
\begin{equation}
\begin{split}
\label{eq:GammaDefApp}
\sum_{\zeta,\zeta'} |\langle \bm{j} \cdot \bm{\Phi}\rangle|^2 \equiv&
    \Gamma_{n,a}(n',a',\bm{k}) + h \Gamma_{n,a}^{(1)}(n',a',\bm{k})\\
    =& \Gamma_{n,a}^{h}(n',a',\bm{k}),
\end{split}
\end{equation}
the sum over $h$ can be done explicitly,
\begin{equation}
\begin{split}
\frac{\D\dot{w}_{n,a,n',a'}}{\D^3k} = &
    \frac{q^2 e^2}{4\pi} \frac{1}{2\pi} \frac{\delta(E-E'-\omega)}{\omega}\\
    \times&\left|\frac{2\pi}{L}\delta(p_z-p_z'-k_z) \right|^2
2  \Gamma_{n,a}(n',a',\bm{k}).
\end{split}
\end{equation}
%

%====================================================================
\section{Evaluation of phase space} \label{sec:PhaseSpace}
%====================================================================
In this appendix we show how the conservation laws constrain the parameter space related to the photon emission. The conservation of the space components of momentum has been included by integrating the corresponding $\delta$ functions and by writing the momentum and angular momentum of the photon in terms of the momentum and angular momentum of the quarks. The nontrivial part is translating the conditions for the conservation of energy as inequalities for the quantum numbers of the splitting process: $n$, $a$, $n'$ and $a'$.

As mentioned in the main text, the energy conservation $E-E'-\omega=0$ can be written as
\begin{equation}\label{eq:AppEnerCons}
\sqrt{a^2+p_y^2} -\sqrt{b^2+(p_y-c)^2} = -\Delta,
\end{equation}
where we defined
\begin{subequations}
\begin{align}
a^2 =& 2|q eB|n + M^2,& b^2 =& 2|q eB|n' + M^2,\\
c =& \omega\cos\theta,& \Delta =& (n-a-n'-a')\Omega-\omega.
\end{align}
\end{subequations}
Equation (\ref{eq:AppEnerCons}) admits solutions for $p_y$ in the following cases:
\begin{equation}\label{eq:AppConditions}
\begin{cases}
-\bar{s} \Delta> \bar{s} P,\,-\bar{s}\Delta<-|c| & \text{No roots}\\
|c|<-\bar{s}\Delta<\bar{s} P & \text{2 roots }p^{0\mp}_y\\
-|c|<-\Delta<|c| & \text{1 root }p^{0-}_y
\end{cases}
\end{equation}
where we defined $\bar{s}=$sign$(a-b)$, and $P$ is the maximum or minimum value of $f(p_y)=\sqrt{a^2+p_y^2} -\sqrt{b^2+(p_y-c)^2}$ depending on $\bar{s}$ and it is given by
\begin{equation}
P = (a-b)\sqrt{1+\frac{c^2}{(a-b)^2}}.
\end{equation}
Clearly the conditions in (\ref{eq:AppConditions}) are inequalities between the quantum numbers of the splitting process $q_i\,\to\, q_f\, + \gamma$. With straightforward but lengthy calculations, we can use the inequalities in (\ref{eq:AppConditions}) to split the sums in the rate (\ref{eq:FinalResult}) into the regions of the parameters where the energy conservation can be satisfied.

First of all, we use the following shorthand notation for Eq.~(\ref{eq:FinalResult}):
\begin{equation}
\begin{split}
\mathcal{R}_f =& \int_{-\infty}^\infty dp_y \sum_{n,n'} \sum_{a,a'} \delta(E-E'-\omega) W(p_y)\\
=& \sum_{n,n'} \sum_{a,a'} \sum_{p^0_y} W(p^0_y).
\end{split}
\end{equation}
Then, we split between three cases: $a>b$, $a=b$, and $a<b$, which correspond respectively to
$n'<n$, $n'=n$, and $n'>n$. In the summation this separation becomes:
\begin{equation}
\mathcal{R}_f = \sum_{n} \left[\sum_{n'<n} + \sum_{n'=n} +\sum_{n'>n}\right]
    \sum_{a,a'} \sum_{p^0_y} W(p^0_y).
\end{equation}
At this point it is easy to translate the inequalities (\ref{eq:AppConditions}) into intervals for the quantum number $a'$ whose boundary depends on the fixed parameters $\omega$, $\Omega$, and $\theta$ and on the other quantum numbers that are being summed: $n$, $n'$, and $a$. We call $R_1$ the interval of $a'$ for which the energy conservation has exactly one solution. We call $R_2^+$ and $R_2^-$ the intervals of $a'$ for which the energy conservation has exactly two solutions, respectively, for $n'< n$ and for $n'> n$. Furthermore, we notice that, in the case $n'\leq n$, the obtained inequalities for $a'$ can only be satisfied for $a\geq a^+_{\rm min}$, whereas in the case $n'> n$ are satisfied only for $a\geq a^-_{\rm min}$.

In conclusion, for $\Omega>0$ to include only the kinematically allowed regions, we can split the summation in the following regions:
\begin{align}
\label{eq:Rf_constrainedKinematics}
\mathcal{R}_f =& \sum_{n=0}^\infty  \Bigg[ \sum_{n'<n} \sum_{a=a^+_{\rm min}}^{\rho_R}\left(\sum_{a'\in R_2^+}^{0\leq a' \leq\rho_\Omega} + \sum_{a'\in R_1}^{0\leq a' \leq\rho_\Omega} \right) \nonumber\\
&+  \sum_{n'=n} \sum_{a=a^+_{\rm min}}^{\rho_R}\sum_{a'\in R_1}^{0\leq a' \leq\rho_\Omega}\\
 &+  \sum_{n'>n} \sum_{a=a^-_{\rm min}}^{\rho_R}\left(\sum_{a'\in R_1}^{0\leq a' \leq\rho_\Omega} + \sum_{a'\in R_2^-}^{0\leq a' \leq\rho_\Omega} \right)
\Bigg]\sum_{p^0_y} W(p^0_y) \nonumber
\end{align}
where
\begin{align}
a^-_{\rm min}=& \text{Max}\left\{0,n-n'-\frac{\omega-P}{\Omega} \right\},\\
a^+_{\rm min}=& \text{Max}\left\{0,n-n'-\left(1+|\cos\theta|\right)\frac{\omega}{\Omega}\right\},
\end{align}
and
\begin{align}
R_1:&\, (1 - |\cos\theta|)\frac{\omega}{\Omega} < a'-n'+n-a < (1 +|\cos\theta|)\frac{\omega}{\Omega},\\
R_2^+:&\, \frac{\omega-P}{\Omega} < a'-n'+n-a < (1 -|\cos\theta|)\frac{\omega}{\Omega},\\
R_2^-:&\, (1 +|\cos\theta|)\frac{\omega}{\Omega} < a'-n'+n-a < \frac{\omega-P}{\Omega}.
\end{align}
The sums over $p^0_y$ include the two terms $p^0_y=p^{0-}_y$ and $p^0_y=p^{0+}_y$ in the regions $R_2^+$ and $R_2^-$ and only $p^0_y=p^{0-}_y$ for $R_1$, where $p^{0\pm}_y$ are given in Eq.~(\ref{eq:pyRoots}).
Notice that in the absence of rotation solutions with $n'\geq n$ are not allowed. In the case of small rotation considered in this work, transitions to higher Landau level numbers are allowed even though their contribution is usually negligible. The equations given above are valid for finite positive rotation only, for $\Omega<0$ the regions become
\begin{align}
R_1:&\, (1 + |\cos\theta|)\frac{\omega}{\Omega} < a'-n'+n-a < (1 -|\cos\theta|)\frac{\omega}{\Omega},\\
R_2^+:&\, (1 -|\cos\theta|)\frac{\omega}{\Omega} < a'-n'+n-a < \frac{\omega-P}{\Omega},\\
R_2^-:&\, \frac{\omega-P}{\Omega} < a'-n'+n-a < (1 +|\cos\theta|)\frac{\omega}{\Omega},
\end{align}
and
\begin{align}
a^-_{\rm min}=& \text{Max}\left\{0,n-n'-\left(1-|\cos\theta|\right)\frac{\omega}{\Omega} \right\},\\
a^+_{\rm min}=& \text{Max}\left\{0,n-n'-\frac{\omega-P}{\Omega}\right\}.
\end{align}
Note that $\Omega$ being negative, the upper limits of the intervals $R_1$, $R_2^\pm$ are reduced by rotation, which can also become negative, meaning that no solutions are allowed for those values of $n$, $n'$, $a$, $\omega$, and $\theta$. As a result, the allowed phase space is overall greatly reduced, which is one of the ways the radiation is suppressed by rotation.

The case of a nonrotating plasma is discussed in Sec.~\ref{sec:nonrotating}. 
For vanishing rotation $\Omega=0$, the sums over $a$ and $a'$ are factorized and can be done separately. As mentioned, the energy is conserved only if $n>n'$. In the special case $\theta=0$, the energy conservation admits only one finite solution given by
\begin{equation}
p^0_y(\theta=0) = \frac{2a^2 b^2-a^4-b^4+4a^2 c^2}{4(a^2-b^2)c}.
\end{equation}
Otherwise, the energy conservation admits solutions only when $\omega\leq (a-b)\sqrt{1+c^2/(a-b)^2}$, which translates into an upper limit for $n'$, that is $n'<n'_{\rm max}$, where
\begin{equation}
\begin{split}
n'_{\rm max} =& n+\frac{\omega}{2|q_f eB|} \Big[ \omega(1-\cos^2\theta)\\
    &\left.-2\sqrt{(M_f^2+2n|q_f eB|)(1-\cos^2\theta)}\right].
\end{split}
\end{equation}
Only when $\omega= P=(a-b)\sqrt{1+c^2/(a-b)^2}$ the energy conservation admits the two solutions $p^{0\pm}_y$ in Eq.~(\ref{eq:pyRoots}), otherwise it only has the solution $p^{0-}_y$.
In conclusion, the sums in the rate $\mathcal{R}_f$ in Eq.~(\ref{eq:FinalResultNoOmega}) for vanishing rotation can be decomposed into
\begin{equation}
\label{eq:Rf_constrainedKinematicsNoRotation}
\begin{split}
\mathcal{R}_f\left(\Omega=0\right) = \sum_{n=0}^\infty  & \sum_{\substack{n'=0 \\ n'<n}}^{n'_{\rm max}} \left[ W(p^{0-}_y) +
    W(p^{0+}_y) \delta_{\omega-P,0}\right].
\end{split}
\end{equation}
%

%====================================================================
\section{Reflection Symmetry} \label{sec:Reflection}
%====================================================================
In this appendix, we discuss the angular symmetries possessed by the rotating synchrotron radiation. In particular, we show that these symmetries allow us to restrict the angular integral in the average yields~(\ref{eq:IntYields1}) and in the flow coefficient~(\ref{eq:v2Integral}) from $\theta$ between 0 and $2\pi$ to $\theta$ between 0 and $\pi/2$.

We consider two reflection transformations: the reflection symmetry in the $yz$ plane (the reaction plane), and a reflection in the $xz$ plane, to which $\bm{B}$ and $\bm{\Omega}$ are orthogonal. We want to study how these reflections change the photon spectrum in Eq.~(\ref{eq:FinalResult}), see also the $\Gamma$ function in Eq.~(\ref{eq:Gamma}). Notice that those quantities depend on $\theta$ only by simple trigonometric functions: $\sin\theta$ and $\cos\theta$, and the reflections under consideration only change $\cos\theta$ and $\sin\theta$ up to a sign. Furthermore, notice that to properly convert coordinates between the original cylindrical coordinates in Eq.~(\ref{eq:Gamma}) and the laboratory hyperbolic coordinates used in Eq.~(\ref{eq:FinalResult}), every appearance of the function $\sin\theta$ in Eq.~(\ref{eq:Gamma}) must be replaced by $|\sin\theta|$, that in turn ensures the invariance under the reflections.

Consider first the reflection symmetry in the $yz$ plane. Performing this reflection sends $\cos\theta$ to $\cos\theta$, leaving all the quantities in Eq.~(\ref{eq:FinalResult}) unaffected. For this reason, we can restrict the angular integration in $\theta$ from 0 to $\pi$.

The second reflection symmetry is in the $xz$ plane, and sends $\cos\theta$ to $-\cos\theta$. Showing the invariance for this case requires a more careful analysis of Eq.~(\ref{eq:FinalResult}). Consider first how $p^0_y$ changes under reflection. When the energy conservation admits two roots, they take the following two values:
\begin{equation}
p^{0\mp}_y = \frac{c(b^2-a^2+c^2-\Delta^2) \mp \textrm{sgn}(c \Delta) \sqrt{\Delta^2\alpha}}{2(c^2-\Delta^2)},
\end{equation}
with
\begin{equation}
\alpha = a^4+(b^2+c^2-\Delta^2)^2-2a^2(b^2-c^2+\Delta^2).
\end{equation}
and where $a^2=2n|q eB|+M^2$, $b^2=2n'|q eB|+M^2$, $c=\omega\cos\theta$, and $\Delta=(n-a-n'-a')\Omega-\omega$.
It is clear that under reflection $c\to-c$ and that $p^{0\mp}_y \to -p^{0\mp}_y$. Similarly, when the energy conservation admits only the root $p^{0-}_y$, under reflection the root transforms as $p^0_y \to -p^0_y$. In the last kinematic condition, when $a=b$ the root is given by
\begin{equation}
\begin{split}
p^0_y  =& \frac{c(c^2-\Delta^2)-\text{sgn}(c)\Delta\sqrt{(c^2-\Delta^2)(4a^2+c^2-\Delta^2)}}{2(c^2-\Delta^2)}\\
\to& -p^0_y .
\end{split}
\end{equation}
In conclusion, in all cases, the reflection acts on the root as $p^0_y \to -p^0_y$.
From this result, it immediately follows that $p_y' = p_y - \omega\cos\theta \to -p_y'$. Similarly, we obtain $E \to E$ and $E' \to E'$.
The denominator of Eq.~(\ref{eq:FinalResult}) also contains a $\cos\theta$ but it remains unaffected by the reflection,
\begin{equation}
\begin{split}
D=&|(E-m\Omega) \omega\cos\theta - p^0_y (\omega - (m-m')\Omega)|\\
 &\to |- (E-m\Omega) \omega\cos\theta + p^0_y (\omega - (m-m')\Omega)|=D ,
\end{split}
\end{equation}
because of the modulus.
Finally, we reflect the function $\Gamma_{n,a}$ given in Eq.~(\ref{eq:Gamma}) after changing coordinates. Notice that, in addition to replacing $\sin\theta$ with $|\sin\theta|$, the momenta $p_z$ and $p_z'$ in Eq.~(\ref{eq:Kappas}) become $p_y$ and $p_y'$ in the lab frame.  The only $K$-coefficients linear in $p_y$ or $p_y'$ are $\overline{K_2K_4}$ and $\overline{K_1K_4}$; under reflection these acquire minus signs. From Eqs. (\ref{eq:Gamma}) and (\ref{eq:GammaDefApp}), one sees that, in $\Gamma_{n,a}$, the only terms multiplying $\cos\theta$ are $\overline{K_2K_4}$ and $\overline{K_1K_4}$; these are thus unchanged. On the contrary, in $\Gamma_{n,a}^{(1)}$, these are the terms which are not multiplied by $\cos\theta$. So $\Gamma_{n,a}^{(1)}$ acquires a minus sign. Overall, under this transformation, the reflection is equivalent to a flip on the photon helicity,
\begin{equation}
    \Gamma_{n,a}^h \to \Gamma_{n,a}^{-h} .
\end{equation}
In conclusion, thanks to these reflection symmetries, in the angular integral we only need to evaluate $\theta$ from 0 to $\frac{\pi}{2}$, up to a change in helicity. As Eq.~(\ref{eq:FinalResult}) contains the sum over the helicities, and the $\Gamma_{n,a}^{(1)}$ is not contributing, we showed that the angular integral in the average yields~(\ref{eq:IntYields1}) and in the flow coefficient~(\ref{eq:v2Integral}) can be restricted from $\theta$ between 0 to $2\pi$ to $\theta$ between 0 to $\pi/2$.

%%%%%%%%%%%%%%%%%%%%%%%%%%%%%%%%
%\bibliographystyle{apsrev4-2}
%\bibliography{biblio}

\begin{thebibliography}{68}%
	\makeatletter
	\providecommand \@ifxundefined [1]{%
		\@ifx{#1\undefined}
	}%
	\providecommand \@ifnum [1]{%
		\ifnum #1\expandafter \@firstoftwo
		\else \expandafter \@secondoftwo
		\fi
	}%
	\providecommand \@ifx [1]{%
		\ifx #1\expandafter \@firstoftwo
		\else \expandafter \@secondoftwo
		\fi
	}%
	\providecommand \natexlab [1]{#1}%
	\providecommand \enquote  [1]{``#1''}%
	\providecommand \bibnamefont  [1]{#1}%
	\providecommand \bibfnamefont [1]{#1}%
	\providecommand \citenamefont [1]{#1}%
	\providecommand \href@noop [0]{\@secondoftwo}%
	\providecommand \href [0]{\begingroup \@sanitize@url \@href}%
	\providecommand \@href[1]{\@@startlink{#1}\@@href}%
	\providecommand \@@href[1]{\endgroup#1\@@endlink}%
	\providecommand \@sanitize@url [0]{\catcode `\\12\catcode `\$12\catcode
		`\&12\catcode `\#12\catcode `\^12\catcode `\_12\catcode `\%12\relax}%
	\providecommand \@@startlink[1]{}%
	\providecommand \@@endlink[0]{}%
	\providecommand \url  [0]{\begingroup\@sanitize@url \@url }%
	\providecommand \@url [1]{\endgroup\@href {#1}{\urlprefix }}%
	\providecommand \urlprefix  [0]{URL }%
	\providecommand \Eprint [0]{\href }%
	\providecommand \doibase [0]{https://doi.org/}%
	\providecommand \selectlanguage [0]{\@gobble}%
	\providecommand \bibinfo  [0]{\@secondoftwo}%
	\providecommand \bibfield  [0]{\@secondoftwo}%
	\providecommand \translation [1]{[#1]}%
	\providecommand \BibitemOpen [0]{}%
	\providecommand \bibitemStop [0]{}%
	\providecommand \bibitemNoStop [0]{.\EOS\space}%
	\providecommand \EOS [0]{\spacefactor3000\relax}%
	\providecommand \BibitemShut  [1]{\csname bibitem#1\endcsname}%
	\let\auto@bib@innerbib\@empty
	%</preamble>
	\bibitem [{\citenamefont {Bazavov}\ \emph {et~al.}(2012)\citenamefont {Bazavov}
		\emph {et~al.}}]{Bazavov:2011nk}%
	\BibitemOpen
	\bibfield  {author} {\bibinfo {author} {\bibfnamefont {A.}~\bibnamefont
			{Bazavov}} \emph {et~al.},\ }\href
	{https://doi.org/10.1103/PhysRevD.85.054503} {\bibfield  {journal} {\bibinfo
			{journal} {Phys. Rev. D}\ }\textbf {\bibinfo {volume} {85}},\ \bibinfo
		{pages} {054503} (\bibinfo {year} {2012})},\ \Eprint
	{https://arxiv.org/abs/1111.1710} {arXiv:1111.1710 [hep-lat]} \BibitemShut
	{NoStop}%
	\bibitem [{\citenamefont {Borsanyi}\ \emph {et~al.}(2014)\citenamefont
		{Borsanyi}, \citenamefont {Fodor}, \citenamefont {Hoelbling}, \citenamefont
		{Katz}, \citenamefont {Krieg},\ and\ \citenamefont
		{Szabo}}]{Borsanyi:2013bia}%
	\BibitemOpen
	\bibfield  {author} {\bibinfo {author} {\bibfnamefont {S.}~\bibnamefont
			{Borsanyi}}, \bibinfo {author} {\bibfnamefont {Z.}~\bibnamefont {Fodor}},
		\bibinfo {author} {\bibfnamefont {C.}~\bibnamefont {Hoelbling}}, \bibinfo
		{author} {\bibfnamefont {S.~D.}\ \bibnamefont {Katz}}, \bibinfo {author}
		{\bibfnamefont {S.}~\bibnamefont {Krieg}},\ and\ \bibinfo {author}
		{\bibfnamefont {K.~K.}\ \bibnamefont {Szabo}},\ }\href
	{https://doi.org/10.1016/j.physletb.2014.01.007} {\bibfield  {journal}
		{\bibinfo  {journal} {Phys. Lett. B}\ }\textbf {\bibinfo {volume} {730}},\
		\bibinfo {pages} {99} (\bibinfo {year} {2014})},\ \Eprint
	{https://arxiv.org/abs/1309.5258} {arXiv:1309.5258 [hep-lat]} \BibitemShut
	{NoStop}%
	\bibitem [{\citenamefont {Arsene}\ \emph {et~al.}(2005)\citenamefont {Arsene}
		\emph {et~al.}}]{BRAHMS:2004adc}%
	\BibitemOpen
	\bibfield  {author} {\bibinfo {author} {\bibfnamefont {I.}~\bibnamefont
			{Arsene}} \emph {et~al.} (\bibinfo {collaboration} {BRAHMS}),\ }\href
	{https://doi.org/10.1016/j.nuclphysa.2005.02.130} {\bibfield  {journal}
		{\bibinfo  {journal} {Nucl. Phys. A}\ }\textbf {\bibinfo {volume} {757}},\
		\bibinfo {pages} {1} (\bibinfo {year} {2005})},\ \Eprint
	{https://arxiv.org/abs/nucl-ex/0410020} {arXiv:nucl-ex/0410020} \BibitemShut
	{NoStop}%
	\bibitem [{\citenamefont {Adcox}\ \emph {et~al.}(2005)\citenamefont {Adcox}
		\emph {et~al.}}]{PHENIX:2004vcz}%
	\BibitemOpen
	\bibfield  {author} {\bibinfo {author} {\bibfnamefont {K.}~\bibnamefont
			{Adcox}} \emph {et~al.} (\bibinfo {collaboration} {PHENIX}),\ }\href
	{https://doi.org/10.1016/j.nuclphysa.2005.03.086} {\bibfield  {journal}
		{\bibinfo  {journal} {Nucl. Phys. A}\ }\textbf {\bibinfo {volume} {757}},\
		\bibinfo {pages} {184} (\bibinfo {year} {2005})},\ \Eprint
	{https://arxiv.org/abs/nucl-ex/0410003} {arXiv:nucl-ex/0410003} \BibitemShut
	{NoStop}%
	\bibitem [{\citenamefont {Adams}\ \emph {et~al.}(2005)\citenamefont {Adams}
		\emph {et~al.}}]{STAR:2005gfr}%
	\BibitemOpen
	\bibfield  {author} {\bibinfo {author} {\bibfnamefont {J.}~\bibnamefont
			{Adams}} \emph {et~al.} (\bibinfo {collaboration} {STAR}),\ }\href
	{https://doi.org/10.1016/j.nuclphysa.2005.03.085} {\bibfield  {journal}
		{\bibinfo  {journal} {Nucl. Phys. A}\ }\textbf {\bibinfo {volume} {757}},\
		\bibinfo {pages} {102} (\bibinfo {year} {2005})},\ \Eprint
	{https://arxiv.org/abs/nucl-ex/0501009} {arXiv:nucl-ex/0501009} \BibitemShut
	{NoStop}%
	\bibitem [{\citenamefont {Shuryak}(1978)}]{Shuryak:1978ij}%
	\BibitemOpen
	\bibfield  {author} {\bibinfo {author} {\bibfnamefont {E.~V.}\ \bibnamefont
			{Shuryak}},\ }\href {https://doi.org/10.1016/0370-2693(78)90370-2} {\bibfield
		{journal} {\bibinfo  {journal} {Phys. Lett. B}\ }\textbf {\bibinfo {volume}
			{78}},\ \bibinfo {pages} {150} (\bibinfo {year} {1978})}\BibitemShut
	{NoStop}%
	\bibitem [{\citenamefont {David}(2020)}]{David:2019wpt}%
	\BibitemOpen
	\bibfield  {author} {\bibinfo {author} {\bibfnamefont {G.}~\bibnamefont
			{David}},\ }\href {https://doi.org/10.1088/1361-6633/ab6f57} {\bibfield
		{journal} {\bibinfo  {journal} {Rept. Prog. Phys.}\ }\textbf {\bibinfo
			{volume} {83}},\ \bibinfo {pages} {046301} (\bibinfo {year} {2020})},\
	\Eprint {https://arxiv.org/abs/1907.08893} {arXiv:1907.08893 [nucl-ex]}
	\BibitemShut {NoStop}%
	\bibitem [{\citenamefont {Gale}\ \emph {et~al.}(2022)\citenamefont {Gale},
		\citenamefont {Paquet}, \citenamefont {Schenke},\ and\ \citenamefont
		{Shen}}]{Gale:2021emg}%
	\BibitemOpen
	\bibfield  {author} {\bibinfo {author} {\bibfnamefont {C.}~\bibnamefont
			{Gale}}, \bibinfo {author} {\bibfnamefont {J.-F.}\ \bibnamefont {Paquet}},
		\bibinfo {author} {\bibfnamefont {B.}~\bibnamefont {Schenke}},\ and\ \bibinfo
		{author} {\bibfnamefont {C.}~\bibnamefont {Shen}},\ }\href
	{https://doi.org/10.1103/PhysRevC.105.014909} {\bibfield  {journal} {\bibinfo
			{journal} {Phys. Rev. C}\ }\textbf {\bibinfo {volume} {105}},\ \bibinfo
		{pages} {014909} (\bibinfo {year} {2022})},\ \Eprint
	{https://arxiv.org/abs/2106.11216} {arXiv:2106.11216 [nucl-th]} \BibitemShut
	{NoStop}%
	\bibitem [{\citenamefont {Paquet}\ \emph {et~al.}(2016)\citenamefont {Paquet},
		\citenamefont {Shen}, \citenamefont {Denicol}, \citenamefont {Luzum},
		\citenamefont {Schenke}, \citenamefont {Jeon},\ and\ \citenamefont
		{Gale}}]{Paquet:2015lta}%
	\BibitemOpen
	\bibfield  {author} {\bibinfo {author} {\bibfnamefont {J.-F.}\ \bibnamefont
			{Paquet}}, \bibinfo {author} {\bibfnamefont {C.}~\bibnamefont {Shen}},
		\bibinfo {author} {\bibfnamefont {G.~S.}\ \bibnamefont {Denicol}}, \bibinfo
		{author} {\bibfnamefont {M.}~\bibnamefont {Luzum}}, \bibinfo {author}
		{\bibfnamefont {B.}~\bibnamefont {Schenke}}, \bibinfo {author} {\bibfnamefont
			{S.}~\bibnamefont {Jeon}},\ and\ \bibinfo {author} {\bibfnamefont
			{C.}~\bibnamefont {Gale}},\ }\href
	{https://doi.org/10.1103/PhysRevC.93.044906} {\bibfield  {journal} {\bibinfo
			{journal} {Phys. Rev. C}\ }\textbf {\bibinfo {volume} {93}},\ \bibinfo
		{pages} {044906} (\bibinfo {year} {2016})},\ \Eprint
	{https://arxiv.org/abs/1509.06738} {arXiv:1509.06738 [hep-ph]} \BibitemShut
	{NoStop}%
	\bibitem [{\citenamefont {McLerran}\ and\ \citenamefont
		{Toimela}(1985)}]{McLerranToimela1985}%
	\BibitemOpen
	\bibfield  {author} {\bibinfo {author} {\bibfnamefont {L.}~\bibnamefont
			{McLerran}}\ and\ \bibinfo {author} {\bibfnamefont {T.}~\bibnamefont
			{Toimela}},\ }\href {https://doi.org/10.1103/PhysRevD.31.545} {\bibfield
		{journal} {\bibinfo  {journal} {Phys. Rev. D}\ }\textbf {\bibinfo {volume}
			{31}},\ \bibinfo {pages} {545} (\bibinfo {year} {1985})}\BibitemShut
	{NoStop}%
	\bibitem [{\citenamefont {Kapusta}\ \emph {et~al.}(1991)\citenamefont
		{Kapusta}, \citenamefont {Lichard},\ and\ \citenamefont
		{Seibert}}]{KapustaLichardSeibert1991}%
	\BibitemOpen
	\bibfield  {author} {\bibinfo {author} {\bibfnamefont {J.~I.}\ \bibnamefont
			{Kapusta}}, \bibinfo {author} {\bibfnamefont {P.}~\bibnamefont {Lichard}},\
		and\ \bibinfo {author} {\bibfnamefont {D.}~\bibnamefont {Seibert}},\ }\href
	{https://doi.org/10.1103/PhysRevD.44.2774} {\bibfield  {journal} {\bibinfo
			{journal} {Phys. Rev. D}\ }\textbf {\bibinfo {volume} {44}},\ \bibinfo
		{pages} {2774} (\bibinfo {year} {1991})}\BibitemShut {NoStop}%
	\bibitem [{\citenamefont {Srivastava}\ and\ \citenamefont
		{Sinha}(1994)}]{SrivastavaSinha1994}%
	\BibitemOpen
	\bibfield  {author} {\bibinfo {author} {\bibfnamefont {D.~K.}\ \bibnamefont
			{Srivastava}}\ and\ \bibinfo {author} {\bibfnamefont {B.}~\bibnamefont
			{Sinha}},\ }\href {https://doi.org/10.1103/PhysRevLett.73.2421} {\bibfield
		{journal} {\bibinfo  {journal} {Phys. Rev. Lett.}\ }\textbf {\bibinfo
			{volume} {73}},\ \bibinfo {pages} {2421} (\bibinfo {year}
		{1994})}\BibitemShut {NoStop}%
	\bibitem [{\citenamefont {Turbide}\ \emph {et~al.}(2004)\citenamefont
		{Turbide}, \citenamefont {Rapp},\ and\ \citenamefont
		{Gale}}]{TurbideRappGale2004}%
	\BibitemOpen
	\bibfield  {author} {\bibinfo {author} {\bibfnamefont {S.}~\bibnamefont
			{Turbide}}, \bibinfo {author} {\bibfnamefont {R.}~\bibnamefont {Rapp}},\ and\
		\bibinfo {author} {\bibfnamefont {C.}~\bibnamefont {Gale}},\ }\href
	{https://doi.org/10.1103/PhysRevC.69.014903} {\bibfield  {journal} {\bibinfo
			{journal} {Phys. Rev. C}\ }\textbf {\bibinfo {volume} {69}},\ \bibinfo
		{pages} {014903} (\bibinfo {year} {2004})},\ \Eprint
	{https://arxiv.org/abs/hep-ph/0308085} {arXiv:hep-ph/0308085} \BibitemShut
	{NoStop}%
	\bibitem [{\citenamefont {Chatterjee}\ \emph {et~al.}(2006)\citenamefont
		{Chatterjee}, \citenamefont {Frodermann}, \citenamefont {Heinz},\ and\
		\citenamefont {Srivastava}}]{ChatterjeeEtAl2006}%
	\BibitemOpen
	\bibfield  {author} {\bibinfo {author} {\bibfnamefont {R.}~\bibnamefont
			{Chatterjee}}, \bibinfo {author} {\bibfnamefont {E.~S.}\ \bibnamefont
			{Frodermann}}, \bibinfo {author} {\bibfnamefont {U.}~\bibnamefont {Heinz}},\
		and\ \bibinfo {author} {\bibfnamefont {D.~K.}\ \bibnamefont {Srivastava}},\
	}\href {https://doi.org/10.1103/PhysRevLett.96.202302} {\bibfield  {journal}
		{\bibinfo  {journal} {Phys. Rev. Lett.}\ }\textbf {\bibinfo {volume} {96}},\
		\bibinfo {pages} {202302} (\bibinfo {year} {2006})},\ \Eprint
	{https://arxiv.org/abs/nucl-th/0511079} {arXiv:nucl-th/0511079} \BibitemShut
	{NoStop}%
	\bibitem [{\citenamefont {Shen}\ \emph {et~al.}(2014)\citenamefont {Shen},
		\citenamefont {Heinz}, \citenamefont {Paquet},\ and\ \citenamefont
		{Gale}}]{Shen:2013vja}%
	\BibitemOpen
	\bibfield  {author} {\bibinfo {author} {\bibfnamefont {C.}~\bibnamefont
			{Shen}}, \bibinfo {author} {\bibfnamefont {U.~W.}\ \bibnamefont {Heinz}},
		\bibinfo {author} {\bibfnamefont {J.-F.}\ \bibnamefont {Paquet}},\ and\
		\bibinfo {author} {\bibfnamefont {C.}~\bibnamefont {Gale}},\ }\href
	{https://doi.org/10.1103/PhysRevC.89.044910} {\bibfield  {journal} {\bibinfo
			{journal} {Phys. Rev. C}\ }\textbf {\bibinfo {volume} {89}},\ \bibinfo
		{pages} {044910} (\bibinfo {year} {2014})},\ \Eprint
	{https://arxiv.org/abs/1308.2440} {arXiv:1308.2440 [nucl-th]} \BibitemShut
	{NoStop}%
	\bibitem [{\citenamefont {Adare}\ \emph {et~al.}(2015)\citenamefont {Adare}
		\emph {et~al.}}]{PHENIX_yields_2014}%
	\BibitemOpen
	\bibfield  {author} {\bibinfo {author} {\bibfnamefont {A.}~\bibnamefont
			{Adare}} \emph {et~al.} (\bibinfo {collaboration} {PHENIX}),\ }\href
	{https://doi.org/10.1103/PhysRevC.91.064904} {\bibfield  {journal} {\bibinfo
			{journal} {Phys. Rev. C}\ }\textbf {\bibinfo {volume} {91}},\ \bibinfo
		{pages} {064904} (\bibinfo {year} {2015})},\ \Eprint
	{https://arxiv.org/abs/1405.3940} {arXiv:1405.3940 [nucl-ex]} \BibitemShut
	{NoStop}%
	\bibitem [{\citenamefont {Abdulameer}\ \emph {et~al.}(2024)\citenamefont
		{Abdulameer} \emph {et~al.}}]{PHENIX_yields_2022}%
	\BibitemOpen
	\bibfield  {author} {\bibinfo {author} {\bibfnamefont {N.~J.}\ \bibnamefont
			{Abdulameer}} \emph {et~al.} (\bibinfo {collaboration} {PHENIX}),\ }\href
	{https://doi.org/10.1103/PhysRevC.109.044912} {\bibfield  {journal} {\bibinfo
			{journal} {Phys. Rev. C}\ }\textbf {\bibinfo {volume} {109}},\ \bibinfo
		{pages} {044912} (\bibinfo {year} {2024})},\ \Eprint
	{https://arxiv.org/abs/2203.17187} {arXiv:2203.17187 [nucl-ex]} \BibitemShut
	{NoStop}%
	\bibitem [{\citenamefont {Adare}\ \emph {et~al.}(2012)\citenamefont {Adare}
		\emph {et~al.}}]{PHENIX_v2_2012}%
	\BibitemOpen
	\bibfield  {author} {\bibinfo {author} {\bibfnamefont {A.}~\bibnamefont
			{Adare}} \emph {et~al.} (\bibinfo {collaboration} {PHENIX}),\ }\href
	{https://doi.org/10.1103/PhysRevLett.109.122302} {\bibfield  {journal}
		{\bibinfo  {journal} {Phys. Rev. Lett.}\ }\textbf {\bibinfo {volume} {109}},\
		\bibinfo {pages} {122302} (\bibinfo {year} {2012})},\ \Eprint
	{https://arxiv.org/abs/1105.4126} {arXiv:1105.4126 [nucl-ex]} \BibitemShut
	{NoStop}%
	\bibitem [{\citenamefont {Adare}\ \emph
		{et~al.}(2016{\natexlab{a}})\citenamefont {Adare} \emph
		{et~al.}}]{PHENIX_v2_2016}%
	\BibitemOpen
	\bibfield  {author} {\bibinfo {author} {\bibfnamefont {A.}~\bibnamefont
			{Adare}} \emph {et~al.} (\bibinfo {collaboration} {PHENIX}),\ }\href
	{https://doi.org/10.1103/PhysRevC.94.064901} {\bibfield  {journal} {\bibinfo
			{journal} {Phys. Rev. C}\ }\textbf {\bibinfo {volume} {94}},\ \bibinfo
		{pages} {064901} (\bibinfo {year} {2016}{\natexlab{a}})},\ \Eprint
	{https://arxiv.org/abs/1509.07758} {arXiv:1509.07758 [nucl-ex]} \BibitemShut
	{NoStop}%
	\bibitem [{\citenamefont {Adare}\ \emph
		{et~al.}(2016{\natexlab{b}})\citenamefont {Adare} \emph
		{et~al.}}]{PHENIX_vn_2016}%
	\BibitemOpen
	\bibfield  {author} {\bibinfo {author} {\bibfnamefont {A.}~\bibnamefont
			{Adare}} \emph {et~al.} (\bibinfo {collaboration} {PHENIX}),\ }\href
	{https://doi.org/10.1103/PhysRevC.93.051902} {\bibfield  {journal} {\bibinfo
			{journal} {Phys. Rev. C}\ }\textbf {\bibinfo {volume} {93}},\ \bibinfo
		{pages} {051902} (\bibinfo {year} {2016}{\natexlab{b}})},\ \Eprint
	{https://arxiv.org/abs/1412.1038} {arXiv:1412.1038 [nucl-ex]} \BibitemShut
	{NoStop}%
	\bibitem [{\citenamefont {Shen}\ \emph {et~al.}(2015)\citenamefont {Shen},
		\citenamefont {Heinz}, \citenamefont {Paquet}, \citenamefont {Kozlov},\ and\
		\citenamefont {Gale}}]{Shen:2013cca}%
	\BibitemOpen
	\bibfield  {author} {\bibinfo {author} {\bibfnamefont {C.}~\bibnamefont
			{Shen}}, \bibinfo {author} {\bibfnamefont {U.~W.}\ \bibnamefont {Heinz}},
		\bibinfo {author} {\bibfnamefont {J.-F.}\ \bibnamefont {Paquet}}, \bibinfo
		{author} {\bibfnamefont {I.}~\bibnamefont {Kozlov}},\ and\ \bibinfo {author}
		{\bibfnamefont {C.}~\bibnamefont {Gale}},\ }\href
	{https://doi.org/10.1103/PhysRevC.91.024908} {\bibfield  {journal} {\bibinfo
			{journal} {Phys. Rev. C}\ }\textbf {\bibinfo {volume} {91}},\ \bibinfo
		{pages} {024908} (\bibinfo {year} {2015})},\ \Eprint
	{https://arxiv.org/abs/1308.2111} {arXiv:1308.2111 [nucl-th]} \BibitemShut
	{NoStop}%
	\bibitem [{\citenamefont {Chatterjee}\ \emph {et~al.}(2013)\citenamefont
		{Chatterjee}, \citenamefont {Holopainen}, \citenamefont {Helenius},
		\citenamefont {Renk},\ and\ \citenamefont {Eskola}}]{Chatterjee:2013naa}%
	\BibitemOpen
	\bibfield  {author} {\bibinfo {author} {\bibfnamefont {R.}~\bibnamefont
			{Chatterjee}}, \bibinfo {author} {\bibfnamefont {H.}~\bibnamefont
			{Holopainen}}, \bibinfo {author} {\bibfnamefont {I.}~\bibnamefont
			{Helenius}}, \bibinfo {author} {\bibfnamefont {T.}~\bibnamefont {Renk}},\
		and\ \bibinfo {author} {\bibfnamefont {K.~J.}\ \bibnamefont {Eskola}},\
	}\href {https://doi.org/10.1103/PhysRevC.88.034901} {\bibfield  {journal}
		{\bibinfo  {journal} {Phys. Rev. C}\ }\textbf {\bibinfo {volume} {88}},\
		\bibinfo {pages} {034901} (\bibinfo {year} {2013})},\ \Eprint
	{https://arxiv.org/abs/1305.6443} {arXiv:1305.6443 [hep-ph]} \BibitemShut
	{NoStop}%
	\bibitem [{\citenamefont {van Hees}\ \emph {et~al.}(2011)\citenamefont {van
			Hees}, \citenamefont {Gale},\ and\ \citenamefont {Rapp}}]{vanHees:2011vb}%
	\BibitemOpen
	\bibfield  {author} {\bibinfo {author} {\bibfnamefont {H.}~\bibnamefont {van
				Hees}}, \bibinfo {author} {\bibfnamefont {C.}~\bibnamefont {Gale}},\ and\
		\bibinfo {author} {\bibfnamefont {R.}~\bibnamefont {Rapp}},\ }\href
	{https://doi.org/10.1103/PhysRevC.84.054906} {\bibfield  {journal} {\bibinfo
			{journal} {Phys. Rev. C}\ }\textbf {\bibinfo {volume} {84}},\ \bibinfo
		{pages} {054906} (\bibinfo {year} {2011})},\ \Eprint
	{https://arxiv.org/abs/1108.2131} {arXiv:1108.2131 [hep-ph]} \BibitemShut
	{NoStop}%
	\bibitem [{\citenamefont {Dion}\ \emph {et~al.}(2011)\citenamefont {Dion},
		\citenamefont {Paquet}, \citenamefont {Schenke}, \citenamefont {Young},
		\citenamefont {Jeon},\ and\ \citenamefont {Gale}}]{Dion:2011pp}%
	\BibitemOpen
	\bibfield  {author} {\bibinfo {author} {\bibfnamefont {M.}~\bibnamefont
			{Dion}}, \bibinfo {author} {\bibfnamefont {J.-F.}\ \bibnamefont {Paquet}},
		\bibinfo {author} {\bibfnamefont {B.}~\bibnamefont {Schenke}}, \bibinfo
		{author} {\bibfnamefont {C.}~\bibnamefont {Young}}, \bibinfo {author}
		{\bibfnamefont {S.}~\bibnamefont {Jeon}},\ and\ \bibinfo {author}
		{\bibfnamefont {C.}~\bibnamefont {Gale}},\ }\href
	{https://doi.org/10.1103/PhysRevC.84.064901} {\bibfield  {journal} {\bibinfo
			{journal} {Phys. Rev. C}\ }\textbf {\bibinfo {volume} {84}},\ \bibinfo
		{pages} {064901} (\bibinfo {year} {2011})},\ \Eprint
	{https://arxiv.org/abs/1109.4405} {arXiv:1109.4405 [hep-ph]} \BibitemShut
	{NoStop}%
	\bibitem [{\citenamefont {Linnyk}\ \emph {et~al.}(2014)\citenamefont {Linnyk},
		\citenamefont {Cassing},\ and\ \citenamefont
		{Bratkovskaya}}]{Linnyk:2013wma}%
	\BibitemOpen
	\bibfield  {author} {\bibinfo {author} {\bibfnamefont {O.}~\bibnamefont
			{Linnyk}}, \bibinfo {author} {\bibfnamefont {W.}~\bibnamefont {Cassing}},\
		and\ \bibinfo {author} {\bibfnamefont {E.~L.}\ \bibnamefont {Bratkovskaya}},\
	}\href {https://doi.org/10.1103/PhysRevC.89.034908} {\bibfield  {journal}
		{\bibinfo  {journal} {Phys. Rev. C}\ }\textbf {\bibinfo {volume} {89}},\
		\bibinfo {pages} {034908} (\bibinfo {year} {2014})},\ \Eprint
	{https://arxiv.org/abs/1311.0279} {arXiv:1311.0279 [nucl-th]} \BibitemShut
	{NoStop}%
	\bibitem [{\citenamefont {Monnai}(2014)}]{Monnai:2014kqa}%
	\BibitemOpen
	\bibfield  {author} {\bibinfo {author} {\bibfnamefont {A.}~\bibnamefont
			{Monnai}},\ }\href {https://doi.org/10.1103/PhysRevC.90.021901} {\bibfield
		{journal} {\bibinfo  {journal} {Phys. Rev. C}\ }\textbf {\bibinfo {volume}
			{90}},\ \bibinfo {pages} {021901} (\bibinfo {year} {2014})},\ \Eprint
	{https://arxiv.org/abs/1403.4225} {arXiv:1403.4225 [nucl-th]} \BibitemShut
	{NoStop}%
	\bibitem [{\citenamefont {Acharya}\ \emph {et~al.}(2019)\citenamefont {Acharya}
		\emph {et~al.}}]{ALICE:2018dti}%
	\BibitemOpen
	\bibfield  {author} {\bibinfo {author} {\bibfnamefont {S.}~\bibnamefont
			{Acharya}} \emph {et~al.} (\bibinfo {collaboration} {ALICE}),\ }\href
	{https://doi.org/10.1016/j.physletb.2018.11.039} {\bibfield  {journal}
		{\bibinfo  {journal} {Phys. Lett. B}\ }\textbf {\bibinfo {volume} {789}},\
		\bibinfo {pages} {308} (\bibinfo {year} {2019})},\ \Eprint
	{https://arxiv.org/abs/1805.04403} {arXiv:1805.04403 [nucl-ex]} \BibitemShut
	{NoStop}%
	\bibitem [{\citenamefont {Skokov}\ \emph {et~al.}(2009)\citenamefont {Skokov},
		\citenamefont {Illarionov},\ and\ \citenamefont {Toneev}}]{Skokov:2009qp}%
	\BibitemOpen
	\bibfield  {author} {\bibinfo {author} {\bibfnamefont {V.}~\bibnamefont
			{Skokov}}, \bibinfo {author} {\bibfnamefont {A.~Y.}\ \bibnamefont
			{Illarionov}},\ and\ \bibinfo {author} {\bibfnamefont {V.}~\bibnamefont
			{Toneev}},\ }\href {https://doi.org/10.1142/S0217751X09047570} {\bibfield
		{journal} {\bibinfo  {journal} {Int. J. Mod. Phys. A}\ }\textbf {\bibinfo
			{volume} {24}},\ \bibinfo {pages} {5925} (\bibinfo {year} {2009})},\ \Eprint
	{https://arxiv.org/abs/0907.1396} {arXiv:0907.1396 [nucl-th]} \BibitemShut
	{NoStop}%
	\bibitem [{\citenamefont {Deng}\ and\ \citenamefont
		{Huang}(2012)}]{Deng:2012pc}%
	\BibitemOpen
	\bibfield  {author} {\bibinfo {author} {\bibfnamefont {W.-T.}\ \bibnamefont
			{Deng}}\ and\ \bibinfo {author} {\bibfnamefont {X.-G.}\ \bibnamefont
			{Huang}},\ }\href {https://doi.org/10.1103/PhysRevC.85.044907} {\bibfield
		{journal} {\bibinfo  {journal} {Phys. Rev. C}\ }\textbf {\bibinfo {volume}
			{85}},\ \bibinfo {pages} {044907} (\bibinfo {year} {2012})},\ \Eprint
	{https://arxiv.org/abs/1201.5108} {arXiv:1201.5108 [nucl-th]} \BibitemShut
	{NoStop}%
	\bibitem [{\citenamefont {Basar}\ \emph {et~al.}(2012)\citenamefont {Basar},
		\citenamefont {Kharzeev}, \citenamefont {Kharzeev},\ and\ \citenamefont
		{Skokov}}]{Basar:2012bp}%
	\BibitemOpen
	\bibfield  {author} {\bibinfo {author} {\bibfnamefont {G.}~\bibnamefont
			{Basar}}, \bibinfo {author} {\bibfnamefont {D.}~\bibnamefont {Kharzeev}},
		\bibinfo {author} {\bibfnamefont {D.}~\bibnamefont {Kharzeev}},\ and\
		\bibinfo {author} {\bibfnamefont {V.}~\bibnamefont {Skokov}},\ }\href
	{https://doi.org/10.1103/PhysRevLett.109.202303} {\bibfield  {journal}
		{\bibinfo  {journal} {Phys. Rev. Lett.}\ }\textbf {\bibinfo {volume} {109}},\
		\bibinfo {pages} {202303} (\bibinfo {year} {2012})},\ \Eprint
	{https://arxiv.org/abs/1206.1334} {arXiv:1206.1334 [hep-ph]} \BibitemShut
	{NoStop}%
	\bibitem [{\citenamefont {Muller}\ \emph {et~al.}(2014)\citenamefont {Muller},
		\citenamefont {Wu},\ and\ \citenamefont {Yang}}]{Muller:2013ila}%
	\BibitemOpen
	\bibfield  {author} {\bibinfo {author} {\bibfnamefont {B.}~\bibnamefont
			{Muller}}, \bibinfo {author} {\bibfnamefont {S.-Y.}\ \bibnamefont {Wu}},\
		and\ \bibinfo {author} {\bibfnamefont {D.-L.}\ \bibnamefont {Yang}},\ }\href
	{https://doi.org/10.1103/PhysRevD.89.026013} {\bibfield  {journal} {\bibinfo
			{journal} {Phys. Rev. D}\ }\textbf {\bibinfo {volume} {89}},\ \bibinfo
		{pages} {026013} (\bibinfo {year} {2014})},\ \Eprint
	{https://arxiv.org/abs/1308.6568} {arXiv:1308.6568 [hep-th]} \BibitemShut
	{NoStop}%
	\bibitem [{\citenamefont {Tuchin}(2013{\natexlab{a}})}]{Tuchin:2012mf}%
	\BibitemOpen
	\bibfield  {author} {\bibinfo {author} {\bibfnamefont {K.}~\bibnamefont
			{Tuchin}},\ }\href {https://doi.org/10.1103/PhysRevC.87.024912} {\bibfield
		{journal} {\bibinfo  {journal} {Phys. Rev. C}\ }\textbf {\bibinfo {volume}
			{87}},\ \bibinfo {pages} {024912} (\bibinfo {year} {2013}{\natexlab{a}})},\
	\Eprint {https://arxiv.org/abs/1206.0485} {arXiv:1206.0485 [hep-ph]}
	\BibitemShut {NoStop}%
	\bibitem [{\citenamefont {Tuchin}(2015)}]{Tuchin:2014pka}%
	\BibitemOpen
	\bibfield  {author} {\bibinfo {author} {\bibfnamefont {K.}~\bibnamefont
			{Tuchin}},\ }\href {https://doi.org/10.1103/PhysRevC.91.014902} {\bibfield
		{journal} {\bibinfo  {journal} {Phys. Rev. C}\ }\textbf {\bibinfo {volume}
			{91}},\ \bibinfo {pages} {014902} (\bibinfo {year} {2015})},\ \Eprint
	{https://arxiv.org/abs/1406.5097} {arXiv:1406.5097 [nucl-th]} \BibitemShut
	{NoStop}%
	\bibitem [{\citenamefont {Yee}(2013)}]{Yee:2013qma}%
	\BibitemOpen
	\bibfield  {author} {\bibinfo {author} {\bibfnamefont {H.-U.}\ \bibnamefont
			{Yee}},\ }\href {https://doi.org/10.1103/PhysRevD.88.026001} {\bibfield
		{journal} {\bibinfo  {journal} {Phys. Rev. D}\ }\textbf {\bibinfo {volume}
			{88}},\ \bibinfo {pages} {026001} (\bibinfo {year} {2013})},\ \Eprint
	{https://arxiv.org/abs/1303.3571} {arXiv:1303.3571 [nucl-th]} \BibitemShut
	{NoStop}%
	\bibitem [{\citenamefont {Wang}\ \emph {et~al.}(2020)\citenamefont {Wang},
		\citenamefont {Shovkovy}, \citenamefont {Yu},\ and\ \citenamefont
		{Huang}}]{Wang:2020dsr}%
	\BibitemOpen
	\bibfield  {author} {\bibinfo {author} {\bibfnamefont {X.}~\bibnamefont
			{Wang}}, \bibinfo {author} {\bibfnamefont {I.~A.}\ \bibnamefont {Shovkovy}},
		\bibinfo {author} {\bibfnamefont {L.}~\bibnamefont {Yu}},\ and\ \bibinfo
		{author} {\bibfnamefont {M.}~\bibnamefont {Huang}},\ }\href
	{https://doi.org/10.1103/PhysRevD.102.076010} {\bibfield  {journal} {\bibinfo
			{journal} {Phys. Rev. D}\ }\textbf {\bibinfo {volume} {102}},\ \bibinfo
		{pages} {076010} (\bibinfo {year} {2020})},\ \Eprint
	{https://arxiv.org/abs/2006.16254} {arXiv:2006.16254 [hep-ph]} \BibitemShut
	{NoStop}%
	\bibitem [{\citenamefont {Sun}\ and\ \citenamefont
		{Yan}(2024{\natexlab{a}})}]{Sun:2024vsz}%
	\BibitemOpen
	\bibfield  {author} {\bibinfo {author} {\bibfnamefont {J.}~\bibnamefont
			{Sun}}\ and\ \bibinfo {author} {\bibfnamefont {L.}~\bibnamefont {Yan}},\
	}\href {https://doi.org/10.11804/NuclPhysRev.41.2023CNPC14} {\bibfield
		{journal} {\bibinfo  {journal} {Nucl. Phys. Rev.}\ }\textbf {\bibinfo
			{volume} {41}},\ \bibinfo {pages} {558} (\bibinfo {year}
		{2024}{\natexlab{a}})}\BibitemShut {NoStop}%
	\bibitem [{\citenamefont {Tuchin}(2013{\natexlab{b}})}]{Tuchin:2013apa}%
	\BibitemOpen
	\bibfield  {author} {\bibinfo {author} {\bibfnamefont {K.}~\bibnamefont
			{Tuchin}},\ }\href {https://doi.org/10.1103/PhysRevC.88.024911} {\bibfield
		{journal} {\bibinfo  {journal} {Phys. Rev. C}\ }\textbf {\bibinfo {volume}
			{88}},\ \bibinfo {pages} {024911} (\bibinfo {year} {2013}{\natexlab{b}})},\
	\Eprint {https://arxiv.org/abs/1305.5806} {arXiv:1305.5806 [hep-ph]}
	\BibitemShut {NoStop}%
	\bibitem [{\citenamefont {Bloczynski}\ \emph {et~al.}(2013)\citenamefont
		{Bloczynski}, \citenamefont {Huang}, \citenamefont {Zhang},\ and\
		\citenamefont {Liao}}]{Bloczynski:2012en}%
	\BibitemOpen
	\bibfield  {author} {\bibinfo {author} {\bibfnamefont {J.}~\bibnamefont
			{Bloczynski}}, \bibinfo {author} {\bibfnamefont {X.-G.}\ \bibnamefont
			{Huang}}, \bibinfo {author} {\bibfnamefont {X.}~\bibnamefont {Zhang}},\ and\
		\bibinfo {author} {\bibfnamefont {J.}~\bibnamefont {Liao}},\ }\href
	{https://doi.org/10.1016/j.physletb.2012.12.030} {\bibfield  {journal}
		{\bibinfo  {journal} {Phys. Lett. B}\ }\textbf {\bibinfo {volume} {718}},\
		\bibinfo {pages} {1529} (\bibinfo {year} {2013})},\ \Eprint
	{https://arxiv.org/abs/1209.6594} {arXiv:1209.6594 [nucl-th]} \BibitemShut
	{NoStop}%
	\bibitem [{\citenamefont {Zakharov}(2016{\natexlab{a}})}]{Zakharov:2016mmc}%
	\BibitemOpen
	\bibfield  {author} {\bibinfo {author} {\bibfnamefont {B.~G.}\ \bibnamefont
			{Zakharov}},\ }\href {https://doi.org/10.1140/epjc/s10052-016-4451-8}
	{\bibfield  {journal} {\bibinfo  {journal} {Eur. Phys. J. C}\ }\textbf
		{\bibinfo {volume} {76}},\ \bibinfo {pages} {609} (\bibinfo {year}
		{2016}{\natexlab{a}})},\ \Eprint {https://arxiv.org/abs/1609.04324}
	{arXiv:1609.04324 [nucl-th]} \BibitemShut {NoStop}%
	\bibitem [{\citenamefont {Zakharov}(2016{\natexlab{b}})}]{Zakharov:2016kte}%
	\BibitemOpen
	\bibfield  {author} {\bibinfo {author} {\bibfnamefont {B.~G.}\ \bibnamefont
			{Zakharov}},\ }\href {https://doi.org/10.1134/S0021364016160025} {\bibfield
		{journal} {\bibinfo  {journal} {JETP Lett.}\ }\textbf {\bibinfo {volume}
			{104}},\ \bibinfo {pages} {213} (\bibinfo {year} {2016}{\natexlab{b}})},\
	\Eprint {https://arxiv.org/abs/1607.04314} {arXiv:1607.04314 [hep-ph]}
	\BibitemShut {NoStop}%
	\bibitem [{\citenamefont {Sun}\ and\ \citenamefont
		{Yan}(2024{\natexlab{b}})}]{Sun:2023pil}%
	\BibitemOpen
	\bibfield  {author} {\bibinfo {author} {\bibfnamefont {J.-A.}\ \bibnamefont
			{Sun}}\ and\ \bibinfo {author} {\bibfnamefont {L.}~\bibnamefont {Yan}},\
	}\href {https://doi.org/10.1016/j.physletb.2024.139046} {\bibfield  {journal}
		{\bibinfo  {journal} {Phys. Lett. B}\ }\textbf {\bibinfo {volume} {858}},\
		\bibinfo {pages} {139046} (\bibinfo {year} {2024}{\natexlab{b}})},\ \Eprint
	{https://arxiv.org/abs/2302.07696} {arXiv:2302.07696 [nucl-th]} \BibitemShut
	{NoStop}%
	\bibitem [{\citenamefont {Wang}\ and\ \citenamefont
		{Shovkovy}(2024{\natexlab{a}})}]{Wang:2024gnh}%
	\BibitemOpen
	\bibfield  {author} {\bibinfo {author} {\bibfnamefont {X.}~\bibnamefont
			{Wang}}\ and\ \bibinfo {author} {\bibfnamefont {I.~A.}\ \bibnamefont
			{Shovkovy}},\ }\href {https://doi.org/10.1103/PhysRevD.110.116005} {\bibfield
		{journal} {\bibinfo  {journal} {Phys. Rev. D}\ }\textbf {\bibinfo {volume}
			{110}},\ \bibinfo {pages} {116005} (\bibinfo {year} {2024}{\natexlab{a}})},\
	\Eprint {https://arxiv.org/abs/2407.06271} {arXiv:2407.06271 [hep-ph]}
	\BibitemShut {NoStop}%
	\bibitem [{\citenamefont {Kroth}\ and\ \citenamefont
		{Tuchin}(2026)}]{Kroth:2026kgm}%
	\BibitemOpen
	\bibfield  {author} {\bibinfo {author} {\bibfnamefont {J.~D.}\ \bibnamefont
			{Kroth}}\ and\ \bibinfo {author} {\bibfnamefont {K.}~\bibnamefont {Tuchin}},\
	}\href {https://doi.org/10.48550/arXiv.2602.02746} {\bibfield  {journal}
		{\bibinfo  {journal} {pre-print}\ ,\ \bibinfo {pages} {02746}} (\bibinfo
		{year} {2026})},\ \Eprint {https://arxiv.org/abs/2602.02746}
	{arXiv:2602.02746 [hep-ph]} \BibitemShut {NoStop}%
	\bibitem [{\citenamefont {Adamczyk}\ \emph {et~al.}(2017)\citenamefont
		{Adamczyk} \emph {et~al.}}]{STAR:2017ckg}%
	\BibitemOpen
	\bibfield  {author} {\bibinfo {author} {\bibfnamefont {L.}~\bibnamefont
			{Adamczyk}} \emph {et~al.} (\bibinfo {collaboration} {STAR}),\ }\href
	{https://doi.org/10.1038/nature23004} {\bibfield  {journal} {\bibinfo
			{journal} {Nature}\ }\textbf {\bibinfo {volume} {548}},\ \bibinfo {pages}
		{62} (\bibinfo {year} {2017})},\ \Eprint {https://arxiv.org/abs/1701.06657}
	{arXiv:1701.06657 [nucl-ex]} \BibitemShut {NoStop}%
	\bibitem [{\citenamefont {Adam}\ \emph {et~al.}(2018)\citenamefont {Adam} \emph
		{et~al.}}]{STAR:2018gyt}%
	\BibitemOpen
	\bibfield  {author} {\bibinfo {author} {\bibfnamefont {J.}~\bibnamefont
			{Adam}} \emph {et~al.} (\bibinfo {collaboration} {STAR}),\ }\href
	{https://doi.org/10.1103/PhysRevC.98.014910} {\bibfield  {journal} {\bibinfo
			{journal} {Phys. Rev. C}\ }\textbf {\bibinfo {volume} {98}},\ \bibinfo
		{pages} {014910} (\bibinfo {year} {2018})},\ \Eprint
	{https://arxiv.org/abs/1805.04400} {arXiv:1805.04400 [nucl-ex]} \BibitemShut
	{NoStop}%
	\bibitem [{\citenamefont {Acharya}\ \emph {et~al.}(2020)\citenamefont {Acharya}
		\emph {et~al.}}]{ALICE:2019aid}%
	\BibitemOpen
	\bibfield  {author} {\bibinfo {author} {\bibfnamefont {S.}~\bibnamefont
			{Acharya}} \emph {et~al.} (\bibinfo {collaboration} {ALICE}),\ }\href
	{https://doi.org/10.1103/PhysRevLett.125.012301} {\bibfield  {journal}
		{\bibinfo  {journal} {Phys. Rev. Lett.}\ }\textbf {\bibinfo {volume} {125}},\
		\bibinfo {pages} {012301} (\bibinfo {year} {2020})},\ \Eprint
	{https://arxiv.org/abs/1910.14408} {arXiv:1910.14408 [nucl-ex]} \BibitemShut
	{NoStop}%
	\bibitem [{\citenamefont {Becattini}\ \emph {et~al.}(2024)\citenamefont
		{Becattini}, \citenamefont {Buzzegoli}, \citenamefont {Niida}, \citenamefont
		{Pu}, \citenamefont {Tang},\ and\ \citenamefont {Wang}}]{Becattini:2024uha}%
	\BibitemOpen
	\bibfield  {author} {\bibinfo {author} {\bibfnamefont {F.}~\bibnamefont
			{Becattini}}, \bibinfo {author} {\bibfnamefont {M.}~\bibnamefont
			{Buzzegoli}}, \bibinfo {author} {\bibfnamefont {T.}~\bibnamefont {Niida}},
		\bibinfo {author} {\bibfnamefont {S.}~\bibnamefont {Pu}}, \bibinfo {author}
		{\bibfnamefont {A.-H.}\ \bibnamefont {Tang}},\ and\ \bibinfo {author}
		{\bibfnamefont {Q.}~\bibnamefont {Wang}},\ }\href
	{https://doi.org/10.1142/9789811294679_0005} {\bibfield  {journal} {\bibinfo
			{journal} {Int. J. Mod. Phys. E}\ }\textbf {\bibinfo {volume} {33}},\
		\bibinfo {pages} {2430006} (\bibinfo {year} {2024})},\ \Eprint
	{https://arxiv.org/abs/2402.04540} {arXiv:2402.04540 [nucl-th]} \BibitemShut
	{NoStop}%
	\bibitem [{\citenamefont {Niida}\ and\ \citenamefont
		{Voloshin}(2024)}]{Niida:2024ntm}%
	\BibitemOpen
	\bibfield  {author} {\bibinfo {author} {\bibfnamefont {T.}~\bibnamefont
			{Niida}}\ and\ \bibinfo {author} {\bibfnamefont {S.~A.}\ \bibnamefont
			{Voloshin}},\ }\href {https://doi.org/10.1142/S0218301324300108} {\bibfield
		{journal} {\bibinfo  {journal} {Int. J. Mod. Phys. E}\ }\textbf {\bibinfo
			{volume} {33}},\ \bibinfo {pages} {2430010} (\bibinfo {year} {2024})},\
	\Eprint {https://arxiv.org/abs/2404.11042} {arXiv:2404.11042 [nucl-ex]}
	\BibitemShut {NoStop}%
	\bibitem [{\citenamefont {Buzzegoli}\ \emph
		{et~al.}(2023{\natexlab{a}})\citenamefont {Buzzegoli}, \citenamefont {Kroth},
		\citenamefont {Tuchin},\ and\ \citenamefont
		{Vijayakumar}}]{Buzzegoli:2023vne}%
	\BibitemOpen
	\bibfield  {author} {\bibinfo {author} {\bibfnamefont {M.}~\bibnamefont
			{Buzzegoli}}, \bibinfo {author} {\bibfnamefont {J.~D.}\ \bibnamefont
			{Kroth}}, \bibinfo {author} {\bibfnamefont {K.}~\bibnamefont {Tuchin}},\ and\
		\bibinfo {author} {\bibfnamefont {N.}~\bibnamefont {Vijayakumar}},\ }\href
	{https://doi.org/10.1103/PhysRevD.108.096014} {\bibfield  {journal} {\bibinfo
			{journal} {Phys. Rev. D}\ }\textbf {\bibinfo {volume} {108}},\ \bibinfo
		{pages} {096014} (\bibinfo {year} {2023}{\natexlab{a}})},\ \Eprint
	{https://arxiv.org/abs/2306.03863} {arXiv:2306.03863 [hep-ph]} \BibitemShut
	{NoStop}%
	\bibitem [{\citenamefont {Sokolov}\ and\ \citenamefont
		{Ternov}(1986)}]{Sokolov_Ternov_1986}%
	\BibitemOpen
	\bibfield  {author} {\bibinfo {author} {\bibfnamefont {A.~A.}\ \bibnamefont
			{Sokolov}}\ and\ \bibinfo {author} {\bibfnamefont {I.~M.}\ \bibnamefont
			{Ternov}},\ }\href@noop {} {\emph {\bibinfo {title} {Radiation from
				relativistic electrons}}},\ edited by\ \bibinfo {editor} {\bibfnamefont
		{C.~W.}\ \bibnamefont {Kilmister}}\ (\bibinfo  {publisher} {American
		Institute of Physics},\ \bibinfo {address} {New York},\ \bibinfo {year}
	{1986})\BibitemShut {NoStop}%
	\bibitem [{\citenamefont {Bordovitsyn}(1999)}]{Bordovitsyn_1999}%
	\BibitemOpen
	\bibfield  {author} {\bibinfo {author} {\bibfnamefont {V.~A.}\ \bibnamefont
			{Bordovitsyn}},\ }\href@noop {} {\emph {\bibinfo {title} {Synchrotron
				Radiation Theory and Its Development}}}\ (\bibinfo  {publisher} {WORLD
		SCIENTIFIC},\ \bibinfo {address} {Singapore},\ \bibinfo {year}
	{1999})\BibitemShut {NoStop}%
	\bibitem [{\citenamefont {Casta{\~n}o-Yepes}\ and\ \citenamefont
		{Mu{\~n}oz}(2025)}]{Castano-Yepes:2025zae}%
	\BibitemOpen
	\bibfield  {author} {\bibinfo {author} {\bibfnamefont {J.~D.}\ \bibnamefont
			{Casta{\~n}o-Yepes}}\ and\ \bibinfo {author} {\bibfnamefont {E.}~\bibnamefont
			{Mu{\~n}oz}},\ }\href {https://doi.org/10.48550/arXiv.2509.18219} {\bibfield
		{journal} {\bibinfo  {journal} {arXiv pre-print}\ ,\ \bibinfo {pages}
			{18219}} (\bibinfo {year} {2025})},\ \Eprint
	{https://arxiv.org/abs/2509.18219} {arXiv:2509.18219 [hep-ph]} \BibitemShut
	{NoStop}%
	\bibitem [{\citenamefont {Buzzegoli}\ \emph
		{et~al.}(2023{\natexlab{b}})\citenamefont {Buzzegoli}, \citenamefont {Kroth},
		\citenamefont {Tuchin},\ and\ \citenamefont
		{Vijayakumar}}]{Buzzegoli:2022dhw}%
	\BibitemOpen
	\bibfield  {author} {\bibinfo {author} {\bibfnamefont {M.}~\bibnamefont
			{Buzzegoli}}, \bibinfo {author} {\bibfnamefont {J.~D.}\ \bibnamefont
			{Kroth}}, \bibinfo {author} {\bibfnamefont {K.}~\bibnamefont {Tuchin}},\ and\
		\bibinfo {author} {\bibfnamefont {N.}~\bibnamefont {Vijayakumar}},\ }\href
	{https://doi.org/10.1103/PhysRevD.107.L051901} {\bibfield  {journal}
		{\bibinfo  {journal} {Phys. Rev. D}\ }\textbf {\bibinfo {volume} {107}},\
		\bibinfo {pages} {L051901} (\bibinfo {year} {2023}{\natexlab{b}})},\ \Eprint
	{https://arxiv.org/abs/2209.02597} {arXiv:2209.02597 [hep-ph]} \BibitemShut
	{NoStop}%
	\bibitem [{\citenamefont {Buzzegoli}\ and\ \citenamefont
		{Tuchin}(2023)}]{Buzzegoli:2023yut}%
	\BibitemOpen
	\bibfield  {author} {\bibinfo {author} {\bibfnamefont {M.}~\bibnamefont
			{Buzzegoli}}\ and\ \bibinfo {author} {\bibfnamefont {K.}~\bibnamefont
			{Tuchin}},\ }\href {https://doi.org/10.1007/JHEP12(2023)113} {\bibfield
		{journal} {\bibinfo  {journal} {JHEP}\ }\textbf {\bibinfo {volume}
			{12}}\bibfield  {number} {\bibinfo  {number} { (2023)},\ \bibinfo {pages}
			{113}},\ }\Eprint {https://arxiv.org/abs/2308.10349} {arXiv:2308.10349
		[hep-ph]} \BibitemShut {NoStop}%
	\bibitem [{\citenamefont {Buzzegoli}\ and\ \citenamefont
		{Tuchin}(2024)}]{Buzzegoli:2024nzd}%
	\BibitemOpen
	\bibfield  {author} {\bibinfo {author} {\bibfnamefont {M.}~\bibnamefont
			{Buzzegoli}}\ and\ \bibinfo {author} {\bibfnamefont {K.}~\bibnamefont
			{Tuchin}},\ }\href {https://doi.org/10.1103/PhysRevD.110.076013} {\bibfield
		{journal} {\bibinfo  {journal} {Phys. Rev. D}\ }\textbf {\bibinfo {volume}
			{110}},\ \bibinfo {pages} {076013} (\bibinfo {year} {2024})},\ \Eprint
	{https://arxiv.org/abs/2405.19530} {arXiv:2405.19530 [hep-ph]} \BibitemShut
	{NoStop}%
	\bibitem [{\citenamefont {Buzzegoli}\ \emph {et~al.}(2025)\citenamefont
		{Buzzegoli}, \citenamefont {Tuchin},\ and\ \citenamefont
		{Vijayakumar}}]{Buzzegoli:2025qfl}%
	\BibitemOpen
	\bibfield  {author} {\bibinfo {author} {\bibfnamefont {M.}~\bibnamefont
			{Buzzegoli}}, \bibinfo {author} {\bibfnamefont {K.}~\bibnamefont {Tuchin}},\
		and\ \bibinfo {author} {\bibfnamefont {N.}~\bibnamefont {Vijayakumar}},\
	}\href {https://doi.org/10.1103/PhysRevC.111.054907} {\bibfield  {journal}
		{\bibinfo  {journal} {Phys. Rev. C}\ }\textbf {\bibinfo {volume} {111}},\
		\bibinfo {pages} {054907} (\bibinfo {year} {2025})},\ \Eprint
	{https://arxiv.org/abs/2503.06649} {arXiv:2503.06649 [hep-ph]} \BibitemShut
	{NoStop}%
	\bibitem [{\citenamefont {Kroth}\ and\ \citenamefont
		{Tuchin}(2025)}]{Kroth:2024mfd}%
	\BibitemOpen
	\bibfield  {author} {\bibinfo {author} {\bibfnamefont {J.~D.}\ \bibnamefont
			{Kroth}}\ and\ \bibinfo {author} {\bibfnamefont {K.}~\bibnamefont {Tuchin}},\
	}\href {https://doi.org/10.1103/PhysRevD.111.056004} {\bibfield  {journal}
		{\bibinfo  {journal} {Phys. Rev. D}\ }\textbf {\bibinfo {volume} {111}},\
		\bibinfo {pages} {056004} (\bibinfo {year} {2025})},\ \Eprint
	{https://arxiv.org/abs/2409.20569} {arXiv:2409.20569 [hep-ph]} \BibitemShut
	{NoStop}%
	\bibitem [{\citenamefont {Wang}\ and\ \citenamefont
		{Shovkovy}(2021{\natexlab{a}})}]{Wang:2021ebh}%
	\BibitemOpen
	\bibfield  {author} {\bibinfo {author} {\bibfnamefont {X.}~\bibnamefont
			{Wang}}\ and\ \bibinfo {author} {\bibfnamefont {I.}~\bibnamefont
			{Shovkovy}},\ }\href {https://doi.org/10.1103/PhysRevD.104.056017} {\bibfield
		{journal} {\bibinfo  {journal} {Phys. Rev. D}\ }\textbf {\bibinfo {volume}
			{104}},\ \bibinfo {pages} {056017} (\bibinfo {year} {2021}{\natexlab{a}})},\
	\Eprint {https://arxiv.org/abs/2103.01967} {arXiv:2103.01967 [nucl-th]}
	\BibitemShut {NoStop}%
	\bibitem [{\citenamefont {Arnold}\ \emph {et~al.}(2001)\citenamefont {Arnold},
		\citenamefont {Moore},\ and\ \citenamefont {Yaffe}}]{Arnold:2001ms}%
	\BibitemOpen
	\bibfield  {author} {\bibinfo {author} {\bibfnamefont {P.~B.}\ \bibnamefont
			{Arnold}}, \bibinfo {author} {\bibfnamefont {G.~D.}\ \bibnamefont {Moore}},\
		and\ \bibinfo {author} {\bibfnamefont {L.~G.}\ \bibnamefont {Yaffe}},\ }\href
	{https://doi.org/10.1088/1126-6708/2001/12/009} {\bibfield  {journal}
		{\bibinfo  {journal} {JHEP}\ }\textbf {\bibinfo {volume} {12}}\bibfield
		{number} {\bibinfo  {number} { (2001)},\ \bibinfo {pages} {009}},\ }\Eprint
	{https://arxiv.org/abs/hep-ph/0111107} {arXiv:hep-ph/0111107} \BibitemShut
	{NoStop}%
	\bibitem [{\citenamefont {Kajantie}\ \emph {et~al.}(1986)\citenamefont
		{Kajantie}, \citenamefont {Kapusta}, \citenamefont {McLerran},\ and\
		\citenamefont {Mekjian}}]{Kajantie:1986dh}%
	\BibitemOpen
	\bibfield  {author} {\bibinfo {author} {\bibfnamefont {K.}~\bibnamefont
			{Kajantie}}, \bibinfo {author} {\bibfnamefont {J.~I.}\ \bibnamefont
			{Kapusta}}, \bibinfo {author} {\bibfnamefont {L.~D.}\ \bibnamefont
			{McLerran}},\ and\ \bibinfo {author} {\bibfnamefont {A.}~\bibnamefont
			{Mekjian}},\ }\href {https://doi.org/10.1103/PhysRevD.34.2746} {\bibfield
		{journal} {\bibinfo  {journal} {Phys. Rev. D}\ }\textbf {\bibinfo {volume}
			{34}},\ \bibinfo {pages} {2746} (\bibinfo {year} {1986})}\BibitemShut
	{NoStop}%
	\bibitem [{\citenamefont {Chen}\ \emph {et~al.}(2016)\citenamefont {Chen},
		\citenamefont {Fukushima}, \citenamefont {Huang},\ and\ \citenamefont
		{Mameda}}]{Chen:2015hfc}%
	\BibitemOpen
	\bibfield  {author} {\bibinfo {author} {\bibfnamefont {H.-L.}\ \bibnamefont
			{Chen}}, \bibinfo {author} {\bibfnamefont {K.}~\bibnamefont {Fukushima}},
		\bibinfo {author} {\bibfnamefont {X.-G.}\ \bibnamefont {Huang}},\ and\
		\bibinfo {author} {\bibfnamefont {K.}~\bibnamefont {Mameda}},\ }\href
	{https://doi.org/10.1103/PhysRevD.93.104052} {\bibfield  {journal} {\bibinfo
			{journal} {Phys. Rev. D}\ }\textbf {\bibinfo {volume} {93}},\ \bibinfo
		{pages} {104052} (\bibinfo {year} {2016})},\ \Eprint
	{https://arxiv.org/abs/1512.08974} {arXiv:1512.08974 [hep-ph]} \BibitemShut
	{NoStop}%
	\bibitem [{\citenamefont {Wang}\ and\ \citenamefont
		{Shovkovy}(2021{\natexlab{b}})}]{Wang:2021eud}%
	\BibitemOpen
	\bibfield  {author} {\bibinfo {author} {\bibfnamefont {X.}~\bibnamefont
			{Wang}}\ and\ \bibinfo {author} {\bibfnamefont {I.}~\bibnamefont
			{Shovkovy}},\ }\href {https://doi.org/10.1140/epjc/s10052-021-09650-3}
	{\bibfield  {journal} {\bibinfo  {journal} {Eur. Phys. J. C}\ }\textbf
		{\bibinfo {volume} {81}},\ \bibinfo {pages} {901} (\bibinfo {year}
		{2021}{\natexlab{b}})},\ \Eprint {https://arxiv.org/abs/2106.09029}
	{arXiv:2106.09029 [nucl-th]} \BibitemShut {NoStop}%
	\bibitem [{\citenamefont {Wang}\ and\ \citenamefont
		{Shovkovy}(2024{\natexlab{b}})}]{Wang:2023fst}%
	\BibitemOpen
	\bibfield  {author} {\bibinfo {author} {\bibfnamefont {X.}~\bibnamefont
			{Wang}}\ and\ \bibinfo {author} {\bibfnamefont {I.~A.}\ \bibnamefont
			{Shovkovy}},\ }\href {https://doi.org/10.1103/PhysRevD.109.056008} {\bibfield
		{journal} {\bibinfo  {journal} {Phys. Rev. D}\ }\textbf {\bibinfo {volume}
			{109}},\ \bibinfo {pages} {056008} (\bibinfo {year} {2024}{\natexlab{b}})},\
	\Eprint {https://arxiv.org/abs/2307.07557} {arXiv:2307.07557 [hep-ph]}
	\BibitemShut {NoStop}%
	\bibitem [{\citenamefont {Fousse}\ \emph {et~al.}(2007)\citenamefont {Fousse},
		\citenamefont {Hanrot}, \citenamefont {Lefèvre}, \citenamefont
		{Pélissier},\ and\ \citenamefont {Zimmermann}}]{Fousse2007}%
	\BibitemOpen
	\bibfield  {author} {\bibinfo {author} {\bibfnamefont {L.}~\bibnamefont
			{Fousse}}, \bibinfo {author} {\bibfnamefont {G.}~\bibnamefont {Hanrot}},
		\bibinfo {author} {\bibfnamefont {V.}~\bibnamefont {Lefèvre}}, \bibinfo
		{author} {\bibfnamefont {P.}~\bibnamefont {Pélissier}},\ and\ \bibinfo
		{author} {\bibfnamefont {P.}~\bibnamefont {Zimmermann}},\ }\href
	{https://doi.org/10.1145/1236463.1236468} {\bibfield  {journal} {\bibinfo
			{journal} {ACM Transactions on Mathematical Software}\ }\textbf {\bibinfo
			{volume} {33}},\ \bibinfo {pages} {13} (\bibinfo {year} {2007})},\ \bibinfo
	{note} {\url{https://www.mpfr.org/}}\BibitemShut {NoStop}%
	\bibitem [{\citenamefont {Voloshin}(2010)}]{Voloshin:2010ut}%
	\BibitemOpen
	\bibfield  {author} {\bibinfo {author} {\bibfnamefont {S.~A.}\ \bibnamefont
			{Voloshin}},\ }\href {https://doi.org/10.1103/PhysRevLett.105.172301}
	{\bibfield  {journal} {\bibinfo  {journal} {Phys. Rev. Lett.}\ }\textbf
		{\bibinfo {volume} {105}},\ \bibinfo {pages} {172301} (\bibinfo {year}
		{2010})},\ \Eprint {https://arxiv.org/abs/1006.1020} {arXiv:1006.1020
		[nucl-th]} \BibitemShut {NoStop}%
	\bibitem [{\citenamefont {Abdallah}\ \emph {et~al.}(2022)\citenamefont
		{Abdallah} \emph {et~al.}}]{STAR:2021mii}%
	\BibitemOpen
	\bibfield  {author} {\bibinfo {author} {\bibfnamefont {M.}~\bibnamefont
			{Abdallah}} \emph {et~al.} (\bibinfo {collaboration} {STAR}),\ }\href
	{https://doi.org/10.1103/PhysRevC.105.014901} {\bibfield  {journal} {\bibinfo
			{journal} {Phys. Rev. C}\ }\textbf {\bibinfo {volume} {105}},\ \bibinfo
		{pages} {014901} (\bibinfo {year} {2022})},\ \Eprint
	{https://arxiv.org/abs/2109.00131} {arXiv:2109.00131 [nucl-ex]} \BibitemShut
	{NoStop}%
	\bibitem [{\citenamefont {Li}\ \emph {et~al.}(2019)\citenamefont {Li},
		\citenamefont {Zhou},\ and\ \citenamefont {Zhou}}]{Li:2019yzy}%
	\BibitemOpen
	\bibfield  {author} {\bibinfo {author} {\bibfnamefont {C.}~\bibnamefont
			{Li}}, \bibinfo {author} {\bibfnamefont {J.}~\bibnamefont {Zhou}},\ and\
		\bibinfo {author} {\bibfnamefont {Y.-J.}\ \bibnamefont {Zhou}},\ }\href
	{https://doi.org/10.1016/j.physletb.2019.07.005} {\bibfield  {journal}
		{\bibinfo  {journal} {Phys. Lett. B}\ }\textbf {\bibinfo {volume} {795}},\
		\bibinfo {pages} {576} (\bibinfo {year} {2019})},\ \Eprint
	{https://arxiv.org/abs/1903.10084} {arXiv:1903.10084 [hep-ph]} \BibitemShut
	{NoStop}%
	\bibitem [{\citenamefont {Buzzegoli}\ \emph {et~al.}(2026)\citenamefont
		{Buzzegoli}, \citenamefont {Busuioc}, \citenamefont {Kroth}, \citenamefont
		{Vijayakumar},\ and\ \citenamefont {Tuchin}}]{Buzzegoli:2026sji}%
	\BibitemOpen
	\bibfield  {author} {\bibinfo {author} {\bibfnamefont {M.}~\bibnamefont
			{Buzzegoli}}, \bibinfo {author} {\bibfnamefont {S.}~\bibnamefont {Busuioc}},
		\bibinfo {author} {\bibfnamefont {J.~D.}\ \bibnamefont {Kroth}}, \bibinfo
		{author} {\bibfnamefont {N.}~\bibnamefont {Vijayakumar}},\ and\ \bibinfo
		{author} {\bibfnamefont {K.}~\bibnamefont {Tuchin}},\ }\href
	{https://doi.org/10.48550/arXiv.2602.13044} {\bibfield  {journal} {\bibinfo
			{journal} {arXiv pre-print}\ ,\ \bibinfo {pages} {13044}} (\bibinfo {year}
		{2026})},\ \Eprint {https://arxiv.org/abs/2602.13044} {arXiv:2602.13044
		[hep-ph]} \BibitemShut {NoStop}%
\end{thebibliography}

\end{document}